\documentclass[letterpaper,titlepage,11pt]{article}

\usepackage[T1]{fontenc}
\usepackage[utf8]{inputenc}
\usepackage{lmodern}
\usepackage{enumerate}
\usepackage[english]{babel}
\usepackage{comment}
\usepackage[numbers,sort&compress]{natbib}
\usepackage{bm}
\usepackage[usenames,dvipsnames]{xcolor}
\usepackage[a4paper]{geometry}

\usepackage[colorlinks=true,urlcolor=blue,citecolor=magenta]{hyperref}
\usepackage{amsmath}
\usepackage{amsfonts}
\usepackage{amssymb}
\usepackage{graphicx}
\usepackage[dvipsnames]{xcolor}
\usepackage{mathtools}
\usepackage{simplewick}
\usepackage{hyperref}
\usepackage{multirow}
\usepackage{enumitem}
\usepackage{physics}
\usepackage[compat=1.1.0]{tikz-feynman}
\usepackage{subcaption}
\usepackage{booktabs}
\usepackage{xcolor}

\usepackage{lipsum}

\allowdisplaybreaks[1]

\newcommand{\CJ}{\mathcal{J}}
\newcommand{\CM}{\mathcal{M}}

\newcommand{\CG}{\mathcal{G}}
\newcommand{\CN}{\mathcal{N}}

\newcommand{\CY}{\mathcal{Y}}

\def\mN{\mathcal{N}}

\def\mM{\mathcal{M}}
\def\mO{\mathcal{O}}
\def\mW{\mathcal{W}}
\def\dalpha{\dot{\alpha}}
\def\dbeta{\dot{\beta}}

\newcommand\blfootnote[1]{%
  \begingroup
  \renewcommand\thefootnote{}\footnote{#1}%
  \addtocounter{footnote}{-1}%
  \endgroup
}

\newcommand{\SU}{\text{SU}}
\newcommand{\U}{\text{U}}
\newcommand{\ds}{\displaystyle}

\newcommand{\spa}{\ , \ \ }
\newcommand{\nn}{\nonumber}
\newcommand{\psd}{\psi^\dagger}
\newcommand{\ra}[1]{\renewcommand{\arraystretch}{#1}}
\def\mba{\mathbf{a}}
\def\mbb{\mathbf{b}}
\def\mbc{\mathbf{c}}
\usepackage{etoolbox}
\def\mfm{\mathfrak{m}}

\newcommand{\beq}{\begin{equation}}
\newcommand{\eeq}{\end{equation}}
\newcommand{\bea}{\begin{eqnarray}}
\newcommand{\eea}{\end{eqnarray}}

\def\le{\left(}
\def\ri{\right)}

\hypersetup{   colorlinks=true,  citecolor=blue, linkcolor=blue, urlcolor=red }

\begin{document}

\numberwithin{equation}{section}

\begin{titlepage}
\rightline{\vbox{   \phantom{ghost} }}

 \vskip 1.8 cm
\begin{center}
{\LARGE \bf
Symmetry structure of the interactions in
\\[2mm]
near-BPS corners of  ${\cal N} = 4$ super-Yang-Mills}
\end{center}
\vskip 1 cm

\title{}
\date{\today}
\author{Stefano Baiguera}
\author{Troels Harmark}
\author{Yang Lei}
\author{Nico Wintergerst}

\centerline{\large {{\bf Stefano Baiguera$^1$, Troels Harmark$^1$,}}}
\vskip 0.2cm
\centerline{\large {{\bf Yang Lei$^{1,2,3}$, Nico Wintergerst$^1$}}}

\vskip 1.0cm

\begin{center}
\sl ${}^1$The Niels Bohr Institute, University of Copenhagen,\\
Blegdamsvej 17, DK-2100 Copenhagen \O, Denmark\\[1mm]
\sl ${}^2$Mandelstam Institute for Theoretical Physics, School of Physics, NITheP, and CoE-MaSS, \\
University of the Witwatersrand, 1 Jan Smuts Avenue, Johannesburg, South Africa \\[1mm]
\sl ${}^3$ Kavli Institute for Theoretical Sciences (KITS), \\
University of Chinese Academy of Sciences, 100190 Beijing, P.R.~China
\end{center}

\vskip 1.3cm \centerline{\bf Abstract} \vskip 0.2cm \noindent
We consider limits of ${\cal N} = 4$ super-Yang-Mills (SYM) theory that approach BPS bounds. 
These limits result in non-relativistic near-BPS theories that describe the effective dynamics near the BPS bounds and upon quantization are known as Spin Matrix theories. The near-BPS theories can be obtained by reducing ${\cal N}=4$ SYM on a three-sphere and integrating out the fields that become non-dynamical in the limits. 
We perform the sphere reduction for the near-BPS limit with $\SU(1,2|2)$ symmetry, which has several new features compared to the previously considered cases with $\SU(1,1)$ symmetry, including a dynamical gauge field. We discover a new structure in the classical limit of the interaction term. We show that the interaction term is built from certain blocks that comprise an irreducible representation of the $\SU(1,2|2)$ algebra. Moreover, the full interaction term can be interpreted as a norm in the linear space of this representation, explaining its features including the positive definiteness. This means one can think of the interaction term as a distance squared from saturating the BPS bound. The $\SU(1,1|1)$ near-BPS theory, and its subcases, is seen to inherit these features. These observations point to a way to solve the strong coupling dynamics of these near-BPS theories. 

\blfootnote{ \scriptsize{ \texttt{stefano.baiguera@nbi.ku.dk, harmark@nbi.ku.dk, leiyang@ucas.ac.cn, nwintergerst@gmail.com}} }

\end{titlepage}
\newpage
\tableofcontents

\section{Introduction}

A major role in the development of theoretical physics in the last decades has been played by the AdS/CFT correspondence, the most studied example being the duality between $\mathcal{N}=4$ supersymmetric Yang-Mills (SYM) theory with gauge group $\mathrm{SU}(N)$ and type IIB string theory on $\mathrm{AdS}_5 \times S^5 $ \cite{Maldacena:1997re}.
Many advances towards a quantitative understanding of the conjecture have been made, in particular in the planar limit where the existence of an integrability structure described by spin chains allows to link the two sides of the duality \cite{Minahan:2008hf, Beisert:2003tq}.
However, this approach has the drawback of being completely blind to non-perturbative phenomena, like black holes and D-branes, which are essential for a deeper knowledge of the strongly coupled regime of gravity.

In order to overcome the limitations of the planar limit, one possibility is to approach specific BPS bounds of $\mathcal{N}=4$ SYM, where the effective dynamics becomes non-relativistic and is described upon quantization by a so-called Spin Matrix Theory (SMT) \cite{Harmark:2014mpa}. 
These models generalize the nearest-neighbour spin chain describing the planar limit, while at the same time introducing sufficient simplifications to handle computations at finite number $N$ of colors.
The realm of SMTs has been classified: the discriminating factor between them is the spin symmetry group which depends on the details of the BPS bound under consideration \cite{Harmark:2007px}.
The non-relativistic traits of SMT are encoded by various phenomena, {\sl{e.g.}} the existence of a global $\mathrm{U}(1)$ symmetry associated with particle number conservation. 
From the bulk perspective, they manifest through the target state of the dual string theory description, since it is described by a non-Lorentzian geometry \cite{Harmark:2008gm, Harmark:2017rpg, Harmark:2018cdl, Harmark:2019upf}.
The relation between SMTs from the field theory side and sigma models from the gravity side was recently explored in the Penrose limit \cite{Harmark:2020vll}.

A novel approach to obtain an effective description of the degrees of freedom surviving the near-BPS limits in terms of the Hamiltonian of a lower-dimensional field theory was found in \cite{Harmark:2019zkn}, and then 
pursued in \cite{Baiguera:2020jgy} for all the cases containing $\mathrm{SU}(1,1)$ as a subset of the spin group.
The technique is the following: we put classical $\mathcal{N}=4$ SYM on $\mathbb{R} \times S^3$ and we take a decoupling limit by zooming in towards a BPS bound of the CFT. 
Then we give a prescription to quantize the theory, which is sufficient to extract self-energy corrections only from normal ordering; the result is then identified as a SMT.  
Contrarily to previous methods \cite{Beisert:2003jj, Beisert:2004ry}, this procedure does not require any computation of loop corrections to the dilatation operator; both the techniques give the same result in all the cases considered.

In this paper, we perform another step in the analysis of the existing SMTs by considering more general BPS bounds, of the form
\beq
E \geq S_1 + S_2 + \sum_{i=1}^3 \omega_i Q_i \, ,
\eeq
with $E$ being the energy, $S_1, S_2,$ the angular momenta on the three-sphere, $Q_i$ the Cartan charges associated to $\SU(4)$ R-symmetry and $\omega_i$ the chemical potentials identifying the bound.

The main novelty with respect to the cases considered in \cite{Baiguera:2020jgy} is the appearance of the rotation generator $S_2,$ which was instead turned off in the $\SU(1,1)$ limits; its presence will allow the existence of dynamical gauge degrees of freedom in the near-BPS limit, which is performed in the following way:
\beq
\lambda \rightarrow 0 \, , \qquad
\frac{1}{\lambda} \le E -  S_1 - S_2 - \sum_{i=1}^3 \omega_i Q_i \ri  \,\,\,
\mathrm{finite} \, , \qquad N \,\,\, \mathrm{fixed} \, ,
\eeq
where $\lambda$ is the 't Hooft coupling.

We consider the classical sphere reduction for two near-BPS limits, giving rise to interactions with either $\SU(1,2) \times \mathrm{U}(1)$ or $\SU(1,2|2) \times \mathrm{U}(1)$ invariance.
The first case is characterized by the choice $(\omega_1, \omega_2, \omega_3)=(0,0,0),$ which means that R-symmetry generators do not contribute at all.
For this reason, we will find that the only dynamical degree of freedom is encoded by a field strength.
It is the simplest example where the gauge field plays an active role in the dynamics other than mediating effective interactions.
The second case corresponds to $(\omega_1, \omega_2, \omega_3)=(1,0,0)$\footnote{The ground state of this subsector is also called $\frac{1}{8}$-BPS subsector, and is described by dual giant gravitons \cite{Mandal:2006tk}. 
Other $\frac{1}{8}$-BPS subsectors, which can be identified as Spin Matrix theory decoupling limits, include the SU$(2|3)$  \cite{Beisert:2003ys,Kinney:2005ej} and the PSU$(1,1|2)$ subsectors \cite{Beisert:2007sk,Choi:2018hmj}.}, which allows for the existence of an additional dynamical scalar field and two fermions, giving rise to several new interaction terms.
There are other two relevant near-BPS limits in the same class that one can consider, with symmetry groups $\SU(1,2|1)$ or $\mathrm{P}\SU(1,2|3);$ the former is a subsector of the $\SU(1,2|2)$ case, while we reserve a treatment of the latter for future work. 

Remarkably, it is possible to show that the classical interacting Hamiltonian in such near-BPS limits can be written in a compact way after defining certain fundamental blocks that naturally arise in the sphere reduction method.
These blocks are quadratic in the surviving modes of the theory, and contain information about the angular momenta on $S^3$ through three dependences: an explicit prefactor, the mode expansion of the fields and a $\SU(2)$ Clebsch-Gordan coefficient. 

The transformations of the fundamental blocks under the generators of the spin group show that they comprise an irreducible representation of the $\SU(1,2|2)$ algebra.
We identify the highest weight of such representations and we derive all the other blocks by acting on it with lowering operators. 
In retrospective, we observe that there is a natural algebraic derivation of the similar blocks that were discovered in the $\SU(1,1)$ limits \cite{Baiguera:2020jgy}.
The interactions are written in a manifestly positive-definite form, which also extends the same pattern already observed in the $\SU(1,1)$ limits.
In the present work, we find an interpretaton of the result as a norm in the linear space of the representation identified by the fundamental blocks.
This procedure gives a natural normalization for all the blocks, and can be thought as a way to measure a distance from the saturation of the BPS bound, where the interactions would vanish.

The fundamental blocks in the SU$(1,2|2)$ subsector which we are deriving in this work can be found  to be consistent with earlier work \cite{Grant:2008sk}. 
However, several aspects of our work are worth emphasing.
First of all, our method is able to incorporate all the fermionic interactions of $\mN=4$ SYM theory, which were not explored in \cite{Grant:2008sk}. 
This results in an explicit expression for the Hamiltonian which completely describes the near-BPS dynamics in this $\frac{1}{8}$-BPS subsector.
Moreover, our analysis shows that all the blocks, which correspond to constraints satisfied by BPS states in \cite{Grant:2008sk}, are transforming in specific representations of an $\mN=2$ vector multiplet. 
We expect that this discovery can provide us with novel insights into understanding the BPS operators and their corresponding near-BPS dynamics.

The paper is organized as follows.
In Section \ref{sec:symm_su111_near_BPS_theory} we start from the interacting Hamiltonian of the $\SU(1,1|1)$ near-BPS limit and we show that the decomposition into blocks naturally emerges from an algebraic analysis; then we interpret the result as a norm.
In Section \ref{sect-sphere_red_examples} we perform the classical reduction on the three-sphere to obtain the interacting Hamiltonian for the $\SU(1,2)$ and $\SU(1,2|2)$ near-BPS limits.
We identify the fundamental blocks of the sector, which are summarized at the end of the section.
The reader primarily interested in the symmetry structure of the interactions can jump to Section \ref{sec:Su12-symmetry-nov}, where we interpret the blocks from an algebraic perspective, and we show that the Hamiltonian is invariant under all the symmetries of the theory.
We conclude and discuss future directions in Section \ref{sec:conclusions}.

Technical details are reserved for the appendices: in Appendix \ref{app:CG_properties} we set the notation for sphere reduction and we outline the explicit procedure to solve the sums over Clebsch-Gordan coefficients.
The discussion on the algebra is performed in Appendix \ref{app:su111} for the $\SU(1,1|1)$ case, and in Appendix \ref{sec:SU12algebra} for the $\SU(1,2)$ case and its extensions.

\section{Symmetry of $\SU(1,1|1)$ near-BPS theory}
\label{sec:symm_su111_near_BPS_theory}

In this section we show that one can reinterpret the interacting part of the classical Hamiltonian of the $\SU(1,1|1)$ near-BPS theory found in \cite{Baiguera:2020jgy} in terms of a norm on the linear space of an infinite-dimensional representation of the $\SU(1,1|1)$ algebra. This provides a natural framework that can explain why the interaction has its particular form, including that it is invariant with respect to any $\SU(1,1|1)$ transformations, as well as the relative normalization of the terms in the interaction and the fact that it is positive definite. We begin by briefly reviewing the $\SU(1,1|1)$ near-BPS theory \cite{Baiguera:2020jgy} in Section \ref{sec:su111_theory}, exhibiting its symmetry properties in Section \ref{sec:su111_symmetry} and finally interpretating it as a norm in Section \ref{sec:su111_norm}. One can think of this as a warm-up to the $\SU(1,2|2)$ case considered in Section \ref{sec:Su12-symmetry-nov}, where we show that one can make an analogous interpretation.

\subsection{$\SU(1,1|1)$ near-BPS theory}
\label{sec:su111_theory}

In \cite{Baiguera:2020jgy} the BPS bound 
\begin{equation}
\label{bpsbound_su111}
E \geq S_1 + Q_1 + \frac{1}{2} (Q_2+Q_3) \,,
\end{equation}
was considered.
Starting with the classical Hamiltonian $H$ of $\CN=4$ SYM on a three-sphere, one can perform the limit 
\begin{equation}
\label{Hint_limit_su111}
H_{\rm int} = \lim_{\lambda\rightarrow 0} \frac{H - S_1 - Q_1 - \frac{1}{2} (Q_2+Q_3)}{\lambda} \,,
\end{equation}
giving the classical Hamiltonian \cite{Baiguera:2020jgy}
\begin{equation}
H_{\rm limit} = H_0 + \tilde{g}^2 H_{\rm int} \,,
\end{equation}
with its quadratic part
\begin{equation}
H_0 = \sum_{n=0}^\infty \left[ ( n+1) \tr ( \Phi_n^\dagger \Phi_n ) + \left( n+\frac{3}{2}\right) \tr(\psi_n^\dagger \psi_n ) \right] \,,
\end{equation}
where $\Phi_n$ is complex and $\psi_n$ is Grassmann-valued.
One has the associated Dirac (anti)brackets
\begin{equation}
\label{dirac_brackets}
\Big\{ (\Phi_n)^i{}_j , (\Phi_{n'}^\dagger)^k{}_l \Big\}_D = i \delta_{n,n'} \delta^i _l \delta ^k _ j
\spa \Big\{ (\psi_n)^i{}_j , (\psi_{n'}^\dagger)^k{}_l \Big\}_D = i \delta_{n,n'} \delta^i _l \delta ^k _ j \,.
\end{equation}
The quartic interaction $H_{\rm int}$ is given by
\begin{equation}
\label{su111_interaction}
H_{\rm int} = \frac{1}{2N} \sum_{l=1}^\infty \frac{1}{l} \tr ( \hat{q}_l^\dagger \hat{q}_l ) + \frac{1}{2N} \sum_{l=0}^\infty \tr ( F_l^\dagger F_l ) \,,
\end{equation}
where we defined the blocks
\begin{equation}
\label{hatql}
\hat{q}_l^\dagger = \sum_{n=0}^\infty \left( [ \Phi_{n+l}^\dagger , \Phi_{n} ] + \frac{\sqrt{n+1}}{\sqrt{n+l+1}} \{ \psi^\dagger_{n+l},\psi_{n}\} \right) \,,
\end{equation}
and
\begin{equation}
\label{Fl}
F_l^\dagger = \sum_{n=0}^\infty \frac{1}{\sqrt{n+l+1}}  [ \Phi_n , \psi_{n+l}^\dagger] \,,
\end{equation}
which are $N\times N$ matrices.

Note that
\begin{equation}
\label{hatq0}
\hat{q}_0^\dagger =\hat{q}_0 = \sum_{n=0}^\infty \left( [ \Phi_{n}^\dagger , \Phi_{n} ] +  \{ \psi^\dagger_{n},\psi_{n}\} \right) = 0 \,,
\end{equation}
thanks to the Gauss constraint. Hence we have set to zero the terms in $H_{\rm int}$ found in \cite{Baiguera:2020jgy} from sphere reduction of $\CN=4$ SYM that include $\hat{q}_0$.

\subsection{Realization of $\SU(1,1|1)$ symmetry in interaction}
\label{sec:su111_symmetry}

We can use Dirac (anti)brackets \eqref{dirac_brackets} to represent infinitesimal symmetry transformations in the classical theory. For instance, the number operator is
\begin{equation}
\hat{N} = \sum_{n=0}^\infty \left[  \tr ( \Phi_n^\dagger \Phi_n ) +  \tr(\psi_n^\dagger \psi_n ) \right] \,.
\end{equation}
One can check that $\{ \hat{N}, H_{\rm limit} \}_D = 0$ which means that $\hat{N}$ is conserved. This also means that the $\SU(1,1|1)$ near-BPS theory is non-relativistic, since the $\mathrm{U}(1)$ conservation in absence of antiparticles is incompatible with Poincaré symmetry.

The $\SU(1,1|1)$ symmetry generators are as follows. We have the $\SU(1,1)$ generators
\begin{eqnarray}
\label{eq:representation_L0_on_modes_su111}
 L_0 &=& \tr \sum_{n=0}^{\infty} \left[ \le n + \frac{1}{2} \ri |\Phi_{n} |^2 + \le n+1 \ri  |\psi_n|^2 \right] \, ,   \\ \ds
 L_+ &=& (L_-)^{\dagger} = \tr \sum_{n=0}^{\infty} \left[ \le n+1 \ri \Phi^{\dagger}_{n+1} \Phi_n + \sqrt{(n+1)(n+2)} \psi^{\dagger}_{n+1} \psi_n \right] \, ,
\label{eq:representation_L+-_on_modes_su111}
\end{eqnarray}
the $\U(1)$ generator
\begin{equation}
R = \sum_{n=0}^\infty \frac{1}{2} \tr ( \Phi_n^\dagger \Phi_n ) \, ,
\end{equation}
and the fermionic generators
\begin{equation}
\label{ferm_gen_su111}
Q = \sum_{n=0}^\infty \sqrt{n+1} \, \tr ( \Phi_n^\dagger \psi_n )
\spa
S = \sum_{n=0}^\infty \sqrt{n+1} \, \tr ( \psi_n^\dagger \Phi_{n+1} ) \, .
\end{equation}
We will show below that for any of the above generators $\CG$ we have
\begin{equation}
\label{sym_hint}
\{ \CG , H_{\rm int}\}_D =0 \,,
\end{equation}
which means the interaction term $H_{\rm int}$ is invariant under the $\SU(1,1|1)$ symmetry.
The above $\SU(1,1|1)$ generators obey the $\SU(1,1|1)$ algebra \eqref{su111_algebra} written in Appendix \ref{app:su111}, provided one translates the (anti)commutator between generators $[\CG_1,\CG_2\}=\CG_3$ to Dirac brackets $\{ \CG_1,\CG_2 \}_D = i \CG_3$.

The key to showing \eqref{sym_hint} is to consider the action of the $\SU(1,1|1)$ generators on the blocks \eqref{hatql} and \eqref{Fl}. We introduce here a new notation for these blocks
\begin{equation}
\label{normblock_su111}
(B_0)_l = \frac{1}{\sqrt{l}} \hat{q}_l \spa (B_1)_l = F_l \,,
\end{equation}
where we note that $l\geq 1$ for $(B_0)_l$ while $l\geq 0$ for $(B_1)_l$.
The purpose of this redefinition will be explained below. 
The interaction \eqref{su111_interaction} can now be written
\begin{equation}
\label{H_int_block}
H_{\rm int} = \frac{1}{2N} \left[  \sum_{l=1}^\infty \tr ( (B_0^{\dagger})_l (B_0)_l ) +  \sum_{l=0}^\infty \tr ( (B_1^{\dagger})_l (B_1)_l ) \right] = \frac{1}{2N} \sum_{I,l} \tr ( (B_I^{\dagger})_l (B_I)_l ) \,,
\end{equation}
where the second expression is a short hand notation for the first, with the implicit understanding that $l$ sums from 1 to $\infty$ for $I=0$ and from 0 to $\infty$ for $I=1$.

With $M$ in the linear space of blocks $(B_I^{\dagger})_l$ or $(B_I)_l ,$ we write the
 action of an $\SU(1,1|1)$ generator $\CG$  as
\begin{equation}
\label{general_brac}
\{ \CG , M^j {}_k \}_D = i (\CG_D M)^j{}_k \,.
\end{equation}
We note that the properties of the Dirac brackets give
\begin{equation}
\label{dagger2}
( \CG_D M )^\dagger = - \CG_D^\dagger M^\dagger \,.
\end{equation}
We want to show that $(B_a^{\dagger})_l$ corresponds to an irreducible representation of $\SU(1,1|1)$. To this end, we consider the action of the raising operators associated with the chosen Dynkin diagram. The Dynkin diagram is
\begin{align}
\textstyle \bigotimes \!\!-\!\!\!-\!\! \textstyle \bigotimes
\end{align}
The associated raising operators for these roots are $Q$ and $S$, respectively, and the Dynkin labels are
given by eigenvalues of the Cartan generators $L_0$ and $R$ as
\begin{equation}
[ L_0 + R , L_0 -R ] 
\end{equation}
As reviewed in Appendix \ref{app:su111} the free spectrum of $\SU(1,1|1)$ is in the irreducible representation with the Dynkin labels of the highest weight state being $[1,0]$, corresponding to $L_0=R=\frac{1}{2}$.

We compute
\begin{equation}\label{eq:actiononblocks-SU111}
\begin{array}{c}
(L_0)_D (B_0^{\dagger})_l =  l (B_0^{\dagger})_l \spa R_D (B_0^{\dagger})_l= 0 \, ,
\\[3mm]
(L_0)_D (B_1^{\dagger})_l=  \left( l + \frac{1}{2} \right) (B_1^{\dagger})_l \spa R_D (B_1^{\dagger})_l =   \frac{1}{2}  (B_1^{\dagger})_l \,.
\end{array}
\end{equation}
Thus, the $(B_0^{\dagger})_l$ block has eigenvalues $L_0=l$ and $R=0 ,$ while the $(B_1^{\dagger})_l$ block has eigenvalues $L_0=l+\frac{1}{2}$ and $R=-\frac{1}{2}$.

For the raising operators $Q$ and $S$ we compute
\begin{equation}
\label{su111_blocks_raising}
Q_D (B_0^{\dagger})_l = 0 \spa S_D (B_0^{\dagger})_l = -  \sqrt{l}\, (B_1^{\dagger})_{l-1} \spa Q_D (B_1^{\dagger})_l = - \sqrt{l} \, (B_0^{\dagger})_l \spa S_D (B_1^{\dagger})_l = 0 \,.
\end{equation}
This means that the block $(B_1^{\dagger})_0$ can be interpreted as a highest weight state with eigenvalues $L_0=\frac{1}{2}$, $R=-\frac{1}{2}$ and with corresponding Dynkin labels $[0,1]$. 
The lowering operators $Q^\dagger$ and $S^\dagger$ act as
\begin{equation}
\label{su111_blocks_lowering}
Q^\dagger_D (B_0^{\dagger})_l = -  \sqrt{l}\, (B_1^{\dagger})_l \spa S^\dagger_D (B_0^{\dagger})_l= 0 \spa Q^\dagger_D (B_1^{\dagger})_l= 0 \spa S^\dagger_D (B_1^{\dagger})_l = - \sqrt{l+1} \, (B_0^{\dagger})_{l+1} \,.
\end{equation}
This shows that one can generate all the blocks $(B_0^{\dagger})_l$ and $(B_1^{\dagger})_l$ by repeated use of the lowering operators on the highest weight block $(B_1^{\dagger})_0$. 
Therefore, the blocks $(B_0^{\dagger})_l$ and $(B_1^{\dagger})_l$ comprise an infinite-dimensional irreducible representation of $\SU(1,1|1)$ with highest weight $[0,1]$.

We now show that the interaction \eqref{H_int_block} is invariant when acting with any of the generators $\SU(1,1|1)$ listed in Eqs.~\eqref{eq:representation_L0_on_modes_su111}-\eqref{ferm_gen_su111} or their hermitian conjugates.
In general, we have for a generator $\CG$
\begin{equation}
\label{action_G_H_int}
 \CG_D \sum_{I,l} \tr ( (B^\dagger_I)_l  (B_I)_l  )
= \sum_{I,l} \Big[ \tr ( ( \CG_D (B^\dagger_I)_l  ) (B_I)_l  ) + s \tr (  (B^\dagger_I)_l  \, \CG_D (B_I)_l  ) \Big]  \,,
\end{equation}
where $s=1$ unless $\CG$ is a fermionic operator and $(B^\dagger_I)_l$ is a Grassmann-valued block, in which case $s=-1$. We can find $\CG_D (B_I)_l$ using \eqref{dagger2} 
\begin{equation}
\CG_D (B_I)_l = - ( \CG_D^\dagger (B^\dagger_I)_l ) ^\dagger \,.
\end{equation}
Consider the action of $Q^\dagger$
\begin{align}
\begin{split}
&Q_D^\dagger \sum_{I,l} \tr ( (B^\dagger_I)_l  (B_I)_l  ) = \sum_{l=1}^\infty \text{tr} \Big[ \Big(Q_D^\dagger (B_0^{\dagger})_l \Big) (B_0)_l \Big]  - \sum_{l=0}^\infty \text{tr} \Big[ (B_1^{\dagger})_l  \Big(Q_D^\dagger  (B_1)_l \Big)  \Big] \\
&=  \sum_{l=1}^\infty \text{tr} \Big[ -\sqrt{l}  (B_1^{\dagger})_l   (B_0)_l + (B_1^{\dagger})_l   \Big(\sqrt{l}  (B_0)_l\Big)   \Big]  =0 \,,
\end{split}
\end{align}
where we used \eqref{su111_blocks_raising} and \eqref{su111_blocks_lowering}.
Similarly for $S^\dagger$
\begin{align}\label{eq:SU111-sACTION}
\begin{split}
&S_D^\dagger \sum_{I,l} \tr ( (B^\dagger_I)_l  (B_I)_l  )= \sum_{l=1}^\infty \text{tr} \Big[  (B_0^{\dagger})_l  \Big(S_D^\dagger (B_0)_l \Big) \Big]  -\sum_{l=0}^\infty \text{tr} \Big[ \Big(S_D^\dagger  (B_1^{\dagger})_l  \Big) (B_1)_l  \Big] \\
&=  \sum_{l=1}^\infty \text{tr} \Big[ - (B_0^{\dagger})_l  \Big(\sqrt{l} (B_1^\dagger)_{l-1}  \Big)\Big]+  \sum_{l=0}^\infty \text{tr} \Big[ \sqrt{l+1}  (B_0^{\dagger})_{l+1}  (B_1)_l  \Big] =0 \,,
\end{split}
\end{align}
where the second term of the second line of \eqref{eq:SU111-sACTION} is the same as the first term after shifting $l \to l-1$.
One can similarly show that the actions of $Q$ and $S$ give zero. Hence, we have shown that $H_{\rm int}$ given in \eqref{H_int_block} is invariant under all four fermionic generators. One can now use the associativity of the actions of the generators, along with the $\SU(1,1|1)$ algebra \eqref{su111_algebra}, to conclude that  $H_{\rm int}$ is invariant under the action of all $\SU(1,1|1)$ generators Eqs.~\eqref{eq:representation_L0_on_modes_su111}-\eqref{ferm_gen_su111}.

\subsection{Interaction as a norm in representation space}
\label{sec:su111_norm}

In terms of the blocks, we saw above that we can write the interaction $H_{\rm int}$ as \eqref{H_int_block}. 
As we shall now see, one can interpret this interaction as a norm in the linear space of the relevant representations of $\SU(N)$ and $\SU(1,1|1)$. 
The idea in the following is that we can write the interaction $H_{\rm int}$ as
\begin{equation}
\label{Hint_norm}
H_{\rm int} = \frac{1}{2N} \langle B | B \rangle \,,
\end{equation}
where $\langle B | B \rangle$ is a norm (squared) on a particular linear space. 
In this interpretation,
\begin{equation}
 | B \rangle = \sum_{i,j} \left( \sum_{l=1}^\infty ( (B_0)_l )^i{}_j | ij ; 0,l  \rangle + \sum_{l=0}^\infty ( (B_1)_l )^i{}_j | ij ; 1,l  \rangle   \right) = \sum_{i,j} \sum_{I,l} ( (B_I)_l )^i{}_j | ij ; I,l  \rangle \,.
 \end{equation}
 Here $|ij ; I l \rangle$ are normalized and orthogonal basis vectors in the linear space given by the adjoint representation of $\SU(N)$ times the irreducible representation of the $\SU(1,1|1)$ with highest weight $[0,1]$.
The blocks $((B_I)_l )^i{}_j $ are then interpreted as components of the vector $|B\rangle$ in this linear space, {\sl i.e.}
\begin{equation}
((B_I)_l )^i{}_j  = \langle ij ; I, l  | B \rangle \spa ((B^\dagger_I)_l )^i{}_j = \langle B |  ij ; I,l  \rangle \,.
\end{equation}
One can then find the expression \eqref{H_int_block} by inserting a complete basis
\begin{equation}
\label{norm_int1}
 \langle B | B \rangle = \sum_{i,j,I,l}  \langle B |  ij ; I,l  \rangle \langle ij ; I, l  | B \rangle = \sum_{I,l} \tr ( (B^\dagger_I)_l  (B_I)_l  ) \,.
 \end{equation}
Thus, we see here that the fact that our basis is normalized and orthogonal is underlying the normalization of the expression for $H_{\rm int}$ in \eqref{H_int_block}.

Two important properties of the above norm are that 1) it is positive definite and 2) it is invariant under the symmetry transformations associated with the algebra that the linear space gives a representation of. With respect to 1) we see indeed the positive definiteness explicitly in \eqref{norm_int1}. With respect to 2) one can think of an infinitesimal symmetry transformation
\begin{equation}
|B \rangle \rightarrow |B \rangle + |\delta B\rangle \spa |\delta B\rangle =  \epsilon \CG |B\rangle \,,
\end{equation}
where $\CG$ is a symmetry generator in the algebra and $\epsilon$ is an infinitesimally small number. The invariance of the norm means%
\footnote{Note that for a fermionic operator $\CG$ the second term can acquire a minus since the block/component can be Grassmann valued, in accordance with the optional sign $s$ in \eqref{action_G_H_int}.}
\begin{equation}
\label{norm_change}
\langle \delta B |B \rangle + \langle B |\delta B \rangle = 0 \,.
\end{equation}
The explicit expression for the LHS of \eqref{norm_change} is the RHS of \eqref{action_G_H_int}. Hence, in this way we can interpret the invariance of $H_{\rm int}$ under all the generators of $\SU(1,1|1)$ as the invariance of the norm $\langle B |B\rangle$ under any rotations in the linear space on which its defined.

From the norm interpretation \eqref{Hint_norm} of the interaction term one can observe using \eqref{Hint_limit_su111} that one can think of $H_{\rm int}$ as measuring a distance between the energy $E$ and the charges $S_1 + Q_1 + \frac{1}{2} (Q_2+Q_3)$. This seems natural since $H_{\rm int}$ measures how close we are to the BPS bound \eqref{bpsbound_su111} when $\lambda$ is sufficiently small. This also means that the BPS states that saturates the bound corresponds to zero-norm states for $\langle B|B\rangle$.%
\footnote{Clearly $\langle B|B\rangle=0$ implies that the contribution to $E - (S_1 + Q_1 + \frac{1}{2} (Q_2+Q_3))$ of order $\lambda$ is zero. However, it has not been shown that  $\langle B|B\rangle=0$ implies that the higher order corrections in powers of $\lambda$ are zero. Nevertheless, this is conjectured to be the case \cite{Grant:2008sk}.}

\section{$\SU(1,2|2)$ near-BPS theory from sphere reduction}
\label{sect-sphere_red_examples}

After the review of the $\SU(1,1|1)$ sector and the new interpretation of its interactions as a norm, we move on to more general near-BPS limits by considering bounds where both the angular momenta on the three-sphere are turned on. 
Since this section is technical, the reader interested in the symmetry structure of the interactions can skip the details and jump to the end of this section, where we give the result for the interacting Hamiltonian in the $\SU(1,2|2)$ limit and we identify the fundamental blocks of the theory.
Then, one can naturally proceed to Section \ref{sec:Su12-symmetry-nov}.

In Section \ref{sect-sphere_reduction_general} we briefly review the general procedure to derive the interacting Hamiltonian for a given near-BPS limit from the sphere reduction of classical $\mathcal{N}=4$ SYM action. 
Then we apply the technique in two specific examples.
We start in Section \ref{sect-su12limit} with the $\SU(1,2)$ sector.
Since the field content surviving the decoupling limit is entirely given by the gauge fields, it is an instructive example that shows one of the main novelties with respect to the $\SU(1,1)$ limits \cite{Harmark:2019zkn, Baiguera:2020jgy}.
Then we add scalars and fermions to the theory by considering the generalization to the $\SU(1,2|2)$ case in Section \ref{sect-su122limit}.

\subsection{General procedure}
\label{sect-sphere_reduction_general}

We consider the classical action of ${\cal N} = 4$ super-Yang-Mills theory compactified on a three-sphere
\begin{multline}
\label{theS}
S = \int_{\mathbb{R} \times S^3} \sqrt{-\mathrm{det} \, g_{\mu\nu}} \, \tr \left\lbrace - \frac{1}{4} F_{\mu\nu}^2 - |D_{\mu} \Phi_a|^2 - |\Phi_a|^2 - i \psi^{\dagger}_a \bar{\sigma}^{\mu} D_{\mu} \psi^A + g \sum_{A,B,a} C_{AB}^{a} \psi^A [\Phi_a, \psi^B]
  \right. \\
\left. + g \sum_{A,B,a} \bar{C}^{aAB} \psi^{\dagger}_A [\Phi^{\dagger}_a, \psi^{\dagger}_B]
- \frac{g^2}{2} \sum_{a,b} \le |[\Phi_a, \Phi_b]|^2 + |[\Phi_a, \Phi^{\dagger}_b]|^2 \ri
 \right\rbrace \, .
\end{multline}
The conventions are the following.
We denote with $g$ the Yang-Mills coupling constant, and we canonically normalize the action on the $\mathbb{R} \times S^3$ background with the radius of the three-sphere set to unity. 
The complex scalar fields transform in the $\mathbf{6}$ representation of the R-symmetry group $ \mathrm{SO}(6)\simeq \SU(4),$ and they are defined as linear combinations of the real fields as $\Phi_a = \frac{1}{\sqrt{2}} (\phi_{2a-1} + i \phi_{2a})$ with $a \in \lbrace 1,2,3 \rbrace . $ 
The Weyl fermions $\psi^A$ with $A \in \lbrace 1,2,3,4 \rbrace$ transform in the representation $\mathbf{4}$ of  $\SU(4) .$
The field strength is defined as
\beq
F_{\mu\nu} = \partial_{\mu} A_{\nu} - \partial_{\nu} A_{\mu} + i g [A_{\mu} , A_{\nu}] \, ,
\eeq
and the covariant derivatives $D_{\mu}$ as
\bea
& D_{\mu} \Phi_a = \partial_{\mu} \Phi_a + i g [A_{\mu}, \Phi_a] \, , & \\
& D_{\mu} \psi^A = \nabla_{\mu} \psi^A + i g [A_{\mu}, \psi^A] \, , & 
\eea
where $\nabla_{\mu}$ is the covariant derivative on the three-sphere, {\sl i.e.} it contains the spin connection contribution for the fermions.
The $C^{a}_{AB}$ are Clebsch-Gordan coefficients coupling two $\mathbf{4}$  representations and one $\mathbf{6}$ representation of the R-symmetry group $\SU(4).$
The fields in the action transform in the adjoint representation of the gauge group $\SU(N)$.

A key role in all the near-BPS limits that we are going to discuss is played by the gauge field, because some of its degrees of freedom decouple on-shell and mediate an effective interaction at order $g^2$.
In this context, the auxiliary degrees of freedom are the temporal and longitudinal components of the gauge field.
In order to integrate them out, we take the Coulomb gauge
\begin{equation}
\nabla_i A^i = 0 
\end{equation}
and we consider the quadratic action for the gauge field with the addition of a generic source to keep track of the constraint structure:
\begin{equation}
S_A = \int_{\mathbb{R}\times S^3} \sqrt{-\mathrm{det} \, g_{\mu\nu}} \, \tr(-\frac{1}{4} F_{\mu\nu}^2 - A^\mu j_\mu)
\,.
\end{equation}
At this point, one computes the Hamiltonian via the Legendre transform and enforces the Coulomb gauge by means of a Lagrange multiplier. 
Significant simplifications come from expanding the fields into spherical harmonics on the three-sphere, in such a way that the constraint equations become algebraic.
Solving for the unphysical degrees of freedom, we finally obtain the unconstrained Hamiltonian\footnote{More precisely, the expression $H_A$ is the Hamiltonian density and must be integrated along the time direction in order to obtain the full Hamiltonian of the system. From now on, we will make an abuse of language by referring instead to $H_A$ as the Hamiltonian of the model, omitting the explicit integration of the modes over time.}
\begin{equation}
\label{eq:Ham_freeYM}
H_A = \tr\sum_{J,m,\tilde{m}} \left[ \sum_{\rho = \pm1}\left( \frac{1}{2} |\Pi_{(\rho)}^{Jm\tilde{m}}|^2 + \frac{1}{2} \omega_{A,J}^2 |A_{(\rho)}^{Jm\tilde{m}}|^2+ A_{(\rho)}^{Jm\tilde{m}} j_{(\rho)}^{\dagger\,Jm\tilde{m}}\right) + \frac{1}{8J(J+1)} |j_0^{Jm\tilde{m}}|^2 \right] \,.
\end{equation}
The details of the previous procedure are explained in Section 2.1 of \cite{Baiguera:2020jgy}; the relevant conventions for the decomposition into spherical harmonics are given in Appendix \ref{app:CG_properties}.

The form of the currents in Eq.~\eqref{eq:Ham_freeYM} can now straightforwardly be reconstructed from the full ${\cal N} = 4$ Hamiltonian, and all further interactions can be restored. 
We describe the general technique to extract the effective Hamiltonian in terms of the dynamical modes surviving a given near-BPS limit; then we will apply the method to each case separately.

We follow these steps:
\begin{enumerate}
	\item Identify the propagating modes in a given near-BPS limit from the quadratic classical Hamiltonian.
	\item Derive the form of the currents that couple to the gauge fields.
	\item Integrate out additional non-dynamical modes that give rise to effective interactions in a given near-BPS limit.
	\item Derive the interacting Hamiltonian by taking the limit.
\end{enumerate}
While a similar discussion was done in \cite{Baiguera:2020jgy}, here there are some important differences.
First of all, the near-BPS limits that will be considered in this work turn on both the Cartan generators $S_1, S_2$ associated to rotations on the three-sphere, in contrast with the $\SU(1,1)$ sectors, where only the former was non-vanishing.
This gives BPS bounds of the form
$H \geq S_1 + S_2 + \sum_{i=1}^3 \omega_i Q_i$ where $H$ is the Hamiltonian, $S_1, S_2$ are the angular momenta and $Q_i$, $i \in \lbrace 1,2,3 \rbrace$, are the three R-charges  of $\CN=4$ SYM on $S^3$. The coefficients $\omega_i$ in front of the R-charges are given in the Table \ref{tab:BPS_bounds}.
 
For each case, the near-BPS limit is 
\begin{equation}
\label{newnearBPSlimit}
\lambda \rightarrow 0 \quad \mbox{with} \quad \frac{H - S_1 - S_2 - \sum_{i=1}^3 \omega_i Q_i}{\lambda} \quad \mbox{fixed}  \,,
\end{equation}
and $N$ is held fixed.
We find that the surviving degrees of freedom  are described by an interacting  Hamiltonian $H_{\rm int}$ of the form
\begin{equation}
\label{full_ham_gen}
 H_{\rm int} = \lim_{g \rightarrow 0} \frac{H - S_1 - S_2 - \sum_{i=1}^3 \omega_i Q_i}{g^2 N} \,.
\end{equation}
The consequence of turning on another rotation charge is that the gauge field will possess some dynamical modes which survive the near-BPS limits and will appear in the interacting Hamiltonian, contrary to the $\SU(1,1)$ sectors.

\begin{table}\centering
	\ra{1.3}
	\begin{tabular}{@{}|c|c|c|c|@{}}\toprule[0.1em]
Sectors		& $\SU(1,2)$ & $\SU(1,2|2)$   \\ \midrule[0.05em]
		$\sum_{i=1}^3 \omega_i Q_i$  & 0  & $Q_1$ \\
		\bottomrule[0.1em]
	\end{tabular}
	\caption{List of the combinations of the R-charges  defining the limits of $\mathcal{N}=4$ SYM theory towards BPS bounds $H \geq S_1 + S_2 + \sum_{i=1}^3 \omega_i Q_i. $}
	\label{tab:BPS_bounds}
\end{table}

Now we give a general derivation of the terms of the interacting Hamiltonian mediated by the non-dynamical modes of the gauge field.
Since the gauge fields are neutral under the $\SU(4)$ R-symmetry, all the limits contain at quadratic order the combination (here $\omega_{A,J} \equiv 2J+2$)
\begin{equation}
H_0 - S_1 - S_2 = 
\sum_{J,M}  \sum_{\rho=-1,1} \frac{1}{2} \left(|\Pi_{(\rho)}^{JM} - 2i\tilde{m}A_{(\rho)}^{\dagger\,JM}|^2+ (\omega_{A,J}^2 - 4 \tilde{m}^2) |A_{(\rho)}^{JM}|^2\right)
\, ,
\label{eq:quadratic_gauge_Hamiltonian_su12sectors}
\end{equation}
which must vanish. 
This implies the constraint
\beq
\Pi^{JM}_{(\rho)} - 2 i \tilde{m} A^{\dagger \, JM}_{(\rho)} = 0 
\eeq
for all the non-dynamical\footnote{The choice $\rho=-1,|\tilde{m}|=2J+1$ corresponds exactly to the surviving dynamical modes of the gauge field in all the $\SU(1,2)$ limits. They will contribute to other effective interactions that will be considered separately.} modes of the gauge field, which means that we don't have to consider at the same time $\rho=-1$ and $|\tilde{m}|=J+1.$

If we take the Hamiltonian in Eq.~\eqref{eq:quadratic_gauge_Hamiltonian_su12sectors} and restore the sources as in Eq.~\eqref{eq:Ham_freeYM}, consistency of the constraints with time evolution implies that
\beq
\lbrace H, \Pi^{JM}_{(\rho)} - 2 i \tilde{m} A^{\dagger \, JM}_{(\rho)} \rbrace =
(\omega_{A,J}^2 - 4 \tilde{m}^2) A_{(\rho)}^{\dagger\,Jm\tilde{m}} + j_{(\rho)}^{\dagger\,Jm\tilde{m}} 
= 0 \, ,
\eeq
or equivalently
\beq
A_{(\rho)}^{Jm\tilde{m}} = -\frac{j^{ Jm\tilde{m}}_{(\rho)} }{\omega_{A,J}^2  -4\tilde{m}^2} \, .
\eeq
Putting this further constraint in the Eq.~\eqref{eq:quadratic_gauge_Hamiltonian_su12sectors} plus sources gives
\begin{equation}
H - S_1 -S_2 = \tr \left( \sum_{J,m,\tilde{m}} \frac{1}{8J(J+1)} |j_0^{Jm\tilde{m}}|^2 - \sum_{\rho=\pm1} \sum_{J,m,\tilde{m}}\frac{1}{2(\omega_{A,J}^2 - 4\tilde{m}^2)} |j_{(\rho)}^{Jm\tilde{m}}|^2\right)  \, .
\label{eq:general_gauge_mediated_interaction_su12limits}
\end{equation}
This expression is general and will be used to extract all the gauge-mediated interactions, once the currents $j_0^{JM}, j^{JM}_{(\rho)}$ are identified from the interacting Hamiltonian of $\mathcal{N}=4$ SYM.


\subsection{Dynamical gauge fields - $\SU(1,2)$}
\label{sect-su12limit}

Turning on both of the angular momenta gives rise to theories that contain remnants of the gluons of ${\cal N} = 4$ SYM. 
The specific BPS bound is $H \geq S_1 + S_2,$ giving rise to interactions with $\SU(1,2) \times \U(1)$ symmetry group.

\subsubsection*{Free Hamiltonian and reduction of degrees of freedom}

Since the limit does not involve any R-charge, it is not restrictive to consider only the gauge sector both to determine the dynamical degrees of freedom and the interactions: all the scalars and fermions are trivially forced to zero.
The relevant combination of Hamiltonian and rotation charges at quadratic level is given by Eq.~\eqref{eq:quadratic_gauge_Hamiltonian_su12sectors},
where 
\beq
\omega_{A,J} \equiv 2J+2  \, , \qquad
|m| \leq J + \frac{\rho(1+\rho)}{2} \, \qquad
|\tilde{m}| \leq J - \frac{\rho(1-\rho)}{2} \, . 
\label{eq:constraints_momenta_quadratic_su12}
\eeq
We find a set of constraints by imposing that the expression \eqref{eq:quadratic_gauge_Hamiltonian_su12sectors} vanishes.
Due to the inequalities \eqref{eq:constraints_momenta_quadratic_su12}, non-trivial solutions can be found only when 
\beq
\rho=-1 \, , \qquad
|m| \leq J \, , \qquad
\tilde{m} = \pm (J+1) \, , \qquad
J \geq 0 \, .
\eeq
We denote the two possibilities with
\begin{equation}
A_\pm^{Jm} \equiv A_{(\rho=-1)}^{J,m,\pm (J+1)}\,, \qquad
\Pi_\pm^{Jm} \equiv \Pi_{(\rho=-1)}^{J,m,\pm (J+1)}\,,
\end{equation}
obeying the non-relativistic constraint
\begin{equation}
\label{eq:su12_const}
\Pi_\pm^{Jm} \pm i\omega_{A,J}A_\pm^{\dagger\,Jm} = {\cal O}(g)\,.
\end{equation}
The other constraints are trivial, {\sl{i.e.}} when $\tilde{m} \ne \pm (J+1)$ we have
\beq
A^{J m \tilde{m}}_{(\rho)} = \mathcal{O} (g) \,,\quad\Pi_{(\rho)}^{Jm\tilde{m}} = \mathcal{O} (g) \,.
\eeq
We can explicitly find the right-hand side of the previous constraints by requiring compatibility with the Hamiltonian evolution.
It is simple to show that the only non-trivial bracket involves the currents, since the corresponding term in the Hamiltonian is linear in the gauge field.
Indeed we find  
\begin{equation}
\{H,\Pi_{(\rho)}^{Jm\tilde{m}} - 2 i \tilde{m} A_{(\rho)}^{\dagger\,Jm\tilde{m}}\} \approx  (\omega_{A,J}^2 - 4 \tilde{m}^2 ) A_{(\rho)}^{\dagger\,Jm\tilde{m}} + j_{(\rho)}^{\dagger\,Jm\tilde{m}} \,.
\end{equation}
Hence we impose the vanishing of the RHS as a constraint to have a consistent Hamiltonian evolution.
Since all the other constraints weakly commute with the Hamiltonian, we find
\begin{align}
\label{eq:constraints_gauge_fields_su12}
&A_{(\rho)}^{Jm\tilde{m}} = -\frac{1}{\omega_{A,J}^2 - 4 \tilde{m}^2} j^{ Jm\tilde{m}}_{(\rho)} \,,\quad\Pi_{(\rho)}^{Jm\tilde{m}} = 0\,,\\
&\Pi_\pm^{Jm} \pm i\omega_{A,J}A_\pm^{\dagger\,Jm} = 0 \,.
\label{eq:constraint_dynamical_gauge_fields_su12}
\end{align}
All the dynamical degrees of freedom satisfy the constraint \eqref{eq:constraint_dynamical_gauge_fields_su12}, which relates the momentum with the complex conjugate of the field.
This is a typical feature of non-relativistic theories that goes also in hand with the $\U(1)$ global symmetry responsible for the conservation of particle number.
The same phenomenon was observed in near-BPS limits with $\SU(1,1)$ subgroup  for the dynamical modes of the scalar fields \cite{Harmark:2019zkn,Baiguera:2020jgy}.

In addition, we observe that $A_+$ and $A_-$ are not independent, since the reality of the gauge field implies
\begin{equation}
A_-^{Jm}  = (-1)^{J-m}(A_{+}^{\dagger})^{J,-m}\,.
\label{eq:gauge_field_hermitian_conjugate}
\end{equation}
We can therefore eliminate $A_-^{Jm}$ and express all the results only in terms of the positive mode $A_+^{J m}$ and its hermitian conjugate.

The free part of the Hamiltonian is obtained by using the constraint  \eqref{eq:constraint_dynamical_gauge_fields_su12} inside Eq.~\eqref{eq:general_gauge_mediated_interaction_su12limits}.
However, we are allowed to perform this step only after accounting for the change of Dirac brackets that this constraint entails, which is
\begin{equation}
\{(A_{+})^{J,m}, (A_+^{\dagger})^{J',m'}\}_D = \frac{i}{2\omega_{A,J}} \delta_{J,J'} \delta_{m,m'}\,.
\label{eq:Dirac_bracket_gauge}
\end{equation}
In order to make the bracket canonical, we define
\begin{equation}
A_{Jm} \equiv \sqrt{2\omega_{A,J}} (A_+)^{J m}\,.
\label{eq:definition_gauge_modes_su12}
\end{equation}
This captures the propagating modes in the $\SU(1,2)$ limit. Note that there is a natural mapping from $A_{Jm}$ to a particular component of the gluon field strength. 
Here, we only remark that such a mapping is natural for at least two reasons: first, the energy of the lightest mode $A_{00}$ is $2$, reminiscent of the scaling dimension of a field strength; second, in the near-BPS limit, $A_{Jm}$ is explicitly gauge invariant.

After the redefinition \eqref{eq:definition_gauge_modes_su12}, we obtain
\beq
H_0 = S_1 + S_2 =
\sum_{s=0}^{\infty} \sum_{m=-\frac{s}{2}}^{\frac{s}{2}}  (s+2) \tr |A_{s,m}|^2 \, ,
\eeq
where now the sum runs only over integers $s=2J \in \mathbb{N}.$

In view of the study of Clebsch-Gordan coefficients in the interactions, we introduce the following notation for the labels of momenta:
\beq
(\hat{\mathcal{J}}, \hat{\mathcal{M}}) \equiv \le J, m,J+1, \rho=-1 \ri \, , \qquad
|m| \leq J \, .
\label{eq:saturation_momenta_gauge_su12}  
\eeq
This saturation precisely corresponds to the case where a dynamical gauge field is considered.
In addition, it is understood that every time there is an index $\hat{\CJ}$ referring to a dynamical gauge field, we take eigenvalues $(J,-m,-J-1)$ in the Clebsch-Gordan coefficient when the hermitian conjugate field $A^{\dagger}_{Jm}$ is involved.
This is a consequence of Eq.~\eqref{eq:gauge_field_hermitian_conjugate}.


\subsubsection*{Interactions}

Since $H_0 - S_1 - S_2 = 0$ by construction, we can define the interacting Hamiltonian in the near-BPS limit as
\begin{equation}
\label{Hintdef_su12}
H_\text{int} =  \lim_{g \to 0} \frac{H - S_1 - S_2}{g^2 N}\,.
\end{equation}
We found that the only dynamical modes involve the gauge field; then we have at disposal three terms that give a non-trivial contribution in this limit: 
\begin{itemize}
\item The cubic interaction from which we extract the current $j_0^{JM} $
\begin{equation}
\sum_{J M} \sum_{J_i M_i \rho_i}  
i g  {\cal D}^{JM}_{J_1M_1\rho_1;J_2M_2\rho_2} 
\tr \le \chi^{JM} [\Pi^{J_1M_1}_{(\rho_1)},A^{J_2M_2}_{(\rho_2)}] \ri\,.
\label{eq:interaction_j0current_su12}
\end{equation}
\item The cubic interaction from which we extract the current $j_{(\rho)}^{JM}$
\beq
 \sum_{J_i M_i \rho_i} i g \rho_1(J_1 + 1) {\cal E}_{J_1M_1\rho_1,J_2M_2\rho_2,J_3M_3\rho_3}\tr \le A^{J_1M_1}_{(\rho_1)}[A^{J_2M_2}_{(\rho_2)},A^{J_3M_3}_{(\rho_3)}] \ri  \, .
 \label{eq:interaction_jrhocurrent_su12}
\eeq
\item The quartic gauge self-interaction
\begin{multline}
 - \sum_{JM} \sum_{J_i M_i \rho_i} \frac{g^2}{4} {\cal D}^{JM}_{J_1M_1\rho_1,J_3M_3\rho_3} {\cal D}_{JM,J_2M_2\rho_2,J_4M_4\rho_4}
\tr \le [A^{J_1M_1}_{(\rho_1)},A^{J_2M_2}_{(\rho_2)}][A^{J_3M_3}_{(\rho_3)},A^{J_4M_4}_{(\rho_4)}] \ri \,.
\label{eq:interaction_quarticterm_su12}
\end{multline}
\end{itemize}
After extracting the gauge currents 
from the first and second terms, we will find the corresponding effective Hamiltonian by applying Eq.~\eqref{eq:general_gauge_mediated_interaction_su12limits}.
The contribution fom the quartic interaction is crucial to perform the explicit summation over the Clebsch-Gordan coefficients.
We consider each of the three terms explicitly.

The first interaction is non-vanishing due to momentum conservation only when both the gauge fields in Eq.~\eqref{eq:interaction_j0current_su12} are dynamical and unaligned, {\sl{i.e.}} with opposite eigenvalues $\tilde{m}.$
Using Eq.~\eqref{eq:su12_const}, as well as the symmetry properties of ${\cal D}$, we extract the current
\begin{equation}
j_0^{\dagger \, JM} = - \frac{g}{2} \sum_{J_i m_i} \frac{\omega_{A,J_1}+\omega_{A,J_2}}{\sqrt{\omega_{A,J_1}\omega_{A,J_2}}}{\cal D}^{\hat{\CJ}_2 \hat{\CM}_2}_{JM; \hat{\CJ}_1 \hat{\CM}_1} [A_{J_1m_1},A^\dagger_{J_2m_2}]\,,
\end{equation}
where the normalization \eqref{eq:definition_gauge_modes_su12} is used for the gauge fields.
In accordance with Eq.\eqref{eq:general_gauge_mediated_interaction_su12limits}, this implies the following term in the interacting Hamiltonian:
\begin{equation}
\begin{aligned}
& \frac{1}{N}  \sum_{J M} \sum_{J_i m_i}
\frac{1}{32J(J+1)}
 \frac{(\omega_{A,J_1}+\omega_{A,J_2})(\omega_{A,J_3}+\omega_{A,J_4})}{\sqrt{\omega_{A,J_1}\omega_{A,J_2}\omega_{A,J_3}\omega_{A,J_4}}} \\
& \times  {\cal D}^{\hat{\CJ}_2 \hat{\CM}_2}_{JM; \hat{\CJ}_1 \hat{\CM}_1} {\cal D}_{\hat{\CJ}_3 \hat{\CM}_3}^{JM; \hat{\CJ}_4 \hat{\CM}_4} 
\tr \le [A_{J_1m_1},A^\dagger_{J_2m_2}] [A_{J_3m_3},A^\dagger_{J_4m_4}] \ri \, .
\end{aligned}
\label{eq:term1_interactions_su12}
\end{equation}
We now consider the interactions \eqref{eq:interaction_jrhocurrent_su12} and \eqref{eq:interaction_quarticterm_su12} together.
The sums over Clebsch--Gordan coefficients can be simplified  by means of Eq.~\eqref{eq:e_d_relation}, giving
\begin{multline}
 \sum_{J_i M_i \rho_i}  i g \rho_1(J_1 + 1) {\cal E}_{J_1M_1\rho_1,J_2M_2\rho_2,J_3M_3\rho_3}
\tr \le A^{J_1M_1}_{(\rho_1)}[A^{J_2M_2}_{(\rho_2)},A^{J_3M_3}_{(\rho_3)}] \ri \\
- \sum_{JM\rho} \sum_{J_i M_i \rho_i} \frac{g^2}{8} {\cal E}^{JM\rho}_{J_1M_1\rho_1,J_ 2M_2\rho_2} {\cal E}_{JM\rho,J_3M_3\rho_3,J_4M_4\rho_4}
\tr \le [A^{J_1M_1}_{(\rho_1)},A^{J_2M_2}_{(\rho_2)}][A^{J_3M_3}_{(\rho_3)},A^{J_4M_4}_{(\rho_4)}] \ri \,.
\label{eq:A_int_su_12}
\end{multline}
Before diving into explicit calculations, let us identify the relevant contributions that arise from the above expression.
In the first line, angular momentum conservation implies that at least one of the fields must be at order $g$ in the near-BPS limit, since there is no singlet in a product of three modes $A_{Jm}$ or $A^\dagger_{Jm}.$ 
Due to the total antisymmetry of ${\cal E},$ we can always put for convenience the dynamical modes inside the commutator, and account for other choices by symmetrizing the prefactor. 
We thus obtain for the first line in Eq.~\eqref{eq:A_int_su_12} the possibilities
\begin{multline}
\frac{1}{4} \sum_{J M \rho} \sum_{J_i m_i} i g \frac{\rho \omega_{A,J} - \omega_{A,J_1} - \omega_{A,J_2}}{\sqrt{\omega_{A,J_1}\omega_{A,J_2}}}
\tr \left\lbrace A^{JM}_{(\rho)} \Bigg(2{\cal E}_{JM\rho;J_1,m_1}{}^{J_2,m_2}[A_{J_1m_1},A^\dagger_{J_2m_2}]\right. \\
\left. + {\cal E}_{JM\rho;J_1,m_1;J_2,m_2}[A_{J_1m_1},A_{J_2m_2}] + {\cal E}_{JM\rho}{}^{J_1,m_1;J_2,m_2}[A^\dagger_{J_1m_1},A^\dagger_{J_2m_2}]
\Bigg) \right\rbrace
\label{eq:intermediate_step_su12_cubicterm}
\end{multline}
where we explicitly identified $\rho_1=\rho_2=-1$ for the dynamical modes and we used the normalization \eqref{eq:definition_gauge_modes_su12}.
The alignment of $J_1$ and $J_2$ in the second line of this expression implies that for the non-dynamical field the momenta are
\beq
J=J_1+J_2 \, , \qquad
\tilde{m} =  -(J_1 + J_2 + 2) = -(J + 1) \, .
\eeq
However this is in clear tension with the non-dynamical nature of the field, which requires $|\tilde{m}| \leq J$. The second line of Eq.~\eqref{eq:intermediate_step_su12_cubicterm} thus identically vanishes.

Integrating out the non-dynamical fields by means of  Eq.~\eqref{eq:general_gauge_mediated_interaction_su12limits} gives the following term
\begin{multline}
- \frac{1}{8N} \sum_{J M \rho} \sum_{J_i m_i} 
\left(\prod_i \sqrt{\omega_{A,J_i}}\right)^{-1}
\frac{|\rho|(\rho \omega_{A,J} - \omega_{A,J_1} - \omega_{A,J_2})(\rho \omega_{A,J} - \omega_{A,J_3} - \omega_{A,J_4})}{\omega_{A,J}^2 - 4m^2}\\
\times
{\cal E}_{JM\rho; \hat{\CJ}_1 \hat{\CM}_1}{}^{\hat{\CJ}_2 \hat{\CM}_2}{\cal E}^{JM\rho; \hat{\CJ}_4 \hat{\CM}_4}{}_{\hat{\CJ}_3 \hat{\CM}_3}
\tr \le [A_{J_1m_1},A^\dagger_{J_2m_2}][A_{J_3m_3},A^\dagger_{J_4m_4}] \ri
\,,
\label{eq:term2_interactions_su12}
\end{multline}
where we have multiplied the expression inside the sum by $|\rho|$ so that its summation can be extended to include $\rho = 0.$ 

Next, we consider the second line of Eq.~\eqref{eq:A_int_su_12}. It exclusively contains the dynamical modes, since all the other contributions are higher order in $g$ and vanish in the near-BPS limit. Angular momentum conservation requires the number of $A_{Jm}$ and $A^\dagger_{Jm}$ factors to be the same.
The alignment of the spins implies that the only surviving interaction is
\begin{equation}
-\frac{1}{8N} \sum_{J M \rho} \sum_{J_i m_i} \left(\prod_i \sqrt{\omega_{A,J_i}}\right)^{-1}
{\cal E}_{JM\rho; \hat{\CJ}_1 \hat{\CM}_1}{}^{\hat{\CJ}_2 \hat{\CM}_2}{\cal E}^{JM\rho; \hat{\CJ}_4 \hat{\CM}_4}{}_{\hat{\CJ}_3 \hat{\CM}_3}
\tr \le [A_{J_1m_1},A^\dagger_{J_2m_2}][A_{J_3m_3},A^\dagger_{J_4m_4}] \ri
\,.
\label{eq:term3_interactions_su12}
\end{equation}
In this way, the contributions to the effective Hamiltonian are given by the sum of Eqs.~\eqref{eq:term1_interactions_su12}, \eqref{eq:term2_interactions_su12} and \eqref{eq:term3_interactions_su12}.
Using the relation \eqref{eq:DErelation} derived in Appendix \ref{app:CG_properties}, one can see that by shifting $J \to J-1$ for those contributions with $\rho = 1$, all terms in the sum cancel.
Defining $\Delta J \equiv J_1 - J_2 = J_4 - J_3, $ we find that when $\Delta J \ne 0$ the only contribution arises from the lower boundary of summation 
\beq
\begin{aligned}
& \frac{1}{4N} \sum_{J_i m_i} \frac{\delta^{J_1-J_2}_{J_4-J_3}}{|\Delta J|}\sqrt{\frac{(1+J_1)(1+J_4)}{(1+J_2)(1+J_3)}}\sqrt{\frac{(1+2J_1)(1+2J_4)}{(1+2J_2)(1+2J_3)}} C^{J_2m_2}_{|\Delta J|\Delta m;J_1m_1} C^{J_3m_3}_{|\Delta J|\Delta m; J_4m_4}\\
& \times \tr \le [A_{J_1m_1},A^\dagger_{J_2m_2}][A_{J_3m_3},A^\dagger_{J_4m_4}] \ri \,.
\end{aligned}
\label{eq:result_differentJ_quarticgauge_su12}
\eeq
Here we put a factor of 2 to account for the two possibilities $J_1 > J_2$ and viceversa, due to the symmetries of the interaction.

For $\Delta J = 0$ we find instead
\begin{equation}
 - \frac{1}{8N} \sum_{J_1 m_1}
 \frac{J_1+2}{J_1+1} \,
\tr \le [A_{J_1m_1},A^\dagger_{J_1m_1}] Q^A \ri \, , 
\label{eq:result_equalJ_quarticgauge_su12}
\end{equation}
where $Q^A$ is the $\SU(N)$ charge carried by the gauge field.
Then this term vanishes due to Gauss' law at classical level.

It is convenient to introduce a charge density for the gauge field, defined as
\beq
\mathfrak{q}_{l, \Delta m} \equiv 
\sum_{m_1,m_2} \sum_{s_2=0}^{\infty} C^{\frac{s_2+l}{2}, m_2 + \Delta m}_{\frac{s_2}{2}, m_2; \frac{l}{2}, \Delta m}
\sqrt{\frac{(s_2+1)(s_2+2)}{(s_2+l+1)(s_2+l+2)}}
 [ A^{\dagger}_{s_2 m_2}, A_{s_2+l, m_2+ \Delta m} ] \, .
 \label{eq:charge_density_gauge_su12}
\eeq
In this way we introduced a notation where the sums run only over integers\footnote{The mode expansion of the fields also becomes $A_{Jm} = A_{\frac{s}{2}, m}.$
However, we use for convenience the notation $A_{s,m}$ to refer to the redefined modes, and it is understood that the sum over $s$ only includes integers.} $s_i = 2J_i ,$ and where the case $l=0$ corresponds to the total $\SU(N)$ charge
\beq
\mathfrak{q}_0 \equiv \mathfrak{q}_{l=0, \Delta m=0} = Q^A = \sum_{s=0}^{\infty} \sum_{m=-\frac{s}{2}}^{\frac{s}{2}} [A^{\dagger}_{s,m}, A_{s,m}] \, .
\label{eq:zero_mode_charge_density_gauge_su12}
\eeq 
This simplifies the result to
\beq
 H_{\rm int}  = \frac{1}{2N} \sum_{l=1}^{\infty}\sum_{\Delta m=-\frac{l}{2}}^{\frac{l}{2}} \frac{1}{l} \tr \le \mathfrak{q}^{\dagger}_{l, \Delta m} \mathfrak{q}_{l, \Delta m} \ri \, .
\eeq
The interaction is positive definite and written only in terms of the charge density for the gauge field.


\subsection{Adding scalars and fermions - $\SU(1,2|2)$}
\label{sect-su122limit}

After the warm-up in Section \ref{sect-su12limit}, we move to the BPS bound $H \geq S_1+S_2+Q_1,$ which includes scalars and fermions in the decoupled theory in addition to the same gauge field modes obtained in the $\SU(1,2)$ case. 
The resulting interaction will be invariant under $\SU(1,2|2) \times \U(1)$ group, which means that the sector is supersymmetric.

\subsubsection*{Free Hamiltonian and reduction of the degrees of freedom}

We follow the same procedure of Section \ref{sect-su12limit}, which requires to evaluate the BPS bound $H - S_1-S_2-Q_1 \geq 0$ at quadratic order.
The left-hand side of this inequality reads
\begin{eqnarray}
&& H - S_1 -S_2 - Q_1 = \sum_{J,M} \tr \left( \left| \Pi^{(\Phi) \dagger}_{JM,1} - i (2\tilde{m}+1) \Phi^1_{JM} \right|^2 + \left( \omega_J^2 - (2\tilde{m}+1)^2\right) \Phi^\dagger_{JM,1} \Phi^1_{JM} \right) \nn \\ 
&& + \sum_{J,M} \sum_{a=2,3}   \tr \left( \left| \Pi^{(\Phi) \dagger}_{JM,a} - i 2\tilde{m} \Phi^a_{JM} \right|^2 + \left( \omega_J^2 - 4\tilde{m}^2\right) \Phi^\dagger_{JM,a} \Phi^a_{JM} \right) 
\label{eq:quadratic_Hamiltonian_su122}  \\
&& + \sum_{JM} \sum_{\kappa=\pm1} \left[  \sum_{A=1,4} \le  \omega^{\psi}_J - 2 \tilde{m} - \frac{\kappa}{2} \ri \tr  \le \psd_{JM\kappa,A} \psi^A_{JM\kappa} \ri 
+ \sum_{A=2,3} \le \omega^{\psi}_J - 2 \tilde{m} + \frac{\kappa}{2} \ri \tr \le \psd_{JM\kappa,A} \psi^A_{JM\kappa} \ri  \right]
\nn \\
&& + \sum_{JM} \sum_{\rho=-1,1} \frac{1}{2}\left(|\Pi_{(\rho)}^{Jm\tilde{m}} - 2i\tilde{m}A_{(\rho)}^{\dagger\,Jm\tilde{m}}|^2+ (\omega_{A,J}^2 - (2\tilde{m})^2) |A_{(\rho)}^{Jm\tilde{m}}|^2\right) \nn \, ,
\end{eqnarray}
where we defined
\beq
\omega_J \equiv 2J+1 \, , \qquad
 \omega_{J}^{\psi} \equiv 2J+\frac{3}{2} \, , \qquad
  \omega_{A,J} \equiv 2J+2 \, .
\eeq
We obtain the set of constraints by imposing that Eq.~\eqref{eq:quadratic_Hamiltonian_su122} vanishes.
\begin{itemize}
\item The scalar fields appear only in the first and second line.
Since $\omega_J=2J+1$ and $|\tilde{m}| \leq J$ we see that only for $\tilde{m}=J$ and $a=1$ one can have a surviving mode, which satisfies the non-trivial constraint 
\begin{equation}
(\Pi_1^{(\Phi)})_{J,m,J} + i  \omega_J (\Phi_1)^\dagger_{J,m,J} = \mathcal{O}(g) \, .
\label{eq:non_trivial_constraint_su122}
\end{equation}
\item Moving on to the fermionic part, the relevant conditions are given by the definition of $\omega^{\psi}_J$ and the constraints on momenta: when $\kappa=1$ we have $|\tilde{m}| \leq J,$ while for $\kappa=-1$ we have $|\tilde{m}| \leq J + \frac{1}{2} .$
In this way the surviving degrees of freedom have
\beq
\kappa=-1 \, , \qquad
|m| \leq J \, , \qquad
\tilde{m}=J+\frac{1}{2} \, , \qquad
\mathrm{for} \,\,\, \psi^{A=2,3} \, .
\label{eq:list_surviving_dof_fermions_su122}
\eeq
These modes are unconstrained.
For notational convenience, we define fields with a shift of the R-symmetry index (in the following identity, $A \in \lbrace 0,1,2,3 \rbrace $):
\beq
\zeta^{A}_{J m \tilde{m}} \equiv \psi^{A+1}_{J m \tilde{m}} \, .
\label{eq:redefinition_fermions_su122}
\eeq
\item The gauge sector can be identified by analyzing the last line of Eq.~\eqref{eq:quadratic_Hamiltonian_su122}.
Clearly, these conditions are the same obtained in the $\SU(1,2)$ subsector, see Section \ref{sect-su12limit}.
\end{itemize}
Adding the relations on the non-dynamical modes and requiring compatibility with the Hamiltonian evolution, we find that the entire set of constraint reads
\bea
\label{eq:constraints_gauge_fields_su122}
&& A_{(\rho)}^{Jm\tilde{m}} = -\frac{j^{ Jm\tilde{m}}_{(\rho)}}{\omega_{A,J}^2 - 4 \tilde{m}^2}  \,,\quad\Pi_{(\rho)}^{(J, m, \tilde{m} \ne J+1)} = 0 \, ,  \\
\label{eq:constraint_dynamical_gauge_fields_su122}
&& \Pi_\pm^{Jm} \pm i\omega_{A,J}A_\pm^{\dagger\,Jm} = 0 \,, 
 \\
&& \Phi_{a=2,3}^{Jm\tilde{m}} = 0\,,\quad (\Pi^{(\Phi)}_{a=2,3})_{Jm\tilde{m}}  = 0\,, \quad
\Phi^{1}_{(J,m, \tilde{m} \ne J)} = 0  \, , \quad
(\Pi^{(\Phi)}_{1})_{(J,m, \tilde{m} \ne J)} = 0 \, ,    \\
\label{eq:constraints_scalars_su122}
&& (\Pi_1^{(\Phi)})_{J,m,J} + i  \omega_J (\Phi_1)^\dagger_{J,m,J} = 0 \, ,  \\
\label{eq:constraints_fermions_su122}
&& \zeta_{A=0,3}^{J m \tilde{m}} = 0 \, , \quad
 \zeta_{A=1,2}^{(J, m, \tilde{m} \ne J+\frac{1}{2})} = 0 \,    \, . 
\eea
The non-relativistic dispersion relation for scalars and the non-vanishing of one chirality for each copy of the surviving fermions was also observed from the null reduction of supersymmetric theories \cite{Auzzi:2019kdd}.

We notice that there are two constraints inducing a change in the Dirac brackets: Eq.~\eqref{eq:constraint_dynamical_gauge_fields_su122} and \eqref{eq:constraints_scalars_su122}, which give 
\begin{align}
 \{(A_+)^{J,m,J+1}, (A_+^{\dagger})^{J',m',J'+1}\}_D &= \frac{i}{2\omega_{A,J}} \delta_{J,J'} \delta_{m,m'} \,, 
\label{eq:Dirac_bracket_gauge_su122} \\
\{ (\Phi_1)_{J,m,J} , (\Phi^\dagger_1)_{J',m',J'} \}_D &=  \frac{i}{2\omega_J} \delta_{J,J'} \delta_{m,m'} \, . 
\label{eq:Dirac_bracket_scalar_su122}
\end{align}
In order to make the brackets canonical, we define the dynamical modes surviving the near-BPS limit as
\beq
\Phi_{Jm} = \sqrt{2\omega_J}  \Phi^1_{J,m,J}  \, , \quad
 \zeta^{a}_{JM} \equiv \zeta^{a}_{J,m,J+\frac{1}{2}, \kappa=-1}  \, ,  \quad
   A_{Jm} \equiv \sqrt{2\omega_{A,J}} (A_+)^{J,m,J+1}\,,
 \label{eq:def_dynamical_modes_su122}
\eeq
where due to Eq.~\eqref{eq:redefinition_fermions_su122} we have $a=1,2.$
On the constraint surface the Hamiltonian becomes
\begin{equation}
\label{H0_su122}
H_0 = \sum_{s=0}^{\infty} \sum_{m=-\frac{s}{2}}^{\frac{s}{2}} \left[ \le s+1 \ri \tr |\Phi_{s,m}|^2 
+ \sum_{a=1,2}  \le s + \frac{3}{2} \ri |\zeta^a_{s,m}|^2
+ (s+2) \tr |A_{s,m}|^2 \right]
\, .
\end{equation}
In summary, the field content surviving the $\SU(1,2|2)$ limit is collected into Eq.~\eqref{eq:def_dynamical_modes_su122}: there is a scalar field, a field strength and two fermions with the same chirality.


\subsubsection*{Interactions: general considerations}

The interacting Hamiltonian of the sector is defined by
\beq
\label{Hint_limit_su122}
H_{\rm int} = \lim_{g \rightarrow 0} \frac{H-S_1-S_2-Q_1}{g^2 N} \, .
\eeq
We follow the procedure outlined in Section \ref{sect-sphere_reduction_general}.
Due to the presence of four surviving letters from the original field content of $\mathcal{N}=4$ SYM, we obtain several terms either from integrating out auxiliary modes, or from the original action.
We combine them into three main categories:
\begin{itemize}
\item Terms mediated by the non-dynamical modes of the gauge field via the currents contributing to Eq.~\eqref{eq:general_gauge_mediated_interaction_su12limits}.
Quartic interactions involving only scalar fields, only gauge fields and the mixed combination of two scalars with two gauge fields.
\item Cubic Yukawa terms.
\item Terms containing dynamical gauge fields mediated by non-dynamical scalars or fermions. 
\end{itemize}
The terms listed in the first bullet combine in such a way to explicitly solve the sums of Clebsch-Gordan coefficients over intermediate momenta $(J,m).$ 
We will show that the Hamiltonian organizes into positive definite blocks, which are naturally distinguished by the non-dynamical field mediating the effective interaction\footnote{More precisely, the cases with scalars and gauge fields are also complemented by quartic interactions already appearing in the $\mathcal{N}=4$ SYM action where all the fields are dynamical.}.
For this reason, we will list the interactions following this criterion. 

Given the dynamical modes surviving the $\SU(1,2|2)$ limit and summarized in Eq.~\eqref{eq:def_dynamical_modes_su122}, it is convenient to introduce a specific notation for the saturation of momenta, that will enter all the relevant Clebsch-Gordan coefficients to evaluate.
We define
\bea
\label{eq:notation_index_CJ_CM}
& (\mathcal{J},\mathcal{M}) \equiv (J, m, J) \, , \qquad
|m| \leq J  \, , &  \\
\label{eq:notation_index_barCJ_barCM}
& (\bar{\mathcal{J}}, \bar{\mathcal{M}}) \equiv \le J, m, J+ \frac{1}{2}, \kappa=-1 \ri \, , \qquad
|m| \leq J \, ,  &   \\
\label{eq:notation_index_hatCJ_hatCM}
& (\hat{\mathcal{J}}, \hat{\mathcal{M}}) \equiv \le J, m,J+1, \rho=-1 \ri \, , \qquad
|m| \leq J \, .  &
\eea
These saturations correspond to scalars, fermions and gauge fields, respectively.
Now we are ready to systematically analyze all the interactions.


\subsubsection*{Interactions I: terms mediated by the non-dynamical gauge field}

In order to compute the effective interactions mediated by the non-dynamical gauge field, we need to identify all the currents in the $\mathcal{N}=4$ SYM action and then apply Eq.~\eqref{eq:general_gauge_mediated_interaction_su12limits}.
For this reason, we collect all the terms in the Hamiltonian linear in the fields $\chi^{Jm \tilde{m}}, A_{(\rho)}^{J m \tilde{m}}$ with the momenta $(J,m, \tilde{m})$ running over non-dynamical modes.
We reserve instead the notation $(J_i, m_i, \tilde{m}_i) $ for the dynamical modes, which are responsible for the saturation of momenta in the Clebsch-Gordan coefficients according to Eqs.~\eqref{eq:notation_index_CJ_CM}, \eqref{eq:notation_index_barCJ_barCM}, \eqref{eq:notation_index_hatCJ_hatCM}.

The scalar part reads
\beq
\begin{aligned}
 \sum_{J,m,\tilde{m}} \sum_{J_i, m_i} & \left\lbrace  i g {\cal C}^{\CJ_2 \CM_2}_{\CJ_1 \CM_1; JM}
\tr \le \chi^{JM}\left([(\Phi_1^{\dagger})_{\CJ_2 \CM_2},(\Pi_1^\dagger)_{\CJ_1 \CM_1}] + [\Phi_1^{\CJ_1 \CM_1},\Pi_1^{\CJ_2 \CM_2}]\right) \ri \right.  \\
& \left.  - 4g \sum_{\rho=\pm1} \sqrt{J_1(J_1+1)} {\cal D}^{\CJ_2 \CM_2}_{\CJ_1 \CM_1; JM\rho}
\tr \le A^{JM}_{(\rho)}[\Phi_1^{\CJ_1 \CM_1},(\Phi_1^\dagger)_{\CJ_2 \CM_2}]  \ri \right\rbrace \,,
 \end{aligned}
 \label{eq:scalar_current_su122}
\eeq
while the fermionic term is given by
\beq
\begin{aligned}
& g \sum_{J,m,\tilde{m}} \sum_{J_i, m_i} \sum_{a=1,2}  \mathcal{F}^{\bar{\CJ}_1 \bar{\CM}_1}_{\bar{\CJ}_2 \bar{\CM}_2; JM}  
\tr \le \chi^{JM}  \lbrace (\zeta^{\dagger}_a)_{\bar{\CJ}_1 \bar{\CM}_1} , (\zeta^a)_{\bar{\CJ}_2 \bar{\CM}_2} \rbrace \ri \\
& + g  \sum_{J,m,\tilde{m}} \sum_{J_i, m_i} \sum_{a=1,2} \sum_{\rho=\pm1} \mathcal{G}^{\bar{\CJ}_1 \bar{\CM}_1}_{\bar{\CJ}_2 \bar{\CM}_2; JM \rho}  \tr \le  A^{JM}_{(\rho)} 
\lbrace (\zeta^{\dagger}_a)_{\bar{\CJ}_1 \bar{\CM}_1} , (\zeta^a)_{\bar{\CJ}_2 \bar{\CM}_2} \rbrace  \ri 
 \,.
 \end{aligned}
 \label{eq:fermion_current_su122}
\eeq
The part involving only the gauge fields can be inherited from the $\SU(1,2)$ subsector and is given by the sum of Eqs.~\eqref{eq:interaction_j0current_su12} and 
\eqref{eq:interaction_jrhocurrent_su12}.
Collecting all the currents from such terms, and additionally the ones derived from Eqs.~\eqref{eq:scalar_current_su122} and \eqref{eq:fermion_current_su122}, we find
\begin{align}
\label{eq:total_currentj0_su122}
\begin{split}
& j_0^{\dagger\,Jm\tilde{m}} = \sum_{J_i m_i} \frac{g( \omega_{J_1}+\omega_{J_2})}{2\sqrt{\omega_{J_1}\omega_{J_2}}}  {\cal C}^{\CJ_2 \CM_2}_{\CJ_1 \CM_1; Jm\tilde{m}}  [\Phi_{J_1 m_1},\Phi^\dagger_{J_2 m_2}]  \\
& +  g \sum_{J_i m_i} \sum_{a=1,2}  \mathcal{F}^{\bar{\CJ}_1 \bar{\CM}_1}_{\bar{\CJ}_2 \bar{\CM}_2; JM}  
\lbrace (\zeta^{\dagger}_a)_{J_1 m_1} , (\zeta^a)_{J_2 m_2} \rbrace 
 -  \sum_{J_i m_i} \frac{g}{2} \frac{\omega_{A,J_1}+\omega_{A,J_2}}{\sqrt{\omega_{A,J_1}\omega_{A,J_2}}}{\cal D}^{\hat{\CJ}_2 \hat{\CM}_2}_{JM; \hat{\CJ}_1 \hat{\CM}_1} [A_{J_1 m_1},A^\dagger_{J_2 m_2}]\,,
\\
& j_{(\rho)}^{\dagger\,Jm\tilde{m}} = 
- 2g \sum_{J_i m_i} \sqrt{\frac{J_1(J_1+1)}{\omega_{J_1}\omega_{J_2}}} {\cal D}^{\CJ_2 \CM_2}_{\CJ_1 \CM_1 ; Jm\tilde{m},\rho}[\Phi_{J_1 m_1},\Phi^\dagger_{J_2 m_2}] \\ 
& + g  \sum_{J_i, m_i} \sum_{a=1,2}  \mathcal{G}^{\bar{\CJ}_1 \bar{\CM}_1}_{\bar{\CJ}_2 \bar{\CM}_2; JM \rho} 
\lbrace (\zeta^{\dagger}_a)_{J_1 m_1} , (\zeta^a)_{J_2 m_2} \rbrace  \\
 & +  \frac{ig}{2}  \sum_{J_i m_i} \frac{\rho \omega_{A,J} - \omega_{A,J_1} - \omega_{A,J_2}}{\sqrt{\omega_{A,J_1}\omega_{A,J_2}}}
 {\cal E}_{JM\rho;\hat{\CJ}_1 \hat{\CM}_1}{}^{\hat{\CJ}_2 \hat{\CM}_2} [A_{J_1 m_1},A^\dagger_{J_2 m_2}] \, .
\end{split}
\end{align}
In order to get these expressions, we used the definitions \eqref{eq:def_dynamical_modes_su122} and we applied the constraints \eqref{eq:constraint_dynamical_gauge_fields_su122} and \eqref{eq:constraints_scalars_su122}. 
Using Eq.~\eqref{eq:general_gauge_mediated_interaction_su12limits} with these currents, we compute all the interactions mediated by the non-dynamical gauge field.
One can show by explicit computation that in all the following manipulations, there will always be a condition on momenta such that we can define
\beq
 \Delta J \equiv J_1 - J_2 = J_4 - J_3 \, , \qquad
\Delta m \equiv m_1 - m_2 = m_4 - m_3 \, .
\label{eq:definition_DeltaJ_Deltam}
\eeq
In addition, we will also define
\beq
s_i \equiv 2 J_i \, , \qquad
  l \equiv 2 \Delta J \, , \qquad
 |m_i| \leq \frac{s_i}{2} \, , \qquad
 | \Delta m| \leq \frac{l}{2} \, ,
\label{eq:definition_l_momenta_s}
\eeq
in such a way that the sum over momenta runs only over integer values.
We will express the effective interactions in terms of the data $(l,\Delta m).$

We start with the purely scalar terms, which arise from the square of the first terms in \eqref{eq:total_currentj0_su122}.
In addition, we also add the quartic scalar self-interaction appearing in the $\mathcal{N}=4$ SYM action, which is given by
\begin{equation}
 \frac{g^2}{8}  \sum_{J m \tilde{m}} \sum_{J_i m_i}  \left(\prod_{i=1}^4 \frac{1}{\sqrt{\omega_{J_i}}} \right)
{\cal C}^{\CJ_1 \CM_1}_{\CJ_2 \CM_2; J m\tilde{m}}{\cal C}^{\CJ_4 \CM_4}_{\CJ_3 \CM_3; J m\tilde{m}}
\tr \le [\Phi_{J_4 m_4},\Phi_{J_3 m_3}^\dagger] [\Phi_{J_2 m_2},\Phi_{J_1 m_1}^\dagger] \ri
\end{equation}
Since the interaction is already at order $g^2,$ non-vanishing contributions in the near-BPS limit are achieved only when all the fields are dynamical: for this reason, in the above expression we immediately used the definition \eqref{eq:def_dynamical_modes_su122}.

Combining all the scalar interactions, we find
\beq
\begin{aligned}
\label{eq:Hphisu12}
& \frac{g^2}{8}  \sum_{J m \tilde{m}} \sum_{J_i m_i}  \left(\prod_{i=1}^4 \frac{1}{\sqrt{\omega_{J_i}}} \right)
 \Bigg(\bigg(
1
 + \frac{( \omega_{J_1}+\omega_{J_2})( \omega_{J_3}+\omega_{J_4})}{4J(J+1)}\bigg) {\cal C}^{\CJ_1 \CM_1}_{\CJ_2 \CM_2;Jm\tilde{m}}{\cal C}^{\CJ_4 \CM_4}_{\CJ_3 \CM_3; Jm\tilde{m}}\\
& -\sum_{\rho = \pm 1} \frac{16}{\omega_{A,J}^2 - 4\tilde{m}^2}\sqrt{J_2(J_2+1)J_3(J_3+1)}{\cal D}^{\CJ_1 \CM_1}_{\CJ_2 \CM_2 ; Jm\tilde{m},\rho}{\cal \bar{D}}^{\CJ_4 \CM_4}_{\CJ_3 \CM_3; Jm\tilde{m},\rho}
 \Bigg) \\
 & \times  \tr \le [\Phi_{J_4 m_4},\Phi_{J_3 m_3}^\dagger] [\Phi_{J_2 m_2},\Phi_{J_1 m_1}^\dagger] \ri \\
\end{aligned}
\eeq
Making use of the procedure explained in Appendix \ref{app-crossing_relations_saturated_momenta} and in particular of the identity \eqref{eq:CDrelation}, we obtain for $\Delta J \neq 0$ the expression
\begin{equation}
 \frac{g^2}{4} \sum_{J_i m_i} 
\frac{\delta^{J_1 - J_2}_{J_4 - J_3}}{\Delta J}
C^{J_1 m_1}_{J_2 m_2;\Delta J \Delta m}C^{J_4 m_4}_{J_3 m_3;\Delta J \Delta m}
\tr \le [\Phi_{J_4 m_4},\Phi_{J_3 m_3}^\dagger] 
 [\Phi_{J_2 m_2},\Phi_{J_1 m_1}^\dagger]\ri
\, ,
\label{eq:intermediate_result_purely_scalar_su122}
\end{equation}
where there is a factor of 2 coming from symmetry properties of the $\SU(2)$ Clebsch-Gordan coefficients after exchanging $J_1 \leftrightarrow J_2$ and $J_3 \leftrightarrow J_4,$ and we used the definitions \eqref{eq:definition_DeltaJ_Deltam}.

The case $\Delta J = 0$ is special. Here we find, again using relation \eqref{eq:CDrelation} that implies a cancellation of all terms in the sum with $J > 0$, the result
\beq
\frac{g^2}{4} \sum_{J_1,m_1} \frac{1- 2J_1}{\omega_{J_1}}
\tr \le [\Phi_{J_1 m_1},\Phi_{J_1 m_1}^\dagger] Q^{\Phi}   \ri \, ,
\label{eq:intermediate_result_case0_purely_scalar_su122}
\eeq
where $Q^{\Phi}$ is the scalar $\SU(N)$ charge.
It is the $l=0$ value of the following charge density
\beq
\begin{aligned}
& q_{l,\Delta m} \equiv \sum_{s_2=0}^{\infty} \sum_{m_2=-\frac{s_2}{2}}^{\frac{s_2}{2}} C^{\frac{s_2+l}{2}, m_2+\Delta m}_{\frac{s_2}{2}, m_2; \frac{l}{2},\Delta m} [\Phi^{\dagger}_{s_2 m_2}, \Phi_{s_2+ l, m_2 + \Delta m}] \, , \\
& q_0 \equiv q_{l=0, \Delta m=0} = Q^{\Phi} = \sum_{s=0}^{\infty} \sum_{m=-\frac{s}{2}}^{\frac{s}{2}} [\Phi^{\dagger}_{s,m}, \Phi_{s,m}] \, .
\label{eq:scalar_charge_density_su122}
\end{aligned}
\eeq 
In this way the effective Hamiltonian for the purely scalar term can be compactly written as
\beq
 \frac{1}{8N} \sum_{n=0}^{\infty}  \frac{1-n}{n+1} \tr \le q_0^2 \ri
+ \frac{1}{2N} \sum_{l=1}^{\infty}\sum_{\Delta m=-\frac{l}{2}}^{\frac{l}{2}} \frac{1}{l} \tr \le q^{\dagger}_{l, \Delta m} q_{l, \Delta m} \ri \, .
\label{eq:compact_scalar_four_int_su122}
\eeq
Now we consider the purely fermionic part, arising from the square of the second terms in \eqref{eq:total_currentj0_su122}. We obtain
\beq
\begin{aligned}
\frac{g^2}{2} & \sum_{J m \tilde{m}} \sum_{J_i m_i}   \le \frac{1}{4J(J+1)} \mathcal{F}^{\bar{\CJ}_1 \bar{\CM}_1}_{\bar{\CJ}_2 \bar{\CM}_2; J m \tilde{m}} \mathcal{F}^{\bar{\CJ}_4 \bar{\CM}_4}_{\bar{\CJ}_3 \bar{\CM}_3; J m \tilde{m}}  - \sum_{\rho = \pm 1} \frac{1}{\omega^2_{A,J}- 4\tilde{m}^2} \mathcal{G}^{\bar{\CJ}_1 \bar{\CM}_1}_{\bar{\CJ}_2 \bar{\CM}_2; Jm \tilde{m} \rho} \bar{\mathcal{G}}^{\bar{\CJ}_4 \bar{\CM}_4}_{\bar{\CJ}_3 \bar{\CM}_3; J m \tilde{m} \rho}  \ri \\
& \times
\tr \le \lbrace (\zeta_a^{\dagger})_{J_1 m_1} , (\zeta^a)_{J_2 m_2} \rbrace \lbrace (\zeta_b^{\dagger})_{J_3 m_3} , (\zeta^b)_{J_4 m_4} \rbrace \ri \, .
\end{aligned}
\label{eq:purely_ferm_intermediate_su122}
\eeq
We solve this sum using the trick explained in Appendix \ref{app-crossing_relations_saturated_momenta}, specifically Eq.~\eqref{eq:FGrelation_saturated_momenta}.
This gives the result
\beq
\begin{aligned}
& \frac{1}{8N} \sum_{J_i,m_i} \frac{\delta^{J_1-J_2}_{J_4-J_3}}{|\Delta J|} \sqrt{\frac{2J_2+1}{2J_1+1}} \sqrt{\frac{2J_3+1}{2J_4+1}} C^{J_1 m_1}_{J_2 m_2; |\Delta J|, \Delta m} C^{J_4 m_4}_{J_3 m_3; |\Delta J|, \Delta m} \\
& \times \tr \le \lbrace (\zeta_a^{\dagger})_{J_1 m_1} , (\zeta^a)_{J_2 m_2}  \rbrace \lbrace (\zeta_b^{\dagger})_{J_3 m_3} , (\zeta^b)_{J_4 m_4} \rbrace \ri \, ,
\end{aligned}
\eeq
where a factor of 2 is included to account for the symmetry of the sum when exchanging $ J_1 \leftrightarrow J_2 $ and $J_3 \leftrightarrow J_4 ,$ and we used again the definitions \eqref{eq:definition_DeltaJ_Deltam}.

When $\Delta J=0,$ a similar method allows to obtain
\beq
- \frac{1}{8N} \sum_{J_1,m_1}  \tr \le \lbrace (\zeta_a^{\dagger})_{J_1 m_1} , (\zeta^a)_{J_1 m_1}  \rbrace \,  Q^{\psi}  \ri \, ,
\eeq 
where we recognize the appearance of the $\SU(N)$ fermionic charge.
More precisely, we can again identify it as the $l=0$ value of a fermonic charge density
\beq
\begin{aligned}
& \tilde{q}_{l, \Delta m} \equiv \sum_{a=1,2} \sum_{s_2=0}^{\infty} \sum_{m_2=-\frac{s_2}{2}}^{\frac{s_2}{2}}  C^{\frac{s_2+l}{2}, m_2+\Delta m}_{\frac{s_2}{2}, m_2; \frac{l}{2}, \Delta m}
\sqrt{\frac{s_2+1}{s_2+l+1}}
 \lbrace (\zeta^{\dagger}_a)_{s_2 m_2}, (\zeta^a)_{s_2+l, m_2+\Delta m} \rbrace \, , \\
& \tilde{q}_0 \equiv \tilde{q}_{l=0, \Delta m=0} = Q^{\psi} = \sum_{a=1,2} \sum_{s=0}^{\infty} \sum_{m=-\frac{s}{2}}^{\frac{s}{2}} 
 \lbrace (\zeta^{\dagger}_a)_{s,m}, (\zeta_a)_{s,m} \rbrace \, .
  \label{eq:fermion_charge_density_su122}
  \end{aligned}
\eeq 
Putting everything together, we find the purely fermionic interaction
\beq
- \frac{1}{8N} \sum_{s=0}^{\infty} \sum_{m=-\frac{s}{2}}^{\frac{s}{2}} \tr \le \lbrace (\zeta_a^{\dagger})_{s,m} , (\zeta^a)_{s,m}  \rbrace  \,  \tilde{q}_0 \ri
 + \frac{1}{2N} \sum_{l=1}^{\infty} \sum_{\Delta m=- \frac{l}{2}}^{\frac{l}{2}}
\frac{1}{l} \tr \le \hat{q}_{l, \Delta m}^{\dagger} \hat{q}_{l, \Delta m} \ri  \, . 
\label{eq:compact_fermion_four_int_su122}
\eeq
The purely gauge interactions are simply inherited from Section \ref{sect-su12limit}, the result being
\beq
  - \frac{1}{8N} \sum_{n=0}^{\infty} \frac{n+4}{n+2} \tr \le \mathfrak{q}_0^2 \ri
+ \frac{1}{2N} \sum_{l=1}^{\infty}\sum_{\Delta m=-\frac{l}{2}}^{\frac{l}{2}} \frac{1}{l} \tr \le \mathfrak{q}^{\dagger}_{l, \Delta m} \mathfrak{q}_{l, \Delta m} \ri \, ,
\label{eq:compact_gauge_four_int_su122}
\eeq
with charge density and its $l=0$ mode given by Eqs.~\eqref{eq:charge_density_gauge_su12} and \eqref{eq:zero_mode_charge_density_gauge_su12}.
This concludes the treatment of all the squares of single terms contributing to the currents.

We still need to consider all the mixed products.
In order to explicitly solve the sums over $J$ for such terms, we apply the procedure outlined in Appendix \ref{app-crossing_relations_saturated_momenta}.
Since it is a straightforward application of the same technique used for the previous terms, we directly give the result.

Remarkably, we can collect all of the terms mediated by the non-dynamical gauge field in a compact way by means of a unique total charge density
\beq
\mathbf{Q}_{l, \Delta m} \equiv q_{l, \Delta m} + \tilde{q}_{l, \Delta m} + \mathfrak{q}_{l, \Delta m} \, .
\label{eq:total_charge_density_su122}
\eeq
The part of the interacting Hamiltonian mediated by the non-dynamical gauge fields reads
\beq
\begin{aligned}
H_{\rm int} & = \frac{1}{2N} \sum_{l=1}^{\infty} \sum_{\Delta m=-\frac{l}{2}}^{\frac{l}{2}}
\frac{1}{l} \tr \le \mathbf{Q}_{l, \Delta m}^{\dagger} \mathbf{Q}_{l, \Delta m} \ri 
- \frac{1}{8N} \tr \le (\mathbf{Q}_0)^2 \ri \\
& + \frac{1}{4N} \sum_{s=0}^{\infty} \sum_{m=-\frac{s}{2}}^{\frac{s}{2}} \frac{1}{s+1} 
\tr \le [\Phi^{\dagger}_{s, m}, \Phi_{s, m}] \mathbf{Q}_0 \ri 
- \frac{1}{4N} \sum_{s=0}^{\infty} \sum_{m=-\frac{s}{2}}^{\frac{s}{2}} \frac{1}{s+2} 
\tr \le [A^{\dagger}_{s, m}, A_{s, m}] \mathbf{Q}_0 \ri \, . 
\end{aligned} 
\label{eq:total_interaction_gauge_mediated_su122}
\eeq
In this way, we observe that all the terms involving the $l=0$ modes of the charge density collect into an expression proportional to the total $\SU(N)$ charge of the system, which vanishes due to Gauss' law.
So, the first term contains all the non-vanishing content of the effective interaction mediated by non-dyamical components of the gauge fields.
All the information is encoded inside the total charge density \eqref{eq:total_charge_density_su122}.


\subsubsection*{Interactions II: terms mediated by the non-dynamical fermions}

In the $\SU(1,2|2)$ near-BPS limit the only non-trivial interactions that can be mediated by a non-dynamical fermion are of two species: the Yukawa term and the minimal coupling in Eq.~\eqref{eq:fermion_current_su122}.

We start with general considerations on the Yukawa term in $\mathcal{N}=4$ SYM action.
This interaction is cubic and contains antisymmetric combinations of one scalar and two fermionic fields, with overall linear dependence from the coupling constant $g.$
According to Eq.~\eqref{eq:def_dynamical_modes_su122}, the only non-vanishing contributions in the near-BPS limit occur when the field content is precisely given by the set $ (\Phi_1, \zeta_1, \zeta_2).$

The case where all the fields are dynamical is forbidden by momentum conservation.
A non-trivial interaction is obtained when we take precisely two fields in the set to be dynamical and one to be non-dynamical, giving the desired order $g^2.$ 
Using the properties of the Clebsch-Gordan coefficient $\mathcal{F},$ we collect the possibilities into the manifestly hermitian form 
\beq
 -2 ig \sum_{JM\kappa} \sum_{J_i} (-1)^{-m_1+\tilde{m}_1+\frac{\kappa_1}{2}} \mathcal{F}^{J M \kappa}_{J_1, -M_1,\kappa_1 ; J_2 M_2} \,
\epsilon_{ab} \, 
  \tr \le \zeta^a_{JM\kappa} [(\Phi_1^{\dagger})^{J_2M_2} , \zeta^b_{J_1 M_1 \kappa_1}] \ri + \mathrm{h.c.} 
\label{eq:Yukawa_type_interaction_su_1_2_2}
\eeq
We consider the case where there is a single auxiliary fermion $\zeta^a_{J M \kappa}$ which need to be integrated out.
Momentum conservation gives the conditions
\beq
m = m_2-m_1 \, , \qquad
\tilde{m} = J_2 - J_1 - \frac{1}{2}  \, .
\eeq
When $\kappa=1$ we further need to impose the constraints
\beq
|m| \leq J+\frac{1}{2} \, , \qquad
|\tilde{m}| \leq J \, , \qquad
\left| J_1 - J_2 + \frac{1}{2} \right| \leq J \leq J_1 + J_2 - \frac{1}{2} \, ,
\label{eq:triangle_inequalities_kappa1_su122}
\eeq
while when $\kappa=-1,$ the constraints are 
\beq
|m| \leq J \, , \qquad
\tilde{m} < J + \frac{1}{2} \, , \qquad
|J_1 - J_2| \leq J \leq J_1 + J_2 \, .
\label{eq:triangle_inequalities_kappa-1_su122}
\eeq
Both cases are allowed, which means that we need to sum over $\kappa= \pm 1 .$ 

Before integrating out the fermion explicitly, we consider the contribution from Eq.~\eqref{eq:fermion_current_su122}, where now we take the gauge field to be dynamical.
It can be shown by similar manipulations that momentum conservation allows for both the cases $\kappa = \pm 1.$

In this way the equations of motion for the auxiliary fermion coming from the Yukawa and the minimal coupling terms read
\beq
\begin{aligned}
\zeta^a_{J M \kappa} & = \sum_{J_i m_i \kappa_i \rho_i} 
\frac{-2ig (-1)^{m_1-\tilde{m}_1+\frac{\kappa_1}{2}} \mathcal{F}^{JM\kappa}_{J_1, -M_1, \kappa_1; J_2 M_2}} {\sqrt{2\omega_{J_2}} \le \kappa \omega^{\psi}_J + 2 \tilde{m} + \frac{1}{2} \ri} 
\, \epsilon^{a b} \,
[\Phi^{J_2 m_2}, (\zeta^{\dagger}_b)_{J_1 m_1 \kappa_1}] \\
& + \sum_{J_i m_i \kappa_i \rho_i} \frac{ g \mathcal{G}^{J M \kappa}_{J_1 M_1 \kappa_1; J_2 M_2 \rho_2}}{\sqrt{2\omega_{A,J_2}} \le \kappa \omega^{\psi}_J + 2 
\tilde{m} + \frac{1}{2} \ri}
[A^{J_2 m_2 \rho_2}, (\zeta^a)_{J_1 m_1 \kappa_1}]
 \, ,
\end{aligned}
\label{eq:EOM_fermions_su122}
\eeq
plus the hermitian conjugate.
Putting back this expression inside Eqs.~\eqref{eq:Yukawa_type_interaction_su_1_2_2} and \eqref{eq:fermion_current_su122}, we obtain a total of three contributions.

The first one gives rise to the effective interaction 
\beq
\begin{aligned}
 & - 2g^2 \sum_{a=1,2} \sum_{J m \tilde{m}} \sum_{\kappa=\pm 1} \sum_{J_i, m_i} \frac{(-1)^{m_1-\tilde{m}_1+\frac{\kappa_1}{2}+\tilde{m}_4-m_4+\frac{\kappa_4}{2}}}{\sqrt{\omega_{J_2} \omega_{J_3}} \le \kappa \omega^{\psi}_J + 2 \tilde{m} + \frac{1}{2}\ri}  
\mathcal{F}^{J m \tilde{m} \kappa}_{\bar{\CJ}_1 \bar{\CM}_1; \CJ_2 \CM_2} \mathcal{F}^{J m \tilde{m} \kappa}_{\bar{\CJ}_4 \bar{\CM}_4; \CJ_3 \CM_3} \\
& \times \tr \le  [(\Phi^{\dagger}_1)^{J_3 m_3} , (\zeta^a)_{J_4 m_4}]  [(\zeta^{\dagger}_a)_{J_1 m_1} , (\Phi_1)^{J_2 m_2}] \ri \, ,
\end{aligned}
\label{eq:Yukawa_interaction_su_1_2_2_after_integrating_out}
\eeq
where we introduced the normalization of scalar fields given in Eq.~\eqref{eq:def_dynamical_modes_su122}.

This interaction contains a sum over intermediate momenta and chiralities $\kappa=\pm 1.$
It turns out that after shifting $J \rightarrow J-\frac{1}{2}$ in the terms with $\kappa=1,$ the argument of the sum cancels.
Indeed, a careful analysis of the endpoints of summation specified in Eqs.~\eqref{eq:triangle_inequalities_kappa1_su122} and \eqref{eq:triangle_inequalities_kappa-1_su122} reveals that 
when $J_2 > J_1,$ all the terms in the sum cancel to 0. Instead there is a non-vanishing contribution coming from $\kappa=-1$ and $J= \Delta J,$ as defined in Eq.~\eqref{eq:definition_DeltaJ_Deltam}, when $J_1 \geq J_2 .$
The result reads
\beq
\frac{1}{2N} \sum_{a=1,2} \sum_{J_i m_i} \frac{C^{J_1, m_1}_{\Delta J, \Delta m;J_2 m_2} C^{J_4, m_4}_{J, \Delta m;J_3 m_3}}{\sqrt{(2 J_1 +1)(2J_4+1)}}
 \tr \le  [(\Phi^{\dagger}_1)^{J_3 m_3} , (\zeta^a)_{J_4 m_4}]  [(\zeta^{\dagger}_a)_{J_1 m_1} , (\Phi_1)^{J_2 m_2}] \ri \, .
 \label{eq:contribution1_mediated_nondyn_fermion}
\eeq
The second contribution from Eq.~\eqref{eq:EOM_fermions_su122} produces the term
\beq
\begin{aligned}
& \frac{g^2}{2} \sum_{a=1,2} \sum_{J_i m_i} \sum_{J m \tilde{m}} \sum_{\kappa= \pm 1}  \frac{1}{ \sqrt{\omega_{A,J_2} \omega_{A,J_3}} \le \kappa \omega^{\psi}_J + 2 \tilde{m}+ \frac{1}{2} \ri}  \mathcal{G}^{J m \tilde{m} \kappa;  \hat{\CJ}_2 \hat{\CM}_2}_{\bar{\CJ}_1 \bar{\CM}_1}  \\
& \times \mathcal{G}^{\bar{\CJ}_4 \bar{\CM}_4}_{J m \tilde{m} \kappa; \hat{\CJ}_3 \hat{\CM}_3} 
\tr \le [(\zeta^{\dagger}_a)_{J_4 m_4} , A^{J_3 m_3}] [A^{\dagger}_{J_2 m_2} , \zeta^a_{J_1 m_1}]  \ri \, .
\end{aligned}
\eeq
The previous expression contains both a sum over $J$ and over $\kappa= \pm 1.$
It turns out that shifting $J \rightarrow J-\frac{1}{2}$ in the term of the sum coming from $\kappa=1$ gives a full cancellation with the term having $\kappa=-1,$ thus reducing again the result to a boundary contribution.
In paricular, the only non-vanishing result comes from the non-trivial remainder at the boundary\footnote{Notice that this definition of $\Delta J$ has opposite sign with respect to Eq.~\eqref{eq:definition_DeltaJ_Deltam}. On the other hand, since the regime where the result is non-vanishing corresponds to $J_2 \geq J_1$ or equivalently $\Delta J \geq 0,$ this scenario is the same that we considered for the terms mediated by the non-dynamical gauge field.} $J= \Delta J \equiv  J_2 - J_1 = J_3 -J_4$ in the case with $J_2 \geq J_1.$
In this way we obtain 
\beq
\begin{aligned} 
& \frac{1}{2N} \sum_{a=1,2} \sum_{J_i m_i} \sum_{\Delta J=0}^{\infty} \delta^{J_1-J_2}_{J_4-J_3} 
 \sqrt{\frac{\omega_{J_1} \omega_{J_4}}{\omega_{J_2} \omega_{A,J_2} \omega_{J_3} \omega_{A,J_3}}} \\
 & \times C^{J_2 m_2}_{\Delta J, \Delta m; J_1 m_1} 
C^{J_3 m_3}_{\Delta J, \Delta m; J_4 m_4}
 \tr  \le [(\zeta^{\dagger}_a)_{J_4 m_4} , A^{J_3 m_3}] [A^{\dagger}_{J_2 m_2} , \zeta^a_{J_1 m_1}]  \ri \, .
\end{aligned} 
\label{eq:contribution2_mediated_nondyn_fermion}
\eeq
The last set of interactions involves the mixed terms in eq.~\eqref{eq:EOM_fermions_su122}, that contributes to the effective interaction
\beq
\begin{aligned}
& i g^2 \sum_{J_i m_i \kappa_i \rho_i}  \sum_{J m \tilde{m}} \sum_{\kappa= \pm 1}
\frac{(-1)^{m_1-\tilde{m}_1+\frac{\kappa_1}{2}}}{\sqrt{\omega_{J_2} \omega_{A,J_3}} \le \kappa \omega^{\psi}_J + 2 \tilde{m} + \frac{1}{2} \ri}
 \mathcal{F}^{J m \tilde{m} \kappa}_{\bar{\CJ}_1, -\bar{\CM}_1; \CJ_2 \CM_2}
 \mathcal{G}^{\bar{\CJ}_4 \bar{\CM}_4}_{J m \tilde{m} \kappa; \hat{\CJ}_3 \hat{\CM}_3} \\
& \times  \epsilon^{a b} \,  \tr \le [A_{J_3 m_3}, (\zeta^{\dagger}_a)_{J_4 m_4}] [\Phi^{J_2 m_2}, (\zeta^{\dagger}_b)_{J_1 m_1}]  \ri + 
\mathrm{h.c.}
\end{aligned}
\eeq
Since all the fields are taken to be dynamical, we are using the normalizations \eqref{eq:def_dynamical_modes_su122} and we are putting $\rho_3 = -1$ and $\kappa_1=\kappa_4 = -1.$ 
In principle two possibilities for the gauge field are allowed, {\sl{i.e.}} to take the eigenvalue of momentum $\tilde{m}_3 = \pm (J_3 +1),$ but the analysis of triangle inequalities shows that one of the cases is forbidden.

The double sum over $J, \kappa$ is solved by observing that a shift $J \rightarrow J - \frac{1}{2}$ in the term with $\kappa=1$ makes the argument to vanish.
A careful analysis of the range of summation shows that if $J_1<J_2$ the result vanishes, while if $J_1 \geq J_2,$ we have a non-vanishing contribution coming from the lower boudary of summation:
\beq
\begin{aligned}
- \frac{1}{2N} \sum_{J_i m_i}  
\frac{\delta^{J_2-J_1}_{J_4-J_3}}{\sqrt{\omega_{J_2} \omega_{A,J_3}}}
C^{J_2 m_2}_{\Delta J, \Delta m; J_1 m_1}  C^{J_4 m_4}_{\Delta J, \Delta m; J_3 m_3}  \epsilon^{a b} \,  \tr \le [A_{J_3 m_3}, (\zeta^{\dagger}_a)_{J_4 m_4}] [\Phi^{J_2 m_2}, (\zeta^{\dagger}_b)_{J_1 m_1}]  \ri + \mathrm{h.c.}
\end{aligned}
\label{eq:new_interaction_mixed_su122}
\eeq
In this case we defined $\Delta J \equiv J_2 - J_1 = J_3 - J_4 \geq 0 .$

The sum of Eqs.~\eqref{eq:contribution1_mediated_nondyn_fermion}, \eqref{eq:contribution2_mediated_nondyn_fermion} and \eqref{eq:new_interaction_mixed_su122} nicely combines into a simple expression if we define the following blocks:
\begin{align}\label{eq:Fblock_su122}
	\begin{split}
(F^a)_{l,\Delta m} &\equiv \sum_{s_2=0}^{\infty} \sum_{m_2=-\frac{s_2}{2}}^{\frac{s_2}{2}} 
 C^{\frac{s_2+l}{2}, m_2+\Delta m}_{\frac{l}{2}, \Delta m; \frac{s_2}{2},m_2} 
 \, \epsilon^{a b} \,
\frac{[(\zeta_b)_{s_2+l, m_2+\Delta m}, \Phi^{\dagger}_{s_2, m_2}]}{\sqrt{s_2+l+1}} \, ,
\\
(K^a)_{l,\Delta m} &\equiv  \sum_{s_2=0}^{\infty} \sum_{m_2=-\frac{s_2}{2}}^{\frac{s_2}{2}}
\sqrt{\frac{s_2+1}{(s_2+l+1)(s_2+l+2)}} \,
 C^{\frac{s_2+l}{2},m_2+\Delta m}_{\frac{l}{2}, \Delta m; \frac{s_2}{2},m_2} 
[(\zeta^{\dagger}_{a})_{s_2, m_2}, A_{s_2+l, m_2+\Delta m}] \, .
\end{split}
\end{align}
In this way all the terms mediated by the non-dynamical fermion are given by
\beq
 H_{\rm int} = \frac{1}{2N}
\sum_{a=1,2} \sum_{l=0}^{\infty}  \sum_{\Delta m=-\frac{l}{2}}^{\frac{l}{2}}
 \tr \le (F^{\dagger}_a + K^{\dagger}_a)_{l, \Delta m} (F^a + K^a)_{l, \Delta m}  \ri \, .
\eeq


\subsubsection*{Interactions III: terms mediated by non-dynamical scalar}
\label{sect-gauge_scalar_interactions_su122}

In this subsection we present effective interactions arising 
from integrating out a non-dynamical scalar field from the same kind of terms: the Yukawa contribution \eqref{eq:Yukawa_type_interaction_su_1_2_2} and the minimal coupling in Eq.~\eqref{eq:scalar_current_su122}.

If we consider  Eq.~\eqref{eq:Yukawa_type_interaction_su_1_2_2} with the scalar field $\Phi_1$ being auxiliary with momenta $(J,m,\tilde{m}),$ and the two fermion modes being dynamical with momenta $(J_1, m_1, \tilde{m}_1)$ and $(J_2, m_2, \tilde{m}_2),$ it is easy to determine the following constraints from momentum conservation:
\beq
\tilde{m} = J_1 + J_2 + 1 \, , \qquad
-J \leq \tilde{m} < J \, , \qquad
|J_1-J_2| \leq J \leq J_1 + J_2  \, ,
\eeq 
These are however in contradiction: then the corresponding interaction term vanishes.

The only non-vanishing term arises from the scalar/gauge interaction \eqref{eq:scalar_current_su122}.
We already considered in \eqref{eq:total_interaction_gauge_mediated_su122} all the contributions mediated by non-dynamical modes of the gauge field: here we study instead how to select the auxiliary modes of the scalar to be integrated out.
We immediately note that $\rho=-1$ is fixed for all the gauge fields, since these are dynamical.
The analysis of momentum conservation and of triangle inequalities implies that the effective Hamiltonian obtained by integrating out the scalar field reads
\beq
\begin{aligned}
 &  - 4 g^2 \sum_{J m \tilde{m}} \sum_{J_i m_i} \frac{1}{\omega^2_{J}-(2 \tilde{m}+1)^2} \sqrt{\frac{ J (J+1) J_1 (J_1+1) }{\omega_{J_1} \omega_{A,J_2}  \omega_{A,J_3} \omega_{J_4} }} \\
&\times \mathcal{D}^{J m \tilde{m}}_{\CJ_1, \CM_1; \hat{\CJ}_2, \hat{\CM}_2}
 \mathcal{D}^{\CJ_4, \CM_4}_{J m \tilde{m} 0; \hat{\CJ}_3, \hat{\CM}_3} 
 \tr \le [\Phi^{\dagger}_{J_4 m_4} , A^{J_3 m_3} ]  [A^{\dagger}_{J_2 m_2} , \Phi_{J_1 m_1} ] \ri \, .
\end{aligned}
\label{eq:scalar_gauge_term_after_integrating_out_su122}
\eeq
In order to perform explicitly the summation over the intermediate momenta $J,$ it is crucial to identify the boundaries of the interval where it is defined.
Precisely we have the constraints (coming from triangle inequalities)
\beq
\mathrm{Max} \, \lbrace J_2-J_1+1 , J_1-J_2 \rbrace  \leq J \leq \mathrm{Min} \, \lbrace J_1 + J_2 , J_3 + J_4 \rbrace \, .
\eeq
We observe that the lowest bound is not symmetric under an exchange of $J_1 \leftrightarrow J_2.$
On the other hand, thank to the symmetries of the Clebsch-Gordan coefficients, the sum in Eq.~\eqref{eq:scalar_gauge_term_after_integrating_out_su122} organizes in such a way that the contributions for the two cases $J_1 \geq J_2$ and $J_1 < J_2$ exactly cancel. 
The only non-vanishing term comes from the non-trivial remainder at the boundary $J= \Delta J \equiv  J_2 - J_1 = J_3 -J_4,$ giving
\beq
\begin{aligned}
H_{\rm int} & = \frac{1}{2N} \sum_{J_i, m_i} \sum_{\Delta J=0}^{\infty}
\frac{2\Delta J +1}{\sqrt{\omega_{J_2} \omega_{A,J_2} \omega_{J_3} \omega_{A, J_3}}}
\, \delta^{J_1-J_2}_{J_4-J_3} 
C^{J_2, m_2}_{\Delta J, \Delta m; J_1, m_1} 
C^{J_3, m_3}_{\Delta J, \Delta m; J_4, m_4} \\
& \times \tr \le [\Phi^{\dagger}_{J_4 m_4} , A_{J_3 m_3} ]  [A^{\dagger}_{J_2 m_2} , \Phi_{J_1 m_1} ] \ri \, .
\end{aligned}
\eeq
The result can be written compactly by defining the block
\beq
W_{l, \Delta m} = \sum_{s_2=0}^{\infty} \sum_{m_2=-\frac{s_2}{2}}^{\frac{s_2}{2}}
\sqrt{\frac{l+1}{(s_2+l+1)(s_2+l+2)}}
 C^{\frac{s_2+l}{2}, m_2+\Delta m}_{\frac{s_2}{2},m_2; \frac{l}{2},\Delta m} \, [\Phi^{\dagger}_{s_2 m_2}, A_{s_2+l, m_2 + \Delta m}] \, ,
 \label{eq:W_block_su122}
\eeq
so that we obtain
\beq
H_{\rm int} =
\frac{1}{2N} \sum_{l=0}^{\infty} \sum_{\Delta m=-\frac{l}{2}}^{\frac{l}{2}} 
\tr \le W^{\dagger}_{l, \Delta m}  W_{l, \Delta m}  \ri \, .
\eeq


\subsubsection*{Final result}

Summing all the non-vanishing interactions, we find
\beq
\begin{aligned}
 H_{\rm int} &  = \frac{1}{2N} \sum_{l=1}^{\infty} \sum_{\Delta m=-\frac{l}{2}}^{\frac{l}{2}}
\frac{1}{l} \tr \le \mathbf{Q}_{l, \Delta m}^{\dagger} \mathbf{Q}_{l, \Delta m} \ri 
 + \frac{1}{2N} \sum_{l=0}^{\infty} \sum_{\Delta \mu=-\frac{l}{2}}^{\frac{l}{2}} 
\tr \le W^{\dagger}_{l, \Delta m}  W_{l, \Delta m}  \ri \\
& + \frac{1}{2N} \sum_{a=1,2} \sum_{l=0}^{\infty}  \sum_{\Delta m=-\frac{l}{2}}^{\frac{l}{2}}
 \tr \le (F^{\dagger}_a + K^{\dagger}_a)_{l, \Delta m} (F^a + K^a)_{l, \Delta m}  \ri    \, .
 \end{aligned} 
 \label{eq:final_interacting_Hamiltonian_sphere_red_su122}
\eeq
The result is naturally organized in terms of three classes of blocks, depending on the non-dynamical field mediating the interactions:
\begin{itemize}
\item The charge density 
\beq\label{eq:block-Q}
\mathbf{Q}_{l, \Delta m} \equiv q_{l, \Delta m} + \tilde{q}_{l, \Delta m} + \mathfrak{q}_{l, \Delta m} \, ,
\eeq
whose elemental contributions are defined in Eqs.~\eqref{eq:charge_density_gauge_su12}, \eqref{eq:scalar_charge_density_su122} and \eqref{eq:fermion_charge_density_su122}, contains information about the interactions mediated by the non-dynamical gauge field and its $l=0$ mode corresponds to the $\SU(N)$ charge.
\item The blocks $F^a_{l, \Delta m}$ and $K^a_{l, \Delta m},$ defined in Eq.~\eqref{eq:Fblock_su122} describe the interactions mediated by the non-dynamical fermions.
\item The block $W_{l, \Delta m}$ defined in Eq.~\eqref{eq:W_block_su122} is the only interaction mediated by the non-dynamical scalar. 
\end{itemize}
In terms of these blocks, the interaction is quadratic and positive definite.
This suggests that one can make a similar reinterpretation of the interaction terms as found in Section \ref{sec:symm_su111_near_BPS_theory} for the $\SU(1,1|1)$ near-BPS theory. Indeed, in analogy with \eqref{normblock_su111} one can define the normalized blocks
\beq\label{eq:block-relabel}
(B_0)_{l, \Delta m} \equiv \frac{1}{\sqrt{l}} \mathbf{Q}_{l, \Delta m} \, , \qquad
(B^a_1)_{l, \Delta m} \equiv F^a_{l, \Delta m} + K^a_{l, \Delta m} \, , \qquad
(B_2)_{l, \Delta m} \equiv W_{l, \Delta m} \, ,
\eeq
We show in Section \ref{sec:Su12-symmetry-nov} that they separately transforms in irreducible representations of the $\SU(1,2)$ algebra, and together comprise an irreducible representation of the $\SU(1,2|2)$ algebra. 
In terms of these blocks, the interacting Hamiltonian assumes the form 
\beq 
\label{eq:interaction-SU122}
\begin{aligned}
H_{\rm int} =& \frac{1}{2N}  
\tr \left[ \sum_{l=1}^{\infty} \sum_{\Delta m=-\frac{l}{2}}^{\frac{l}{2}}  (B_0)^{\dagger}_{l, \Delta m}  
(B_0)_{l, \Delta m} 
+ \sum_{l=0}^{\infty} \sum_{\Delta m=-\frac{l}{2}}^{\frac{l}{2}} \sum_{a=1}^2  (B_1^a)^{\dagger}_{l, \Delta m}  (B_1^a)_{l, \Delta m} \right.
\\ &
\left.
+  \sum_{l=0}^{\infty} \sum_{\Delta m=-\frac{l}{2}}^{\frac{l}{2}}   (B_2)^{\dagger}_{l, \Delta m}  (B_2)_{l, \Delta m} 
 \right] \, .
 \end{aligned}
\eeq
We show in Section \ref{sec:Su12-symmetry-nov} that this is a natural form in which one can exhibit the invariance under the $\SU(1,2|2)$ symmetry of the interaction, and that one can interpret it as a norm, extending our observations of Section \ref{sec:symm_su111_near_BPS_theory}.

\section{Symmetry of $\SU(1,2|2)$ near-BPS theory} \label{sec:Su12-symmetry-nov}

In this section we show that the interacting part of the Hamiltonian of the $\SU(1,2|2)$ near-BPS theory has the same type of structure as the $\SU(1,1|1)$ near-BPS theory, as exhibited in Section \ref{sec:symm_su111_near_BPS_theory}. In particular, we show that the interactions, being quartic in the fields, are built out of blocks quadratic in the fields that comprise a particular irreducible representation of the $\SU(1,2|2)$ algebra. Using this block structure, we show that the interaction is invariant with respect to the action of any of the $\SU(1,2|2)$ generators. In connection with this, we show that one can interpret the interaction term as a norm in the linear space of the irreducible representation. Finally, we note that the $\SU(1,1|1)$ near-BPS interaction emerges as a subcase of the $\SU(1,2|2)$ near-BPS interaction by setting certain modes to zero.

In Section \ref{sect-su122limit} we considered the limit \eqref{Hint_limit_su122} of $\CN=4$ SYM on a three-sphere which approaches the BPS bound $E \geq S_1 + S_2 + Q_1$. In this limit, the effective dynamics is described classically by the following $\SU(1,2|2)$ near-BPS Hamiltonian 
\begin{equation}
H_{\rm limit} = H_0 + \tilde{g}^2 H_{\rm int} \,,
\end{equation}
where $H_0$ is given by \eqref{H0_su122} and $H_{\rm int}$ by \eqref{eq:interaction-SU122}. In this section we present the symmetry structure of $H_{\rm int}$ and show that the block form anticipated in \eqref{eq:interaction-SU122} indeed is natural from the point of view of the algebra of $\SU(1,2|2)$. 

Before analyzing the structure of \eqref{eq:interaction-SU122} we change the notation for labelling the states with respect to the $\SU(1,2)$ subalgebra. In Section \ref{sect-su122limit} we used the notation $(J,m)$ associated with spherical harmonics but in this section we find it convenient to use instead the alternative labels $(n,k)$ given by
\begin{align}
\label{dict_nk_labels}
	\begin{split}
		& n_i=J_i-m_i, \qquad k_i = J_i+m_i \, , \\
		&  n = n_2-n_1 =\Delta J-\Delta m, \qquad   k = k_2-k_1=\Delta J+\Delta m \, .
	\end{split}
\end{align}
Note also $l = 2 \Delta J$.
In this notation we have the Dirac (anti-)brackets for the modes of the fields
\begin{equation}
\label{dirac_brackets_su122}
\begin{array}{c} \ds
\Big\{ (\Phi_{n,k})^i{}_j , (\Phi_{n',k'}^\dagger)^m{}_l \Big\}_D = i \delta_{n,n'}\delta_{k,k'} \delta^i _l \delta ^m _ j
\spa \Big\{ (\zeta^a_{n,k})^i{}_j , (\zeta^{a' \, \dagger}_{n',k'})^m{}_l \Big\}_D = i \delta_{a,a'} \delta_{n,n'} \delta_{k,k'} \delta^i _l \delta ^m _ j \,,
\\[2mm] \ds
\Big\{ (A_{n,k})^i{}_j , (A_{n',k'}^\dagger)^m{}_l \Big\}_D = i \delta_{n,n'}\delta_{k,k'} \delta^i _l \delta ^m _ j \,.
\end{array}
\end{equation}
Using the dictionary \eqref{dict_nk_labels} one can translate the expressions \eqref{eq:charge_density_gauge_su12}, \eqref{eq:scalar_charge_density_su122}, \eqref{eq:fermion_charge_density_su122}, \eqref{eq:total_charge_density_su122}, \eqref{eq:Fblock_su122} and \eqref{eq:W_block_su122}, as well as  the blocks defined by \eqref{eq:block-relabel}, to use labelling $(n,k)$. 
With this, the interaction \eqref{eq:interaction-SU122} attains the form 
\begin{align}\label{eq:SU122-inter-block}
	H_{\rm int}= \frac{1}{2N}\sum_{n,k=0}^\infty \tr \left( (1-\delta_{n+k,0}) (B_0^\dagger)_{n,k}(B_0)_{n,k} + \sum_{a=1}^2 (B_1^{a\dagger})_{n,k} (B_1^a)_{n,k}  + (B_2^\dagger)_{n,k} (B_2)_{n,k} \right) \, .
\end{align}
Note that the $(n,k)$ summation labels takes values along non-negative integers and the $(1-\delta_{n+k,0})$ factor is there since there is no block $(B_0^\dagger)_{n,k}$ for $n=k=0$.
One can also translate the $\SU(2)$ Clebsch-Gordan coefficient entering the blocks. In particular, Eq.~\eqref{eq:CGmJexpliti} becomes%
\footnote{Actually, the coefficients \eqref{eq:newCcoef} have appeared in \cite{Grant:2008sk} as the holomorphic constraint satisfied by BPS solutions of the free theory.
After subtle normalization of the constraint, we can identify our blocks as the holomorphic constraints there.
These constraints fix the near-BPS dynamics of the BPS operators.}
\begin{equation}\label{eq:newCcoef}
	\mathfrak{C}^{n_1k_1}_{nk}=
	\sqrt{\frac{(n_1+k_1)!(n+k)!(n_1+n)!(k_1+k)!}{(n_1+k_1+n+k)!n_1!k_1! n!k!}} \, .
\end{equation}
There are several reasons that it is convenient to  use $(n,k)$ to label the blocks. 
For one thing, with the spherical harmonic labels $(J,m)$ we need to sum over $J$ being both integer and half-integer, while also imposing $|m|\leq J$. Another motivation is that  the $(n,k)$ label is adapted to a new basis for SU$(1,2)$ algebra, as shown in Eq.~\eqref{eq:newbasisforSU21}.\footnote{The standard way of thinking about representations of SU$(1,2)$ would be based on the corresponding isotropy group SU$(2)\times \mathrm{U}(1)$, which contains the U$(1)^2$ Cartan generator $(J_0,D)$ to specify the states.
Instead, the basis in \eqref{eq:newbasisforSU21} has the U$(1)^2$ Cartan generators $(D-J_0,D+J_0)$. 
} 
This basis and the $(n,k)$ labels make the derivative descendants explicit, {\sl e.g.}~as for the fields written in \eqref{eq:gaugecomponent-nkdesc}. This could possibly be a clue to how to construct a 2+1 dimensional (semi-)local field theory description of the $\SU(1,2|2)$ near-BPS theory, in analogy with the 1+1 dimensional semi-local description of the $\SU(1,1|1)$ near-BPS theory of  \cite{Baiguera:2020jgy}, although this is not clear at this point.%
\footnote{It is expected \cite{Harmark:2007px} that the SU$(1,2)$ subsector can be effectively described by a quantum field theory in $2+1$ dimensions. 
The fact that the new algebra basis can be considered as a basis to decompose SU$(1,2)$ as two interacting SU$(1,1)$ could offer a clue to this.  Indeed, one could speculate that if we label the coordinates of the 1+2 dimensional field theory as $(t,x_1,x_2)$, then $x_1$ and $x_2$ correspond separately to a SU$(1,1)$ subsector, in accordance with the labelling $(n,k)$ that give the number of derivatives $d_1$ and $d_2$ for these two directions, respectively. However, we will postpone the construction of a possible local field theory description in $2+1$ dimensions to a future work.}

We investigate now the symmetry properties of the expression \eqref{eq:SU122-inter-block} for $H_{\rm int}$. Using the analysis of how the generators of $\SU(1,2|2)$ act on the spectrum in Appendix \ref{sec:SU122-represen}, one can find the generators of $\SU(1,2|2)$ expressed in terms of modes of the fields by requiring consistency with the Dirac (anti-)brackets \eqref{dirac_brackets_su122}. 
We begin by analyzing the action of the $\SU(1,2)$ generators. 
To this end, we will label the letters in SU$(1,2|2)$ subsector as 
\begin{equation}\label{eq:letter-label}
	\mW_I = (\Phi,\zeta^{1,2},\bar{F}_+) \, .
\end{equation}
Note that for $I=2$ the fully accurate definition is $\mW_I^a = \zeta^a$ with $a=1,2$ since the spectrum has two fermionic species.
From \eqref{eq:SU21onstates}, we acquire
the  generators in terms of fields $\mW_I$ (labelled in \eqref{eq:letter-label}) 
as
\begin{equation}\label{eq:SU12-generator-bysusy}
\begin{array}{rl}
L_+ &= \sum_{I=1}^3 \sum_{n,k=0}^\infty \sqrt{(n+1)(n+k+I)} \tr \Big[ \big( \mW_I^\dagger\big)_{n+1,k} \big( \mW_I\big)_{n,k} \Big] \, , \\[2mm]
L_0 &= \sum_{I=1}^3 \sum_{n,k=0}^\infty \left(n+\frac{k+I}{2} \right)\tr\Big[ \big( \mW_I^\dagger\big)_{nk} \big( \mW_I\big)_{n,k} \Big] \, , \\[2mm]
L_- &= \sum_{I=1}^3 \sum_{n,k=0}^\infty \sqrt{n(n+k+I-1)} \tr \Big[\big( \mW_I^\dagger\big)_{n-1,k} \big( \mW_I\big)_{n,k} \Big] \, ,  \\[2mm]
\tilde{L}_+ &= \sum_{I=1}^3 \sum_{n,k=0}^\infty \sqrt{(k+1)(n+k+I)}\tr \Big[ \big( \mW_I^\dagger\big)_{n,k+1} \big( \mW_I\big)_{n,k} \Big] \, , \\[2mm]
\tilde{L}_0 &= \sum_{I=1}^3 \sum_{n,k=0}^\infty \left(k+\frac{n+I}{2} \right) \tr \Big[\big( \mW_I^\dagger\big)_{n,k} \big( \mW_I\big)_{n,k} \Big] \, , \\[2mm]
\tilde{L}_- &= \sum_{I=1}^3 \sum_{n,k=0}^\infty \sqrt{k(n+k+I-1)} \tr \Big[\big( \mW_I^\dagger\big)_{n,k-1} \big( \mW_I\big)_{n,k} \Big] \, , \\[2mm]
J_+ &= \sum_{I=1}^3 \sum_{n,k=0}^\infty \sqrt{k(n+1)} \tr \Big[\big( \mW_I^\dagger\big)_{n+1,k-1} \big( \mW_I\big)_{n,k}  \Big] \, , \\[2mm]
J_- &= \sum_{I=1}^3 \sum_{n,k=0}^\infty \sqrt{n(k+1)} \tr \Big[\big( \mW_I^\dagger\big)_{n-1,k+1} \big( \mW_I\big)_{n,k} \Big] \, ,
\end{array}
\end{equation}
where $I=1,2,3$ stands for scalars, fermions and gauge fields, respectively. Note that when $I=2$ in the above sums one should sum over $a=1,2$, but we have chosen to hide this sum for brevity of the expressions.
We introduce a similar notation for the blocks 
\begin{equation}
(B_I^\dagger)_{n,k} = \Big( ( B_0^\dagger)_{n,k}, ( B_1 ^{1,2} {}^\dagger)_{n,k}, ( B_2^\dagger)_{n,k} \Big) \,.
\end{equation}
However, notice that this is with $I=0,1,2$ instead, thus for the blocks the index $I$ does not run over the same values as for the spectrum. Note also that for $I=1$ the fully accurate version of the above is $(B_I^{a\dagger})_{n,k} = ( B_1 ^{a\dagger})_{n,k}$ for $a=1,2$ since we have two different species of fermionic blocks.

Acting now with the $\SU(1,2)$ generators \eqref{eq:SU12-generator-bysusy} on the blocks we obtain
\begin{equation}\label{eq:SU122-block-I}
	\begin{array}{rl}
		(L_+)_D(B_I^\dagger)_{n,k}  &=  \sqrt{(n+1)(n+k+I)} (B_I^\dagger)_{n+1,k}\,, \\[1mm]
					(\tilde{L}_+)_D(B_I^\dagger)_{n,k}  &=  \sqrt{(k+1)(n+k+I)} (B_I^\dagger)_{n,k+1}\,, \\[1mm]
		(L_0)_D(B_I^\dagger)_{n,k}  &=  \left(n+\frac{k+I}{2} \right) (B_I^\dagger)_{n,k}\,, \\[1mm] 	
		(\tilde{L}_0)_D(B_I^\dagger)_{n,k} &=  \left(k+\frac{n+I}{2} \right) (B_I^\dagger)_{n,k}\,, \\[1mm]
		(L_{-})_D(B_I^\dagger)_{n,k}  &=  \sqrt{n(n+k+I-1)} (B_I^\dagger)_{n-1,k}\,,  \\[1mm]
		(\tilde{L}_{-})_D(B_I^\dagger)_{n,k} &=  \sqrt{k(n+k+I-1)} (B_I^\dagger)_{n,k-1}\,, \\[1mm]
		(J_+)_D(B_I^\dagger)_{n,k}   &= \sqrt{k(n+1)} (B_I^\dagger)_{n+1,k-1}\,, \\[1mm]
		(J_-)_D(B_I^\dagger)_{n,k}   &= \sqrt{n(k+1)} (B_I^\dagger)_{n-1,k+1} \,,
	\end{array}
\end{equation}
where $I=0,1,2$ for these blocks shown in \eqref{eq:interaction-SU122}. We used the same notation \eqref{general_brac} introduced in Section \ref{sec:symm_su111_near_BPS_theory}, but here with respect to the Dirac (anti-)brackets \eqref{dirac_brackets_su122}.
Comparing \eqref{eq:SU122-block-I} with the representations shown in the appendix \ref{sec:SU122-represen} we see that the block $(B_I^\dagger)_{n,k}$ is in the $(p,q)=(0,I-3)$ representation of $\SU(1,2)$, where $I=0,1,2$. Here we are using the $\SU(1,2)$ representation labels $p$ and $q$ defined in Appendix \ref{appen:SU12}. Notice that these block representations are in contrast with those of the spectrum, where one has the $(p,q)=(0,I-3)$ representations of $\SU(1,2)$ with $I=1,2,3$. 

Considering now $H_{\rm int}$ in Eq.~\eqref{eq:SU122-inter-block} we see that the three terms (including sums over $n$ and $k$, and $a$ for the second term) each correspond to a different representation of $\SU(1,2)$. The first term corresponds to the $(p,q)=(0,-3)$, the second to the $(p,q)=(0,-2)$ and the third term to the $(p,q)=(0,-1)$ representation of $\SU(1,2)$. Therefore, the invariance with respect to the $\SU(1,2)$ generators \eqref{eq:SU12-generator-bysusy} should be shown separately for those three terms. This is indeed what one finds.

\begin{figure}
	\centering
	\begin{subfigure}{.5\textwidth}
		\centering
		\includegraphics[trim=3cm 15cm 8cm 4cm,width=0.96\linewidth]{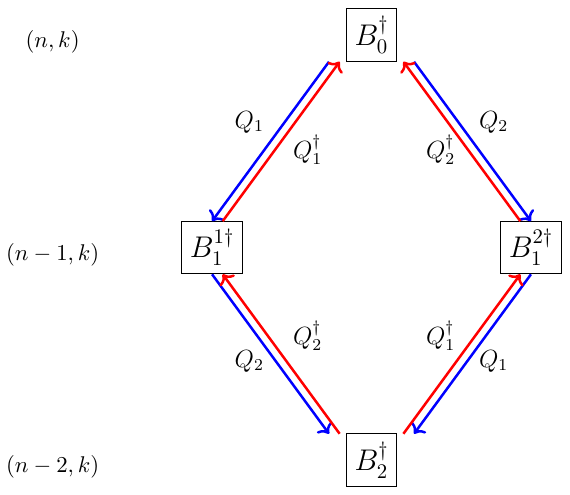}
		\caption{Actions by supercharge generator $Q_i$}
		\label{fig:R3-x-sub}
	\end{subfigure}%
	\begin{subfigure}{.45\textwidth}
		\centering
		\includegraphics[trim=3cm 15cm 9cm 4cm,width=0.96\linewidth]{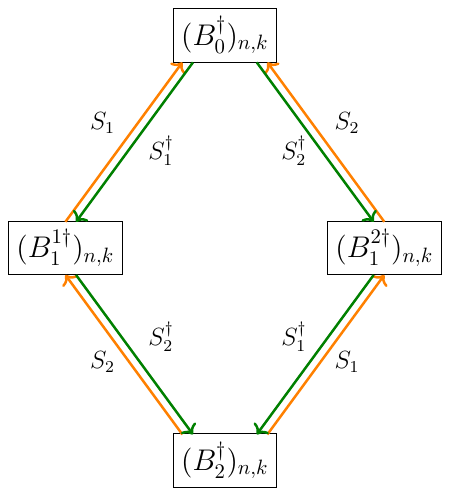}
		\caption{Actions by supercharge generator $S_i$}
		\label{fig:R2-x-sub}
	\end{subfigure}
	\caption{\footnotesize The blocks as an $\mN=2$ vector multiplet. 
		The blue arrows are actions by $Q_i$ while the red arrows are the actions by $Q_i^\dagger$.
		Similarly the orange arrows are the actions of $S_i$ while the green arrows are the actions by $S_i^\dagger$.
		The  $\tilde{Q}_i$ supercharges are not shown in this diagram, which will only differ from this one by changing the $k$ levels, which corresponds to the derivative operator $d_2$. Note that the supercharges $Q_i$ will change the level by $1$ while the supercharges of $S$ kind do not change the levels.}
	\label{fig:diagram}
\end{figure}

Turning to the fermionic generators of $\SU(1,2|2)$ we find
\begin{equation}\label{eq:supercharge-SU122-4}
	\begin{array}{rl}
		Q_1^\dagger &=\sum_{n,k=0}^\infty \sqrt{n+1} \tr (\Phi_{n+1,k}^\dagger \zeta^2_{n,k}) +\sum_{n,k=0}^\infty\sqrt{n+1} \tr( \zeta^{1\dagger}_{n+1,k} A_{n,k}) \, ,	\\[2mm]
		\tilde{Q}_1^\dagger &=\sum_{n,k=0}^\infty \sqrt{k+1} \tr( \Phi_{n,k+1}^\dagger \zeta^2_{n,k} )+ \sum_{n,k=0}^\infty\sqrt{k+1} \tr( \zeta^{1\dagger}_{n,k+1} A_{n,k}) \, , \\[2mm]
		S_1^\dagger &= \sum_{n,k=0}^\infty \sqrt{n+k+1} \tr(\zeta_{n,k}^{2\dagger} \Phi_{n,k} )+\sum_{n,k=0}^\infty \sqrt{n+k+2} \tr(A_{n,k}^\dagger \zeta^1_{n,k}) \, ,	 \\[2mm]
		Q_2^\dagger &= \sum_{n,k=0}^{\infty} \sqrt{n+1} \tr(\Phi_{n+1,k}^\dagger \zeta_{n,k}^1) -\sum_{n,k=0}^\infty \sqrt{n+1} \tr( \zeta^{2\dagger}_{n+1,k} A_{n,k} ) \, , \\[2mm]
		\tilde{Q}_2^\dagger &= \sum_{n,k=0}^{\infty} \sqrt{k+1} \tr( \Phi_{n,k+1}^\dagger \zeta_{n,k}^1) -\sum_{n,k=0}^{\infty} \sqrt{k+1} \tr(\zeta^{2\dagger}_{n,k+1} A_{n,k}) \, , \\[2mm]
		S_2^\dagger &= \sum_{n,k=0}^\infty \sqrt{n+k+1} \tr( \zeta^{1\dagger}_{n,k} \Phi_{n,k}) -\sum_{n,k=0}^\infty \sqrt{n+k+2} \tr(A^\dagger_{n,k} \zeta^{2}_{n,k}) \, .
	\end{array}
\end{equation}
We note that the SU$(1,2)$ generators \eqref{eq:SU12-generator-bysusy} can be worked out as anti-commutators of these using \eqref{eq:anticommu-QtildeQ} and \eqref{eq:anticommu-QS}. 
Acting on the blocks with the above fermionic generators and their hermitian conjugates, we find that the non-zero actions are
\begin{equation}\label{eq:SU122-super-block}
	\begin{array}{rll}
		\ &	
		(Q_1^\dagger)_D (B_1^{1\dagger})_{n,k}  = \sqrt{n+1} (B_0^\dagger)_{n+1,k}  
		\, , &  
		(Q_1^\dagger)_D(B_2^\dagger)_{n,k} = -\sqrt{n+1} (B_1^{2\dagger})_{n+1,k} 
		\, , \\
		\ &	
		(Q_2^\dagger)_D (B_1^{2\dagger})_{n,k} =  -\sqrt{n+1} (B_0^\dagger)_{n+1,k}
		 \, ,  & 
		(Q_2^\dagger)_D (B_2^{\dagger})_{n,k} = -\sqrt{n+1} (B_1^{1\dagger})_{n+1,k} \, , \\
		 &	 (S_1^\dagger)_D(B_0^\dagger)_{n,k} = \sqrt{n+k} (B_1^{1\dagger})_{n,k} \, , &   (S_1^\dagger)_D (B_1^{2\dagger})_{n,k}  = -\sqrt{n+k+1} (B_2^\dagger)_{n,k} \, ,  \\
		\ &	( S_2^\dagger)_D(B_0^\dagger)_{n,k} = \sqrt{n+k} (B_1^{2\dagger})_{n,k} \, , & 
		 (S_2^\dagger)_D (B_1^{1\dagger})_{n,k}  = -\sqrt{n+k+1} (B_2^{\dagger})_{n,k} \, , \\
		\ & (Q_1)_D (B_0^\dagger)_{n,k}  = \sqrt{n} (B_1^{1\dagger})_{n-1,k} \, , &   
		(Q_1)_D(B_1^{2\dagger})_{n,k}  = -\sqrt{n} (B_2^{\dagger})_{n-1,k} \, , \\
		\ & (Q_2)_D(B_0^\dagger)_{n,k}  = - \sqrt{n} (B_1^{2\dagger})_{n-1,k} \, , &   
		(Q_2)_D (B_1^{1\dagger})_{n,k}  = -\sqrt{n} (B_2^{\dagger})_{n-1,k} \, ,\\
		\ & 
		(S_1)_D (B_1^{1\dagger})_{n,k}  =\sqrt{n+k} (B_0^\dagger)_{n,k} 
		\, , \ \  &  
		(S_1)_D (B_2^{\dagger})_{n,k}= -\sqrt{n+k+1} (B_1^{2\dagger})_{n,k} 
		\, , \\
		\ &	
		(S_2)_D (B_1^{2\dagger})_{n,k}  = -\sqrt{n+k}(B_0^\dagger)_{n,k}  
		\, , \ \  &  
		(S_2)_D (B_2^{\dagger})_{n,k} = -\sqrt{n+k+1} (B_1^{1\dagger})_{n,k}
		\, .
	\end{array}
\end{equation}
From the action of the fermionic charges one observes that the blocks $B_I^\dagger$, $I=0,1,2$, form a  $\mN=2$ vector multiplet representation of SU$(1,2|2)$. This is illustrated in Figure \ref{fig:diagram}. 
The structure can be understood as follows. 
As we know, the $\mN=2$ vector multiplet is made of one $\mN=1$ vector multiplet and one $\mN=1$ chiral multiplet.
Basically, $Q_1^\dagger,S_1^\dagger$ are the supercharges relating $(B_2^\dagger, B_1^{2\dagger})$ to form an $\mN=1$ vector multiplet and $(B_0^\dagger, B_1^{1\dagger})$ to be an $\mN=1$ chiral multiplet. 
Similarly, 
$Q_2^\dagger,S_2^\dagger$ are the supercharges relating $(B_2^\dagger, B_1^{1\dagger})$ to form an $\mN=1$ vector multiplet and $(B_0^\dagger, B_1^{2\dagger})$ to be an $\mN=1$ chiral multiplet. 

In addition to the fermionic generators and $\SU(1,2)$ generators, we have also R-symmetry generators that act on the fermionic fields in the spectrum as one can see from Appendix \ref{sec:SU122-represen}. 
The R-symmetry generators are%
\begin{equation}\label{eq:R-charge}
	R_0 =\frac{1}{2}\sum_{n,k=0}^\infty \tr (\zeta^{1\dagger}_{n,k}\zeta^1_{n,k} - \zeta^{2\dagger}_{n,k}\zeta^2_{n,k}) \, , \qquad R_+= R_-^\dagger=\sum_{n,k=0}^\infty \tr ( \zeta^{1\dagger}_{n,k} \zeta^2_{n,k} )
\end{equation}
The R-charges of the blocks $B_0^\dagger$ and $B_2^\dagger$ are zero, while the blocks $B_1^{i\dagger}$ have opposite non-trivial R-charges. These are simply given by
\begin{align}
	 (R_0)_D (B_1^{1\dagger})_{n,k} = -\frac{1}{2} (B_1^{1\dagger})_{n,k}, \quad 	(R_0)_D (B_1^{2\dagger})_{n,k}  = \frac{1}{2} (B_1^{2\dagger})_{n,k} \, .
\end{align}
Finally, in the $\SU(1,2|2)$ algebra one has an additional $U(1)$ generator that we denote by $G$, see Appendix \ref{sec:SU122-represen}. In terms of the fields this generator takes the form
\begin{equation}\label{eq:GU(1)-fields}
G= -\sum_{n,k=0}^\infty \tr \left[ (\mW_1^\dagger)_{n,k} (\mW_1)_{n,k} \right] - \frac{1}{2}	\sum_{a=1,2} \sum_{n,k=0}^\infty \tr \left[ (\mW_2^{a\dagger})_{n,k} (\mW_2^a)_{n,k} \right]
\end{equation}

As we now have presented all the generators of the $\SU(1,2|2)$ algebra, we are ready to discuss the invariance of $H_{\rm int}$, as given in \eqref{eq:SU122-inter-block}, under $\SU(1,2|2)$ transformations. 
Using \eqref{eq:SU122-super-block} it is straightforward to check that when acting with any of the 8 fermionic charges on $H_{\rm int}$ one gets zero. This means that $H_{\rm int}$ is invariant with respect to the action of all 8 fermionic generators of $\SU(1,2|2)$. An example of how this works for a particular fermionic generator is given in  Appendix \ref{sec:symmetry-action}. Since the Dirac (anti-)commutators \eqref{eq:anticommu-QtildeQ} and \eqref{eq:anticommu-QS} reveal the bosonic generators of the $\SU(1,2|2)$ algebra, it follows from the associativity of the Dirac (anti-)brackets that one can obtain the action of the bosonic generators by combining actions of the fermionic generators. 
In this way it is straightforward to see that since the action of all fermionic generators gives zero, it directly follows that also the action of the bosonic generators on $H_{\rm int}$ give zero as well.

It is important to remark that the $\SU(1,2|2)$ near-BPS theory, in addition to the invariance with respect to the $\SU(1,2|2)$ generators, possesses a $\U(1)$ invariance associated with the number operator
\begin{equation}
\hat{N} = \sum_{n,k=0}^\infty \tr \left[  \Phi^\dagger_{n,k} \Phi_{n,k}  + \sum_{a=1}^2  \zeta^{a\dagger}_{n,k} \zeta^a_{n,k} + A^\dagger_{n,k} A_{n,k}  \right] \,.
\end{equation}
One can check that $\hat{N}_D H_{\rm limit}  = 0,$ which means that $\hat{N}$ is conserved and hence the $\SU(1,2|2)$ near-BPS theory is non-relativistic.


We consider now representations with respect to the full $\SU(1,2|2)$ superalgebra. 
Both the spectrum and the blocks correspond to certain representations of this superalgebra. 
We use the following Dynkin diagram of SU$(1,2|2)$
\begin{align} \label{dynkin_su122}
	\bigcirc \!\!-\!\!\!-\!\!  \textstyle \bigotimes \!\!-\!\!\!-\!\!
	\bigcirc \!\!-\!\!\!-\!\! \textstyle \bigotimes
\end{align}
with two bosonic and two fermionic roots. Indeed, the associated raising operators to this Dynkin diagram are $J_-$, $\tilde{Q}_1$, $R_+$ and $S_2$ and the associated lowering operators are the conjugates of these.
The Dynkin labels are defined as $[s_1,r_1,q_1,r_2]$ with
\begin{equation}\label{eq:Dynkin-U1Cartan}
s_1 = 2\tilde{L}_0 - 2 L_0 \spa r_1 = -\frac{2}{3} L_0 + \frac{4}{3}\tilde{L}_0+\frac{1}{3} G - R_0 \spa q_1=2R_0 \spa r_2 = \frac{2}{3} (L_0+\tilde{L}_0) - \frac{1}{3} G - R_0 
\end{equation}
in terms of the Cartan generators $L_0$, $\tilde{L}_0$, $R_0$ and $G$ of $\SU(1,2|2)$.
The highest weight for the spectrum is $|\Phi\rangle$ since $J_-$, $\tilde{Q}_1$, $R_+$ and $S_2$ all are zero when acting on this state. This corresponds to $L_0=\tilde{L}_0=\frac{1}{2}$, $R_0=0$ and $G=-1$ giving the Dynkin labels $[0,0,0,1]$. This defines the irreducible unitary representation of the spectrum, generated by repeated application of the lowering operators on the highest weight. See Appendix  \ref{sec:SU12algebra} for more details on this.

Turning now to the blocks, one finds that the action of $J_-$, $\tilde{Q}_1$, $R_+$ and $S_2$ all are zero on the particular block $(B_1^{1\dagger})_{0,0}$ with Cartan charges $L_0=\tilde{L}_0=R_0=G=\frac{1}{2}$ and Dynkin labels $[0,0,1,0]$.%
\footnote{Note here the subtlety that for $(B_0^\dagger)_{n,k}$ one does not have a block with $n=k=0$ unlike for $(B^{a\dagger}_1)_{n,k}$ and $(B^{\dagger}_2)_{n,k}$. This is consistent with the representation of $\SU(1,2|2)$. The origin of this is that $\mathbf{Q}_{0,0}$ is zero by the singlet constraint.}
 Thus, this defines the irreducible unitary representation of the blocks $(B_I^\dagger)_{n,k}$ by repeated application of the lowering operators on the highest weight $(B_1^{1\dagger})_{0,0}$. The blocks  $(B_I^\dagger)_{n,k}$  are all normalized with respect to this, which means in particular that the expression for $H_{\rm int}$ in Eq.~\eqref{eq:SU122-inter-block} uses normalized blocks.

We saw in Section \ref{sec:su111_norm} for the $\SU(1,1|1)$ near-BPS theory that the quartic interaction $H_{\rm int}$ can be written as \eqref{Hint_norm} where $\langle B | B\rangle$ can be seen as a norm (squared). We show now that this is also true for the $\SU(1,2|2)$ near-BPS theory. 
We write
\begin{equation}\label{eq:ket-SU122}
 | B \rangle = \sum_{n,k=0}^\infty \left[ (1-\delta_{n+k,0}) \ket{(B_0)_{n,k}} + \sum_{a=1}^2 \ket{(B_1^a)_{n,k}} + \ket{(B_2)_{n,k}} \right]
\end{equation}
with 
\begin{equation}
\ket{(B_I^a)_{n,k}} =\sum_{i,j} ( (B_I^a)_{n,k})^i{}_j  |i,j ; I,a,n,k \rangle
\end{equation}
where $I=0,1,2$ and $a$ is either empty (for $I=0,2$) or $a=1,2$ (for $I=1$). In this expression, $|ij ; I, a, n, k \rangle$ are normalized and orthogonal basis vectors in the linear space given by the adjoint representation of $\SU(N)$ times the irreducible representation of the $\SU(1,2|2)$ with highest weight $[0,0,1,0]$. Just like in Section \ref{sec:su111_norm}  this means one can consider the blocks as components of $|B\rangle$ in the linear space of the representation and one finds $\langle B | B \rangle$ by inserting a complete basis, demonstrating that \eqref{Hint_norm} is realized for the $\SU(1,2|2)$ near-BPS theory.

One can further elaborate on this construction by using the fact that we can generate the representation from a highest weight. However, for consistency of notations, we should work in the conjugate representations of blocks for the norm interpretation, which means that it is actually a lowest weight. 
Let us start with the lowest weight state $\ket{(B_1^2)_{0,0}}$ that is annihilated by the lowering operators $J_+$, $\tilde{Q}_1^\dagger$, $R_-$ and $S_2^\dagger$ with respect to the Dynkin diagram \eqref{dynkin_su122}.
The supermultiplet ground states are created using the corresponding raising operators
\begin{align}\label{eq:Raising-Su122}
	\begin{split}
		& S_1 \ket{(B_1^2)_{0,0}} =\ket{(B_2)_{0,0}},\qquad Q_2 \ket{(B_1^2)_{0,0}} =   \ket{(B_0)_{0,0}} \\
		&R_-\ket{(B_1^2)_{0,0}} =-\ket{(B_1^1)_{0,0}}, \qquad Q_1 \ket{(B_1^1)_{0,0}} =- \ket{(B_0)_{1,0}} \, .
	\end{split}
\end{align}
In this way, all the lowest weight states in the three distinct $\SU(1,2)$ representations are generated. 
The conjugate descendants are generated by applying the $L_-$ and $\tilde{L}_-$ operators, which are anti-commutators of supercharges \eqref{eq:anticommu-QS}.
The results simply read
\begin{align}
	\begin{split}
		\ket{(B_1^i)_{n,k}} &= \frac{1}{\sqrt{n ! k! (n+k)!}} L_-^{n} \tilde{L}_-^{k} \ket{(B_1^i)_{0,0}} \\
		\ket{(B_2)_{n,k}} &= \frac{1}{\sqrt{n ! k! (n+k+1)!}} L_-^{n} \tilde{L}_-^{k}\ket{(B_2)_{0,0}} \\
		\ket{(B_0)_{n,k}} &= \frac{1}{\sqrt{(n-1) ! k! (n+k)!}} L_-^{n-1} \tilde{L}_-^{k} \ket{(B_0)_{1,0}} \, .
	\end{split}
\end{align}
Each of these correspond to the conjugate of a $\SU(1,2)$ representation, being the $(p,q)=(0,-3),(0,-2),(0,-1)$ representations, respectively.

As a final remark, we note that the norm structure of SU$(1,1|1)$ Hamiltonian can be acquired straightforwardly from the Hamiltonian \eqref{eq:SU122-inter-block} of the SU$(1,2|2)$ theory with the state defined as \eqref{eq:ket-SU122}.
This is due to the fact that the letters of SU$(1,1|1)$ consist of an $\mN=1$ chiral multiplet including their corresponding descendants, which transform as a subset of the $\mN=2$ vector multiplet of the  SU$(1,2|2)$ theory.

\section{Conclusion and outlook}
\label{sec:conclusions}

In this paper we derived by sphere reduction and integrating out non-dynamical modes the classical interacting Hamiltonian for the $\SU(1,2|2)$ near-BPS limit of $\mathcal{N}=4$ SYM. We demonstrated that the interactions have a symmetry structure that allow them to be written on a convenient block form
\beq
	H_{\rm int}= \frac{1}{2N}\sum_{n,k=0}^\infty \tr \left( (1-\delta_{n+k,0}) (B_0^\dagger)_{n,k}(B_0)_{n,k} + \sum_{a=1}^2 (B_1^{a\dagger})_{n,k} (B_1^a)_{n,k}  + (B_2^\dagger)_{n,k} (B_2)_{n,k} \right) \, .
	\label{eq:H_int_conclusions_su122}
\eeq
These blocks correspond to a particular representation of the $\SU(1,2|2)$ superalgebra. Furthermore, we have shown that the full structure of $H_{\rm int}$ arises from considering it as a norm in the linear space spanned associated with the representations of the blocks, explaining the normalization of the terms in $H_{\rm int}$, its positive-definite form and its invariance under $\SU(1,2|2)$ transformations. 
We have also worked this out in the $\SU(1,1|1)$ case where $H_{\rm int}$ takes the reduced form 
\begin{equation}
H_{\rm int}  = \frac{1}{2N} \tr \left[  \sum_{l=1}^\infty  (B^{\dagger}_0)_l (B_0)_l  +  \sum_{l=0}^\infty  (B^{\dagger}_1)_l (B_1)_l  \right] \, .
\end{equation}
This can be seen as a subcase of the $\SU(1,2|2)$ expression \eqref{eq:H_int_conclusions_su122}.

The algebraic analysis performed in this paper justifies the underlying structures of the interactions.
One way to obtain them is to perform the classical sphere reduction \cite{Ishiki:2006rt}, which is a technical but systematic procedure to obtain a definite result.
On the other hand, it turns out that the Hamiltonian \eqref{eq:H_int_conclusions_su122} is basically the consequence of symmetry requirements with minimal inputs such as: 
1) the interactions arise from quadratic blocks involving the brackets of the fields and 2) the Hamiltonian is quadratic in these blocks. 
These conditions are then sufficient to identify the highest weight of the corresponding representation in order to obtain all the other interactions (up to constant shifts) with the action of lowering operators.
This is not only a more convenient way to proceed, but it also gives a unique prescription to find the quartic interactions.

The positive-definite structure of Eq.~\eqref{eq:H_int_conclusions_su122} points to a way that one can solve the strong-coupling limit. Sending the coupling to infinity means that only configurations with $H_{\rm int}=0$ survive, corresponding to having zero norm in the above-mentioned norm interpretation.  
This is only possible if all the blocks vanish
\begin{equation}\label{eq:vacuum-constraint}
	(B^\dagger_0)_{n,k} =0 \spa (B^{a\dagger}_1)_{n,k} =0 \spa (B^\dagger_2)_{n,k} =0 \, ,
\end{equation}
with $a=1,2$, including both bosonic and fermionic constraints. 
Thus, one should look for configurations that obey this infinite set of constraint equations. 
Given such a configuration one can then study small fluctuations around it, both in the classical and the quantum regime, to reveal the strong coupling dynamics. 

The bosonic part of the constraints \eqref{eq:vacuum-constraint} were derived in \cite{Grant:2008sk}.
Configurations that obey the constraints should be identified with $\frac{1}{8}$-BPS states \cite{Mandal:2006tk,Grant:2008sk}. 
An interesting novelty of our approach is that we find these constraints are transforming in specific representations of SU$(1,2|2)$ and the one-loop Hamiltonian exactly consists of these constraint. This should help in understanding the space of solutions. 


The interactions obtained from sphere reduction are classical; in order to quantize the system, we need to promote the Hamiltonian to an operator. With this, one should get $\SU(1,2|2)$ Spin Matrix theory \cite{Harmark:2014mpa}.
In the $\SU(1,1)$ near-BPS limits, the prescription was simply to directly promote the classical result to the quantum level, and it was possible to show that the normal ordering terms were sufficient to give the self-energy corrections in terms of harmonic numbers \cite{Harmark:2019zkn, Baiguera:2020jgy}. 
In this way, the result matched the one-loop corrections of the dilatation operator computed in quantum $\mathcal{N}=4$ SYM restricted to the sector of interest.
For the $\SU(1,2)$ limits considered here, it is interesting to understand if quantization will leave unchanged the positivity and what will happen to the interpretation of the interactions as part of a norm.
Since we expect that quantization should only change the zero-point energy and that supersymmetry should preserve the positivity of the Hamiltonian, it is interesting to investigate in detail how the $\SU(1,2|2)$ generators organize in order to obtain such structure.
We leave this topic for a future investigation.

One would expect that the structure in Eq.~\eqref{eq:H_int_conclusions_su122} can be derived from a local formulation, along the lines of \cite{Baiguera:2020jgy}: the effective Hamiltonians corresponding to limits preserving the $\SU(1,1)$ spin group were found to arise from $(1+1)-$dimensional field theories on the circle with non-canonical commutation relations.
In the case of supersymmetry, a superfield formulation was also found.
In the more general cases considered in this paper, there are two parameters containing informations about the three-sphere: the momenta $(J,m)$ or the analog quantum numbers $(n,k),$ which point towards the existence of a local description of the theories on a $(2+1)-$dimensional manifold. 
Finding a local interpretation of the interacting Hamiltonian would allow to identify novel field theories whose properties are interesting to analyze by themselves, being non-trivial realizations of the non-relativistic symmetries.

It is interesting to consider the implications of the present work in relation to the gravitational and string theory side of AdS/CFT duality. As exhibited in \cite{Harmark:2017rpg, Harmark:2018cdl, Harmark:2019upf, Harmark:2020vll} this involves non-relativistic string theories with a $\U(1)-$Galilean target space.
Depending from the number of colours of the gauge group $\SU(N),$ we can make the following observations:
\begin{itemize}
\item In the strict planar limit $N = \infty$ one can use the description one the field theory side in terms of integrable spin chains. Using a long-wavelength limit \cite{Kruczenski:2003gt,Harmark:2008gm} one obtains a sigma-model that can be compared directly to the non-relativistic string theories of \cite{Harmark:2017rpg, Harmark:2018cdl, Harmark:2019upf, Harmark:2020vll} which are found by taking the equivalent near-BPS limits of the string theory side that we have explored here for $\mathcal{N}=4$ SYM. 
This would extend the $\SU(2)$ case considered in  \cite{Harmark:2008gm}. One can find the spectra of these theories by employing integrability techniques.
\item In the planar limit where $N$ is large but not strictly infinite, one can study $1/N$ corrections from both sides of the duality. 
Following \cite{Harmark:2019upf}, it is possible to extend the classical connection between the non-relativistic sigma models and the continuum limit of the SMT spin chain to include the interactions between strings.  
\item In the case of finite $N,$ non-perturbative effects play a major role. This is related to non-perturbative states on the string side, such as D-branes and black holes.
It is the case where the analysis performed in the present paper is particularly relevant, since the presence of a positive-definite structure of the interactions allows to study the strongly-coupled regime by setting to zero the individual blocks.
This regime was considered in the $\SU(2)$ case in \cite{Harmark:2016cjq} where a direct relation between the large $g$ behavior of $\SU(2)$ SMT and dual interacting Giant Gravitons were found. It would be highly interesting if one can extend these studies to the cases considered here.
\end{itemize}

One of our future goals is to obtain and explore the PSU$(1,2|3)$ near-BPS theory, and the corresponding PSU$(1,2|3)$ Spin Matrix theory. 
Obtaining the $\SU(1,2|2)$ near-BPS theory and understanding its symmetry structure have given new insights into the possible near-BPS limits of $\CN=4$ SYM theory. However, the PSU$(1,2|3)$ near-BPS limit gives the maximally possible Spin Matrix theory that one can obtain. Thus, the PSU$(1,2|3)$ near-BPS theory includes all previously considering near-BPS theories as subcases. 
New types of interactions will be present in the PSU$(1,2|3)$ near-BPS theory since it includes an interacting triplet of scalar fields. In addition, this is the only case in which one has a Dirac equation for the fermions that is important for understanding the number of fermionic degrees of freedom. 
As a subcase it should include the PSU$(1,1|2)$ near-BPS theory previously considered in \cite{Baiguera:2020jgy}. It would be highly interesting to see if the symmetry structure of the interactions found in this work can extend to the full PSU$(1,2|3)$ case, as well as the PSU$(1,1|2)$ case.

One potential application of PSU$(1,2|3)$ SMT is to understand the microscopic states of $\frac{1}{16}$-BPS black hole in AdS$_5$ spacetime. 
It is a long term question \cite{Kinney:2005ej}  to understand how the states of $\mN=4$ SYM can capture the AdS$_5$ black hole entropy \cite{Chong:2005hr,Chong:2005da,Kunduri:2006ek}, following the original technique developed for the static magnetically charged case in $\mathrm{AdS}_4 \times S^7 $ \cite{Benini:2015eyy}.
The work \cite{Chang:2013fba} tried to resolve this problem by solving a BPS cohomology equation, whose solutions to a few low orders are found to be consistent with multi-gravitons. 
On the contrary, recent progress \cite{Cabo-Bizet:2018ehj,Choi:2018hmj, Benini:2018ywd} enlightened by \cite{Hosseini:2017mds} showed the superconformal index of $\mN=4$ SYM can successfully reproduce the black hole entropy by considering the saddles with complex chemical potentials \footnote{It is also worth to mention that recent work \cite{Murthy:2020rbd} showed the superconformal index actually interpolates between black hole phase at large charge and multigraviton phase in the small charge regime.}.
As a decoupling limit, Spin Matrix theory in PSU$(1,2|3)$ subsector might provide simpler insights to resolve this problem. 
Besides, the supersymmetric AdS$_5$ black holes have an Extremal Vanishing Horizon limit, where an AdS$_3$ factor decouples \cite{Johnstone:2013eg}.
It was shown in \cite{Goldstein:2019gpz} that such decoupling can be understood as emergence of 2d Cardy formula from the 4d superconformal index.
However, it is still unclear how operators in the UV $\mN=4$ SYM can organize themselves into a representation of the infinite dimensional Virasoro algebra in the IR. 
We expect constructing an interacting field theory describing the PSU$(1,2|3)$ subsector could help understanding the dynamics here. Another interesting phenomenon that one can hope to observe is the emergence of dual D-branes in the form of giant gravitons, similarly to the investigation done in \cite{Harmark:2016cjq}.
 
Another type of application of this work is towards the understanding of the dilatation operator of $\CN=4$ SYM theory. 
We have observed that the Spin Matrix Theory Hamiltonian consists of blocks transforming in specific representations of spin group. 
It is interesting to study whether such structures can help to reorganize the quantum corrections of dilatation operator of $\mN=4$ SYM beyond the decoupling limit \cite{Beisert:2004ry}. 
Especially, the higher loop corrections of dilatation operators are notoriously difficult to calculate. 
Some of the achievements can be found in \cite{Beisert:2003tq,Beisert:2003jj}. 
A possible mechanism like organization of higher loop terms by global symmetry would be very handy for practical calculations. 

Finally, one could contemplate connections to factorizations of 4d SCFT partition functions.
The free limit of partition function  in 4d SCFT can be factorized into holomorphic blocks, which are the partition functions in $D_2 \times T^2$ \cite{Nieri:2015yia,Gadde:2020bov}. 
The blocks have been shown to be related to 2d elliptic genera \cite{Chen:2014rca,Poggi:2017kut, Nieri:2015dts} of the corresponding worldsheet theory.
The geometric picture is that one can acquire the $\mM^4$ manifold by glueing two solid $\mathbb{T}^3$ torus by  an SL$(3,\mathbb{Z})$ group element. 
This is known as the Heegaard decomposition. 
The group element then relates the complex structures of two solid $\mathbb{T}^3$ torus.
Such factorization structure can also be observed in $\mN=4$ SYM \cite{Goldstein:2020yvj}.  
It is interesting to know whether such factorization structure can survive in the one-loop level with the Spin Matrix theory coupling turned on and also what is the corresponding effect on the 2d elliptic genera. 
Also following the spirit of \cite{Brunner:2016nyk, Longhi:2019hdh,Gadde:2020bov}, which studied the SL$(3,\mathbb{Z})$ modularity of 4d supersymmetric partition function, we can ask how the SL$(3,\mathbb{Z})$ modularity can be affected considering the Spin Matrix theory is not protected by supersymmetry.
The answers to these questions might shed light on the structures of 4d conformal field theories beyond supersymmetry.

\section*{Acknowledgements}

We thank Chi-ming Chang, Niels Obers, Gerben Oling and Huajia Wang for useful discussions. We acknowledge support from the Independent Research Fund Denmark grant number DFF-6108-00340 “Towards a deeper understanding of black holes with non-relativistic holography”. YL is also supported by the South African Research Chairs Initiative of the Department of Science and Technology and the National Research Foundation, and by the UCAS program of special research associate and by the internal funds of the KITS.

\appendix



\section{Properties of spherical harmonics and Clebsch-Gordan coefficients}
\label{app:CG_properties}

We refer the reader to the Appendices A and B of \cite{Baiguera:2020jgy}, where the following conventions are stated: decomposition of the fields into spherical harmonics on the three-sphere, expression of the Cartan charges in terms of the modes, list of the weights under $\SU(4)$ R-symmetry of all the letters in $\mathcal{N}=4$ SYM theory, interacting Hamiltonian of the theory.
In the present paper, we follow exactly the same notation.

Other details on the algebra and the quantum numbers of the letters will be discussed in Appendices \ref{app:su111} and \ref{sec:SU12algebra}.
The relevant terms from the interacting Hamiltonian are written in Section \ref{sect-sphere_red_examples} for all the specific cases considered.
In the following, we define the Clebsch-Gordan coefficients that give rise to interactions between spherical harmonics and find several relations between them: they are crucial to perform the explicit computations of the sphere reduction.


\subsection{Definition of the Clebsch-Gordan coefficients}

We report the explicit definitions of the Clebsch-Gordan coefficients, given {\sl{e.g.}} in \cite{Ishiki:2006rt}.
\beq
\mathcal{C}^{J_1 M_1}_{J_2 M_2; JM} = 
\sqrt{\frac{(2J+1)(2J_2+1)}{2J_1+1}}
C^{J_1 m_1}_{J_2 m_2; J m} C^{J_1 \tilde{m}_1}_{J_2 \tilde{m}_2; J \tilde{m}} \, ,
\label{eq:app_definitionClebsch_C}
\eeq
\beq
\begin{aligned}
\mathcal{D}^{J_1 M_1}_{J_2 M_2 \rho_2; JM \rho} & = 
(-1)^{\frac{\rho_2+\rho}{2} +1}
\sqrt{3(2J_2+1)(2J_2+2 \rho_2^2 +1)(2J+1)(2J+2 \rho^2 +1)} \\
& \times C^{J_1,m_1}_{Q_2,m_2; Q,m} C^{J_1, \tilde{m}_1}_{\tilde{Q}_2, \tilde{m}_2 ; \tilde{Q},\tilde{m}}
\begin{Bmatrix}
Q_2 & \tilde{Q}_2 & 1 \\
Q & \tilde{Q} & 1 \\
J_1 & J_1 & 0  
\end{Bmatrix} \, ,
\label{eq:app_definitionClebsch_D}
 \end{aligned}
\eeq
\beq
\begin{aligned}
 \mathcal{E}_{J_1 M_1 \rho_1; J_2 M_2 \rho_2; JM \rho} & =  \sqrt{6(2J_1+1)(2J_1 + 2 \rho_1^2 +1)(2J_2+1)(2J_2+2 \rho_2^2 +1)(2J+1)(2J+2 \rho^2 +1)} \\
& \times (-1)^{-\frac{\rho_1+\rho_2+\rho+1}{2}} 
\begin{Bmatrix}
Q_1 & \tilde{Q}_1 & 1 \\
Q_2 & \tilde{Q}_2 & 1 \\
Q & \tilde{Q} & 1 
\end{Bmatrix} 
\begin{pmatrix}
Q_1 & Q_2 & Q \\
m_1 & m_2 & m
\end{pmatrix}
\begin{pmatrix}
\tilde{Q}_1 & \tilde{Q}_2 & \tilde{Q} \\
\tilde{m}_1 & \tilde{m}_2 & \tilde{m}
\end{pmatrix} \, ,
\label{eq:app_definitionClebsch_E}
\end{aligned}
\eeq
\beq
\begin{aligned}
\mathcal{F}^{J_1 M_1 \kappa_1}_{J_2 M_2 \kappa_2; JM} = & (-1)^{\tilde{U}_1+U_2+J+\frac{1}{2}} \sqrt{(2J+1)(2J_2+1)(2J_2+2)}  \\
& \times C^{U_1, m_1}_{U_2, m_2; J,m} C^{\tilde{U}_1,\tilde{m}_1}_{\tilde{U}_2,\tilde{m}_2; J,\tilde{m}}
\begin{Bmatrix}
U_1 & \tilde{U}_1 & \frac{1}{2} \\
\tilde{U}_2 & U_2 & J
\end{Bmatrix} \, ,
\label{eq:app_definitionClebsch_F}
\end{aligned}
\eeq
\beq
\begin{aligned}
\mathcal{G}^{J_1 M_1 \kappa_1}_{J_2 M_2 \kappa_2; JM \rho} = & (-1)^{\frac{\rho}{2}} \sqrt{6(2J_2+1)(2J_2+2)(2J+1)(2J+2 \rho^2 +1)} \\
& \times C^{U_1,m_1}_{U_2,m_2; Q,m} C^{\tilde{U}_1, \tilde{m}_1}_{\tilde{U}_2, \tilde{m}_2 ; \tilde{Q},\tilde{m}}
\begin{Bmatrix}
U_1 & \tilde{U}_1 & \frac{1}{2} \\
U_2 & \tilde{U}_2 & \frac{1}{2} \\
Q & \tilde{Q} & 1 
\end{Bmatrix} \, ,
\label{eq:app_definitionClebsch_G}
\end{aligned}
\eeq
where we defined the quantities
\beq
U \equiv J + \frac{\kappa+1}{4} \spa
\tilde{U} \equiv J + \frac{1-\kappa}{4} \spa
Q \equiv J+ \frac{\rho(\rho+1)}{2} \spa
\tilde{Q} \equiv J + \frac{\rho(\rho-1)}{2} \, .
\label{eq:app_labels_harmonics}
\eeq
Properties of 9-j and 6-j Wigner symbols were used to write the coefficient $\mathcal{F}$ in this form, but the expression is still completely general.

The previous Clebsch-Gordan coefficients can be defined in terms of specific contractions of spherical harmonics on the three-sphere.
This representation is convenient to keep track of momentum conservation and also to obtain specific crossing relations between different kinds of harmonics ({\sl{e.g.}} scalar and vector ones), as we will show below.
Precisely, we have
\beq
\mathcal{C}^{J_1 M_1}_{J_2 M_2; JM} = \int_{S^3} d \Omega \, (\mathcal{Y}_{J_1 M_1})^* \, \mathcal{Y}_{J_2 M_2} \mathcal{Y}_{JM} \, , 
\label{eq:app_integral_repr_C}
\eeq
\beq
\mathcal{D}^{J_1 M_1}_{J_2 M_2 \rho_2; JM \rho} = \int_{S^3} d \Omega \, (\mathcal{Y}_{J_1 M_1})^* \, \mathcal{Y}^{\rho_2}_{J_2 M_2 i} \mathcal{Y}^{\rho}_{JM i} \, , 
\label{eq:app_integral_repr_D}
\eeq
\beq
\mathcal{E}_{J_1 M_1 \rho_1; J_2 M_2 \rho_2; JM \rho} = \int_{S^3} d \Omega \, \epsilon_{ijk} \mathcal{Y}^{\rho_1}_{J_1 M_1 i} \mathcal{Y}^{\rho_2}_{J_2 M_2 j} \mathcal{Y}^{\rho}_{JM k} \, , 
\label{eq:app_integral_repr_E}
\eeq
\beq
\mathcal{F}^{J_1 M_1 \kappa_1}_{J_2 M_2 \kappa_2; JM} = \int_{S^3} d \Omega \, (\mathcal{Y}^{\kappa_1}_{J_1 M_1 \alpha})^*  \, \mathcal{Y}^{\kappa_2}_{J_2 M_2 \alpha} \mathcal{Y}_{JM} \, , 
\label{eq:app_integral_repr_F}
\eeq
\beq
\mathcal{G}^{J_1 M_1 \kappa_1}_{J_2 M_2 \kappa_2; JM \rho} = \int_{S^3} d \Omega \, \sigma^i_{\alpha \beta} (\mathcal{Y}^{\kappa_1}_{J_1 M_1 \alpha})^*  \, \mathcal{Y}^{\kappa_2}_{J_2 M_2 \beta} \mathcal{Y}^{\rho}_{JMi} \, ,
\label{eq:app_integral_repr_G}
\eeq
where $\mathcal{Y}^*$ denotes the complex conjugate of the harmonics $\mathcal{Y}.$

Notice that these Clebsch-Gordan coefficients always contain three sets of indices $(J_i, M_i),$ with possible other labels $(\kappa_i, \rho_i)$ specifying the spinorial or vectorial component of the harmonics.
We will adopt the following convention: whenever any set of indices is higher, it means that the corresponding spherical harmonic appears with the complex conjugation, {\sl{e.g.}} we define
\beq
\mathcal{E}^{J_1 M_1 \rho_1}{}_{J_2 M_2 \rho_2; JM \rho} = \int_{S^3} d \Omega \, \epsilon_{ijk} (\mathcal{Y}^{\rho_1}_{J_1 M_1 i})^* \, \mathcal{Y}^{\rho_2}_{J_2 M_2 j} \mathcal{Y}^{\rho}_{JM k} \, .
\eeq


\subsection{General crossing relations}

We derive a general crossing relation that relates quadratic combinations of the $\mathcal{D}$ and $\mathcal{E}$ Clebch-Gordan coefficients, and it is used to compute the effective Hamiltonian in the limits analyzed in Section \ref{sect-sphere_red_examples}. 
The starting point is a particular product of four vector spherical harmonics
\begin{equation}
\label{eq:fourvecint}
I = \int d\Omega \, \epsilon_{ijk} \epsilon_{kln} \CY^{\rho_1}_{J_1M_1i} \CY^{\rho_2}_{J_2M_2j} \CY^{\rho_3}_{J_3M_3l}\CY^{\rho_4}_{J_4M_4n}\,.
\end{equation}
Since the block $\epsilon_{ijk}Y^{\rho_1}_{J_1M_1i}Y^{\rho_2}_{J_2M_2j}$ transforms as a vector under spatial rotations, we can decompose it into vector harmonics, with
\begin{equation}
V_k^{(12)} \equiv \epsilon_{ijk} \CY^{\rho_1}_{J_1M_1i}\CY^{\rho_2}_{J_2M_2j} \equiv \sum_{J,M,\rho} V^{(12),(\rho)}_{JM} \CY^\rho_{JMk}\,.
\end{equation}
The coefficients of the linear expansion are nothing but the Clebsch-Gordan coefficients $\mathcal{E},$ as defined in Eq.~\eqref{eq:app_definitionClebsch_E}. Indeed we find
\begin{equation}
V^{(12),(\rho)}_{JM} = \int d\Omega \, V_k^{(12)} (\CY^\rho_{JMk})^* = \int d\Omega \, \epsilon_{ijk}
(\CY^\rho_{JMk})^* \, \CY^{\rho_1}_{J_1M_1i} \CY^{\rho_2}_{J_2M_2j} = {\cal E}^{JM\rho}{}_{J_1M_1\rho_1;J_2M_2\rho_2}\,,
\end{equation}
as follows from the normalization of vector spherical harmonics. 
Consequently, we find for the integral \eqref{eq:fourvecint}
\begin{equation}
I = \int d \Omega \, V_k^{(12)} V_k^{(34)} = \sum_{J,M,\rho} {\cal E}^{JM\rho}{}_{J_1M_1\rho_1;J_2M_2\rho_2} {\cal E}_{JM\rho;J_3M_3\rho_3;J_4M_4\rho_4}\,.
\label{eq:app_first_identity_integralI}
\end{equation}
On the other hand, another way to write Eq.~\eqref{eq:fourvecint} is the following comes from the properties of the Levi-Civita symbol:
\begin{equation}
I = \int d\Omega \, \le \CY^{\rho_1}_{J_1M_1i} \CY^{\rho_2}_{J_2M_2j}\CY^{\rho_3}_{J_3M_3i} \CY^{\rho_4}_{J_4M_4j}-\CY^{\rho_1}_{J_1M_1i}\CY^{\rho_2}_{J_2M_2j}\CY^{\rho_3}_{J_3M_3j}\CY^{\rho_4}_{J_4M_4i} \ri \,.
\end{equation}
If we insert a complete set of scalar spherical harmonics,
we get
\beq
\begin{aligned}
I & = 
\sum_{J,M} \int d \Omega\,\CY_{JM} \CY^{\rho_1}_{J_1M_1i} \CY^{\rho_3}_{J_3M_3i} \int d\Omega'\, \CY_{JM} \CY^{\rho_2}_{J_2M_2j} \CY^{\rho_4}_{J_4M_4j}\\
& -\sum_{J,M} \int d \Omega\, \CY_{JM} \CY^{\rho_1}_{J_1M_1i} \CY^{\rho_4}_{J_4M_4i} \int d \Omega'\,\CY_{JM} \CY^{\rho_2}_{J_2M_2j} \CY^{\rho_3}_{J_3M_3j}\,.
\end{aligned}
\eeq
After recognizing the appearance of the ${\cal D}$-type Clebsch--Gordan coefficient defined in \eqref{eq:app_integral_repr_D}, and combining with Eq.~\eqref{eq:app_first_identity_integralI}, we find the relation
\beq
\begin{aligned}
\label{eq:e_d_relation}
& \sum_{J,M,\rho} {\cal E}^{JM\rho}{}_{J_1M_1\rho_1;J_2M_2\rho_2} {\cal E}_{JM\rho;J_3M_3\rho_3;J_4M_4\rho_4} = \\
& = \sum_{J,M} \le {\cal D}^{JM}_{J_1M_1\rho_1;J_3M_3\rho_3} {\cal D}_{JM;J_2M_2\rho_2;J_4M_4\rho_4} - {\cal D}^{JM}_{J_1M_1\rho_1;J_4M_4\rho_4} {\cal D}_{JM;J_2M_2\rho_2;J_3M_3\rho_3} \ri \,.
\end{aligned}
\eeq


\subsection{Crossing relations at saturated angular momenta}
\label{app-crossing_relations_saturated_momenta}

The equation \eqref{eq:e_d_relation} derived above relates Clebsch-Gordan coefficients for arbitrary values of the quantum numbers characterizing the harmonics.

From now on, we will consider instead crossing relations arising from particular saturations of the Clebsch-Gordan coefficients, specifically the assignments of quantum numbers given by Eq.~\eqref{eq:notation_index_CJ_CM}, \eqref{eq:notation_index_barCJ_barCM} and \eqref{eq:notation_index_hatCJ_hatCM}.
While all the special cases are characterized by an unspecified value of $ -J \leq m \leq J,$ there is a saturation of the $\tilde{m}$ eigenvalue and this information will be sufficient to provide non-trivial conditions between Clebsch-Gordan coefficients.

This analysis represents a generalization of Appendix C in \cite{Baiguera:2020jgy}, where the factor of $\SU(2)$ Clebsch-Gordan coefficient containing the unfixed momentum $m$ is collected in front of all the relevant expressions.
We will see that it is possible to express all the coefficients in terms of the purely scalar one, which reads
\beq
{\cal C}^{\CJ_1 \CM_1}_{\CJ_2 \CM_2; J m \tilde{m}} 
= \sqrt{\frac{(2J+1)(2J_1)!(2J_2+1)!}{(J_1+J_2-J)!(J_1 +J_2+J+1)!}}  \,
 C^{J_1 m_1}_{J_2 m_2;Jm}\,.
\label{eq:C_saturated_momenta}
\eeq
We specialize the definitions \eqref{eq:app_definitionClebsch_C}--\eqref{eq:app_definitionClebsch_G} to the case of saturated momenta. 
We define the following quantities
\beq
\Delta J \equiv J_1 - J_2 \, , \qquad
\Delta m \equiv m_1 - m_2 \, .
\label{eq:app_definition_DeltaJ_Deltam}
\eeq 
In particular, an important role is played by this specific
 $\SU(2)$ Clebsch-Gordan coefficient with saturated momenta
\begin{equation}
	C_{J_2 m_2;\Delta J,\Delta m}^{J_1 m_1} = (-1)^{2J_2-J_1+m_1}\sqrt{ \frac{(2J_2)! (2 \Delta J)!(J_1-m_1)!(J_1+m_1)!}{(2J_1)!(J_2-m_2)!(J_2+m_2)!(\Delta J-\Delta m)!(\Delta J+\Delta m)!}} \, ,
	\label{eq:CGmJexpliti}
\end{equation}
since it enters explicitly all the fundamental blocks defined in Section \ref{sect-sphere_red_examples}.

The following relations hold:
\begin{itemize}
\item Clebsch-Gordan coefficient $\mathcal{D}$
\begin{equation}
{\cal D}^{\CJ_1 \CM_1}_{\CJ_2 \CM_2; Jm\tilde{m}, \rho=-1} =  -\frac{i}{2} (J_1+J_2-J) \sqrt{\frac{(J+1+\Delta J)(J+1-\Delta J)}{J_2(J_2+1)(J+1)(2J+1)}}  
{\cal C}^{\CJ_1 \CM_1}_{\CJ_2 \CM_2; J m \tilde{m}}\,,
\label{eq:app_Dminus_saturated}
\end{equation}
\begin{equation}
{\cal D}^{\CJ_1 \CM_1}_{\CJ_2 \CM_2; J m \tilde{m}, \rho=1} =  \frac{i}{2} (J+J_1+J_2+2) \sqrt{\frac{(J+1+\Delta J)(J+1-\Delta J)}{J_2(J_2+1)(J+1)(2J+3)}} 
\mathcal{C}^{\CJ_1 \CM_1}_{\CJ_2 \CM_2;J+1, m \tilde{m}}\,.
\label{eq:app_Dplus_saturated}
\end{equation}
\item Clebsch-Gordan coefficient $\mathcal{E}$
\beq
{\cal E}_{JM,\rho=-1; \hat{\CJ}_1 \hat{\CM}_1}{}^{\hat{\CJ}_2 \hat{\CM}_2}
= (-1)^{m_1-J}
\sqrt{\frac{(J+1+\Delta J)(J+1-\Delta J)}{(J+1)(2J+1)}} \,
{\cal C}^{\CJ_2 \CM_2}_{\CJ_1 \CM_1; J m \tilde{m}}
\,,
\label{eq:app_Eminus_saturated}
\eeq
\beq
 {\cal E}_{JM,\rho=0;\hat{\CJ}_1 \hat{\CM}_1}{}^{\hat{\CJ}_2 \hat{\CM}_2}  = (-1)^{m+m_1-J+1}
\frac{|\Delta J|}{\sqrt{J(J+1)}}
 {\cal C}^{\CJ_2 \CM_2}_{\CJ_1 \CM_1; J m \tilde{m}}
\,,
\label{eq:app_E0_saturated}
\eeq
\beq
{\cal E}_{JM,\rho=1; \hat{\CJ}_1 \hat{\CM}_1}{}^{\hat{\CJ}_2 \hat{\CM}_2} =
(-1)^{m_1-J+1}
\sqrt{\frac{(J+1+\Delta J)(J+1-\Delta J)}{(J+1)(2J+3)}}
{\cal C}^{\CJ_2 \CM_2}_{\CJ_1 \CM_1; J+1, m \tilde{m}} \, .
\label{eq:app_Eplus_saturated}
\eeq
\item Clebsch-Gordan coefficient $\mathcal{F}$
\beq
\mathcal{F}^{\bar{\CJ}_1 \bar{\CM}_1}_{\bar{\CJ}_2 \bar{\CM}_2; J m \tilde{m}}  = 
\mathcal{C}^{\CJ_1 \CM_1}_{\CJ_2 \CM_2; J m \tilde{m}}
\eeq
\item Clebsch-Gordan coefficient $\mathcal{G}$
\beq
 \mathcal{G}^{\bar{\CJ}_1 \bar{\CM}_1}_{\bar{\CJ}_2 \bar{\CM}_2 ;J m \tilde{m}, \rho=-1} =
 i \sqrt{\frac{(J+\Delta J+1) (J-\Delta J+1)}{(J+1) (2 J+1)}}
   \mathcal{C}^{\CJ_1 \CM_1}_{\CJ_2 \CM_2; J m \tilde{m}} \, ,
\eeq
\beq
 \mathcal{G}^{\bar{\CJ}_1 \bar{\CM}_1}_{\bar{\CJ}_2 \bar{\CM}_2 ;J m \tilde{m}, \rho=1} =
 -i \sqrt{\frac{(J+\Delta J+1) (J-\Delta J+1)}{(J+1) (2 J+3)}}
   \mathcal{C}^{\CJ_1 \CM_1}_{\CJ_2 \CM_2; J+1, m \tilde{m}} \, .
\eeq
\end{itemize}
We explain the general technique to explicitly solve the sum over $J$ for the terms of the effective Hamiltonian mediated by the non-dynamical gauge field.
The method to solve the sums for the other interactions is directly explained in Section \ref{sect-sphere_red_examples}.


\subsubsection*{Interactions mediated by non-dynamical gauge field}

While there are many terms contributing to the result in Eq.~\eqref{eq:total_interaction_gauge_mediated_su122}, all the sums over intermediate quantum numbers that involve terms mediated by non-dynamical modes of the gauge field can be solved applying the same trick.
We explain in full generality the technique, and then we will show the explicit application in some specific examples; the other cases can be treated similarly.

The common feature of all the interactions mediated by the non-dynamical gauge field is that their contribution arises from evaluating the expression \eqref{eq:general_gauge_mediated_interaction_su12limits}.
In the case of terms involving only scalar fields, gauge fields or mixed products of them, it is supplemented by specific quartic interactions appearing in the original $\mathcal{N}=4$ SYM action.
This does not happen for fermions, since there are no such quartic terms in the action.

The formula \eqref{eq:general_gauge_mediated_interaction_su12limits} contains a sum over three contributions: the square of the current $j^{J m \tilde{m}}_0$ and the square of the current $j^{J m \tilde{m}}_{(\rho)},$ for the values $\rho= \pm 1.$
These currents are always linear in the Clebsch-Gordan coefficients.
It can be shown\footnote{We verified that this statement is true for all the blocks entering Eq.~\eqref{eq:general_gauge_mediated_interaction_su12limits}, but in this Appendix we will show the details only for some specific cases. The procedure is straightforward but repetitive.} that all such blocks can be put into the form
\beq
\begin{aligned}
 H_{\rm int} & =  \tr \left( \sum_{J,m,\tilde{m}} \frac{1}{8J(J+1)} |j_0^{Jm\tilde{m}}|^2 - \sum_{\rho=\pm1} \sum_{J,m,\tilde{m}}\frac{1}{2(\omega_{A,J}^2 - 4\tilde{m}^2)} |j_{(\rho)}^{Jm\tilde{m}}|^2\right)  = \\
& =  \sum_{J=J_{\rm min}(\rho)}^{J=J_{\rm max}(\rho)} \sum_{m, \tilde{m}}
\left[ ( f_{(-1)} (J_i, J) + f_{(0)} (J_i, J)) \, \mathcal{C}^{\CJ_1 \CM_1}_{\CJ_2 \CM_2; J m \tilde{m}} \mathcal{C}^{\CJ_4 \CM_4}_{\CJ_3 \CM_3; J m \tilde{m}} \right. \\ 
& \left. + f_{(1)} (J_i, J)  \, \mathcal{C}^{\CJ_1 \CM_1}_{\CJ_2 \CM_2; J+1, m \tilde{m}} \mathcal{C}^{\CJ_4 \CM_4}_{\CJ_3 \CM_3; J+1, m \tilde{m}}  \right]  
 \,  \tr \le [\mathcal{W}_{J_1 m_1}, \mathcal{W}_{J_2 m_2}  \rbrace  [ \mathcal{W}_{J_3 m_3}, \mathcal{W}_{J_4 m_4} \rbrace \ri \, .
\end{aligned}
\label{eq:app_structure_sums_terms_mediated_gauge_field}
\eeq
Here we denoted with $\mathcal{W} \in \lbrace \Phi, \Phi^{\dagger}, \zeta_a, \zeta^{\dagger}_a, A, A^{\dagger} \rbrace$ a generic field in the set of the dynamical modes \eqref{eq:def_dynamical_modes_su122}, with $[\cdot,\cdot \rbrace $ the (anti)commutator between fields (depending from the statistics) and with $f_{(-1)}, f_{(0)}, f_{(1)}$ some functions of the momenta $J, J_i.$
Specifically, the subscript refers to the contributions from $\rho= \pm 1,$ while the case $\rho=0$ corresponds to the square of the current $j^{J m \tilde{m}}_0 .$ 

The sums over $(m, \tilde{m})$ are trivial because they correspond to solving constraints in the form of delta functions arising from momentum conservation.
The endpoints of the sum over $J$ are determined from triangle inequalities and generically depend from the value of $\rho.$ 
This allows to apply the following trick to solve the sums:
\begin{enumerate}
\item Consider the term with $\rho=1$ in the sum and perform a shift $J \rightarrow J-1.$
This step has two consequences: the first one is to bring the corresponding quadratic term in the last line of Eq.~\eqref{eq:app_structure_sums_terms_mediated_gauge_field}
to the form
\beq
\mathcal{C}^{\CJ_1 \CM_1}_{\CJ_2 \CM_2; J m \tilde{m}} \mathcal{C}^{\CJ_4 \CM_4}_{\CJ_3 \CM_3; J m \tilde{m}} \, ,
\label{eq:app_common_prefactor_calC_terms}
\eeq
which is the same combination multiplying the functions $f_{(-1)}, f_{(0)}.$
The second consequence is the shift of the endpoints of summation for the term with $\rho=1.$
\item At this point we can collect the common factor \eqref{eq:app_common_prefactor_calC_terms} from all the terms in the argument of the sum \eqref{eq:app_structure_sums_terms_mediated_gauge_field}, which now involves the combination
\beq
\mathfrak{f} (J,J_i) \equiv f_{(-1)} (J_i, J) + f_{(0)} (J_i, J) + f_{(1)} (J_i, J+1) \, .
\label{eq:app_combination_functions_f_int_ham}
\eeq
\item Notice that for all the terms mediated by the non-dynamical modes of the gauge field in the interacting Hamiltonian, we obtain 
\beq
\mathfrak{f} (J,J_i) = 0 \, .
\eeq
Then all the non-vanishing part of the sums over $J$ can only arise from the boundaries, if the endpoints of summation after the shift of the $\rho=1$ term are not the same.
\end{enumerate}
There are few more remarks on this procedure.
The first caveat is that the case $\Delta J=0$ is special, because the lower boundary of summation is $J=0$ and the term corresponding to $\rho=0$ has a singular prefactor.
This implies that we need to start the corresponding summation from $J>0,$ but the rest of the procedure works in the same way.

The second remark is that for the terms involving four scalars, four gauge fields or two scalars and two gauge fields, the combination \eqref{eq:app_combination_functions_f_int_ham} does not vanish.
On the other hand, the $\mathcal{N}=4$ SYM action contains an additional quartic contribution for each of these terms.
After adding such terms, which also contain the quadratic combination \eqref{eq:app_common_prefactor_calC_terms}, we remarkably find that the improved function $\mathfrak{f}(J,J_i)$ vanishes. 
This allows to solve exactly these summations using the same strategy. 
We show how the procedure works in practice in few cases.

\textbf{Example 1: quartic scalar term.}
We start from the purely scalar interaction and we define the quantities
\begin{equation}
\label{eq:ACDdef}
{\cal A}_{J_2 m_2,J_3 m_3;Jm\tilde{m}}^{J_1 m_1,J_4 m_4} = \bigg(
1
+ \frac{( \omega_{J_1}+\omega_{J_2})( \omega_{J_3}+\omega_{J_4})}{4J(J+1)}\bigg) {\cal C}^{\CJ_1 \CM_1}_{\CJ_2 \CM_2; Jm\tilde{m}}{\cal C}^{\CJ_4 \CM_4}_{\CJ_3 \CM_3;J m \tilde{m}}
\end{equation}
\begin{equation}
{\cal B}_{J_2 m_2,J_3 m_3;Jm\tilde{m}\rho}^{J_1 m_1,J_4 m_4} = - \frac{16}{\omega_{A,J}^2 - 4\tilde{m}^2}\sqrt{J_2(J_2+1)J_3(J_3+1)}{\cal D}^{\CJ_1 \CM_1}_{\CJ_2 \CM_2; J m \tilde{m},\rho}{\cal \bar{D}}^{\CJ_4 \CM_4}_{\CJ_3 \CM_3; J m \tilde{m},\rho}\,.
\end{equation}
Simple algebraic manipulations and the application of Eqs.~\eqref{eq:app_Dminus_saturated}, \eqref{eq:app_Dplus_saturated} give rise to the relation
\begin{equation}
\label{eq:CDrelation}
 {\cal A}_{J_2 m_2,J_3 m_3;Jm\tilde{m}}^{J_1 m_1,J_4 m_4} 
 + {\cal B}_{J_2 m_2,J_3 m_3;J m \tilde{m},\rho=-1}^{J_1 m_1,J_4 m_4} + {\cal B}_{J_2 m_2,J_3 m_3;J-1,m \tilde{m},\rho=1}^{J_1 m_1,J_4 m_4} = 0 \,,
\end{equation}
valid for $J \geq 1$.
The combination in the left-hand side of \eqref{eq:CDrelation} is the same as in the interacting term \eqref{eq:Hphisu12}, except that the term with $\rho=1$ is shifted.
Using the definitions in Eq.~\eqref{eq:app_definition_DeltaJ_Deltam}, we need to 
distinguish two cases as follows:
\begin{itemize}
\item In the case $|\Delta J| < |\Delta m|,$ the lower bound of summation $J_\text{min}$  depends on $\rho$.
From the triangle inequalities, we find that when $\rho \ne 1,$  $J_\text{min} = |\Delta{m}|.$
When $\rho=1,$ the definition of the vector spherical harmonics implies instead that $J_\text{min} = |\Delta{m}|-1.$ 
Then the shift $J \rightarrow J-1 $ in the last term above gives
\begin{equation}
\sum_{J \geq |\Delta{m}|} \left(
{\cal A}_{J_2 m_2,J_3 m_3;J,\Delta m,\Delta J}^{J_1 m_1,J_4 m_4} + {\cal B}_{J_2 m_2,J_3 m_3;J,\Delta m,\Delta J,\rho=-1}^{J_1 m_1,J_4 m_4} + {\cal B}_{J_2 m_2,J_3 m_3;J-1,\Delta m,\Delta J,\rho=1}^{J_1 m_1,J_4 m_4}
\right) = 0,
\end{equation}
after applying Eq.~\eqref{eq:CDrelation}. 
\item In the case $|\Delta{J}| \geq |\Delta{m}|,$ we find that $J_\text{min} = |\Delta{J}|$ for all the terms in \eqref{eq:Hphisu12}. Shifting $J \to J-1$ thus only cancels the terms in the sum with $J > |\Delta{J}|$, leaving the final expression
\beq
\begin{aligned}
& \sum_{J \geq |\Delta{J}|} \left(
{\cal A}_{J_2 m_2,J_3 m_3;J,\Delta m,\Delta J}^{J_1 m_1,J_4 m_4} + {\cal B}_{J_2 m_2,J_3 m_3;J,\Delta m,\Delta J,\rho=-1}^{J_1 m_1,J_4 m_4} + {\cal B}_{J_2 m_2, J_3 m_3;J,\Delta m,\Delta J,\rho=1}^{J_1 m_1,J_4 m_4}
\right)\\
& = - {\cal B}_{J_2 m_2,J_3 m_3;|\Delta J|-1,\Delta m, \Delta J,\rho=1}^{J_1 m_1,J_4 m_4} \,,
\end{aligned}
\eeq
where we have used Eq.~\eqref{eq:CDrelation} once more.
\end{itemize}
The evaluation of the last term gives the result \eqref{eq:intermediate_result_purely_scalar_su122} when $\Delta J \ne 0.$
The limiting case $\Delta J=0$ works in the same way, except that the $\mathcal{A}$ term is singular in the lower bound of summation $J=0.$
For this reason, it must be excluded from the sum and we find that the result is 
\beq
{\cal B}_{J_2 m_2,J_3 m_3;J=0,m=0, \tilde{m}=0,\rho=-1}^{J_1 m_1,J_4 m_4} \, , 
\eeq
which corresponds to Eq.~\eqref{eq:intermediate_result_case0_purely_scalar_su122}.

This computation represents a generalization of the similar technique explained in Appendix C of \cite{Baiguera:2020jgy}.
In particular, when the eigenvalue of momenta saturates to $m_i = -J_i,$ we obtain Eq.(C.22) of that paper.

\textbf{Example 2: quartic term with gauge fields.}
As a second example of the procedure, we consider the case containing only gauge fields, which is a new contribution of the $\SU(1,2)$ near-BPS limits.

Along the lines of the purely scalar case, we define
\begin{equation}
{\cal M}_{J_2 m_2,J_3 m_3;Jm\tilde{m}}^{J_1 m_1,J_4 m_4} = \frac{(\omega_{A,J_1} + \omega_{A,J_2})(\omega_{A,J_3}+\omega_{A,J_4})}{4J(J+1)}{\cal D}^{\hat{\CJ}_1 \hat{\CM}_1}_{JM; \hat{\CJ}_2 \hat{\CM}_2} {\cal D}_{\hat{\CJ}_4 \hat{\CM}_4}^{JM; \hat{\CJ}_3 \hat{\CM}_3}\,,
\end{equation}
\beq
\begin{aligned}
{\cal N}_{J_2 m_2,J_3 m_3;Jm\tilde{m}\rho}^{J_1 m_1,J_4 m_4} 
= & - \left(|\rho|\frac{(\rho \omega_{A,J} + \omega_{A,J_1} + \omega_{A,J_2})(\rho \omega_{A,J} + \omega_{A,J_3} + \omega_{A,J_4})}{\omega_{A,J}^2 - 4m^2} - 1\right) \\
& \times {\cal E}_{JM\rho;\hat{\CJ}_2 \hat{\CM}_2}{}^{\hat{\CJ}_1 \hat{\CM}_1}{\cal E}^{JM\rho; \hat{\CJ}_3 \hat{\CM}_3}{}_{\hat{\CJ}_4 \hat{\CM}_4}\,,
\end{aligned}
\eeq
where now $\rho \in \lbrace -1,0,1 \rbrace$ because we already included the contribution from the quartic term in the gauge fields appearing in the original $\mathcal{N}=4$ SYM action, as explained in Section \ref{sect-su12limit}.

Using the expressions \eqref{eq:app_Eminus_saturated}, \eqref{eq:app_E0_saturated} and \eqref{eq:app_Eplus_saturated} for the saturated Clebsch-Gordan coefficients, we obtain by direct computation that
\begin{equation}
\label{eq:DErelation}
{\cal M}_{J_2 m_2,J_3 m_3;Jm\tilde{m}}^{J_1 m_1,J_4 m_4} + 
{\cal N}_{J_2 m_2,J_3 m_3;J-1, m \tilde{m},\rho=1}^{J_1 m_1,J_4 m_4}
+{\cal N}_{J_2 m_2,J_3 m_3;Jm\tilde{m},\rho=0}^{J_1 m_1,J_4 m_4}
+{\cal N}_{J_2 m_2,J_3 m_3;Jm\tilde{m},\rho=-1}^{J_1 m_1,J_4 m_4} = 0\,.
\end{equation}
The block in the left-hand side is precisely the same appearing in the interacting Hamiltonian containing only dynamical gauge fields, which is given by the sum of Eqs.~\eqref{eq:term1_interactions_su12}, \eqref{eq:term2_interactions_su12} and \eqref{eq:term3_interactions_su12}.
The only difference is that the term with $\rho=1$ is shifted.
In order to perform the sum over $J,$ we need to distinguish the two cases:
\begin{itemize}
\item When $|\Delta J| < |\Delta m|,$ after shifting $J \rightarrow J-1$ in the term with $\rho=1$ we find that the argument of the sum is precisely given by Eq.\eqref{eq:DErelation}, and the endpoints of summation coincide for all the terms.
Then the total result is vanishing.
\item When $|\Delta J| \geq |\Delta m|,$ the lower boundary of summation coincides between all the terms and it is $J_{\rm min}= |\Delta J|.$
After the shift $J \rightarrow J-1$ in the term with $\rho=1,$ and using Eq.~\eqref{eq:DErelation}, we find that the sum reduces to a single contribution
\beq
\hspace{-12mm}
\begin{aligned}
\sum_{J \geq J_\text{min} } & \le
{\cal M}_{J_2 m_2,J_3 m_3;Jm\tilde{m}}^{J_1 m_1,J_4 m_4} +
{\cal N}_{J_2 m_2,J_3 m_3;J\,m\tilde{m},\rho=1}^{J_1 m_1,J_4 m_4}
+{\cal N}_{J_2 m_2,J_3 m_3;Jm\tilde{m},\rho=0}^{J_1 m_1,J_4 m_4}
+{\cal N}_{J_2 m_2,J_3 m_3;Jm\tilde{m},\rho=-1}^{J_1 m_1,J_4 m_4} 
\ri = \\
& = - {\cal N}_{J_2 m_2,J_3 m_3;|\Delta{J}|-1, \Delta{m},\Delta{J},\rho=1}^{J_1 m_1,J_4 m_4} \, .
\end{aligned}
\eeq
\end{itemize}
When evaluating the last expression with $\Delta J \ne 0$ explicitly, we obtain precisely the result in Eq.~\eqref{eq:result_differentJ_quarticgauge_su12}.
The case $\Delta J=0$ is special because the $\mathcal{M}$ term is singular at the boundary value $J=0,$ and then its sum needs to start instead from $J>0.$ 
Taking this into account, we find that the only non-vanishing contribution comes from 
\beq
{\cal N}_{J_2 m_2,J_3 m_3;J=0,m=0, \tilde{m}=0,\rho=-1}^{J_1 m_1,J_4 m_4} 
+ {\cal N}_{J_2 m_2,J_3 m_3;J=0,m=0, \tilde{m}=0,\rho=0}^{J_1 m_1,J_4 m_4}  
\, , 
\eeq
giving precisely the result in Eq.~\eqref{eq:result_equalJ_quarticgauge_su12}.

\textbf{Example 3: quartic fermionic term.}
We conclude the set of the examples with the purely fermionic term.
In this case there isn't any quartic term appearing directly in the $\mathcal{N}=4$ SYM action, and the effective interaction is completely mediated by the non-dynamical modes of the gauge field.
We define the blocks
\bea
& \mathcal{P}^{J_1 m_1;J_4 m_4}_{J_2 m_2; J_3 m_3; J m \tilde{m}} = \frac{1}{8J(J+1)} \mathcal{C}^{\CJ_1 \CM_1}_{\CJ_2 \CM_2; J m \tilde{m}} \mathcal{C}^{\CJ_4 \CM_4}_{\CJ_3 \CM_3; J m \tilde{m}} \, ,  & \\
& \mathcal{Q}^{J_1 m_1;J_4 m_4}_{J_2 m_2; J_3 m_3; J m \tilde{m}, \rho=1} = -  \frac{1}{8(J+1)(2J+3)} \mathcal{C}^{\CJ_1 \CM_1}_{\CJ_2 \CM_2; J+1, m, \tilde{m}} \mathcal{C}^{\CJ_4 \CM_4}_{\CJ_3 \CM_3; J+1, m, \tilde{m}} \, , & \\
& \mathcal{Q}^{J_1 m_1;J_4 m_4}_{J_2 m_2; J_3 m_3; J m \tilde{m}, \rho=-1} =  -  \frac{1}{8(J+1)(2J+1)} \mathcal{C}^{\CJ_1 \CM_1}_{\CJ_2 \CM_2; J m \tilde{m}} \mathcal{C}^{\CJ_4 \CM_4}_{\CJ_3 \CM_3; J m \tilde{m}} \, . &
\eea
It can be shown that
\beq
\mathcal{P}^{J_1 m_1;J_4 m_4}_{J_2 m_2; J_3 m_3; J m \tilde{m}}  +
 \mathcal{Q}^{J_1 m_1;J_4 m_4}_{J_2 m_2; J_3 m_3; J-1, m, \tilde{m}, \rho=1} +
  \mathcal{Q}^{J_1 m_1;J_4 m_4}_{J_2 m_2; J_3 m_3; J m \tilde{m}, \rho=-1} = 0 \, .
  \label{eq:FGrelation_saturated_momenta}
\eeq
We recognize that, except for the shift of the $\rho=1$ term, the left-hand side of the previous expression corresponds to Eq.~\eqref{eq:purely_ferm_intermediate_su122}.
Distinguishing the different cases where we compare $\Delta J$ and $\Delta m,$ it is possible to show that a shift $J \rightarrow J-1$ in the term with $\rho=1$ leaves a non-vanishing remainder in the sum only when $|\Delta J| \geq |\Delta m|.$
The result is 
\beq
\begin{aligned}
& \sum_{J \geq J_{\rm min}}  \le \mathcal{P}^{J_1 m_1;J_4 m_4}_{J_2 m_2; J_3 m_3; J, \Delta m, \Delta J}  +
 \mathcal{Q}^{J_1 m_1;J_4 m_4}_{J_2 m_2; J_3 m_3; J, \Delta m, \Delta J, \rho=1} +
  \mathcal{Q}^{J_1 m_1;J_4 m_4}_{J_2 m_2; J_3 m_3; J, \Delta m, \Delta J, \rho=-1} \ri = \\
  & = - \mathcal{Q}^{J_1 m_1;J_4 m_4}_{J_2 m_2; J_3 m_3; \Delta J-1, \Delta m, \Delta J, \rho=1} \, .
\end{aligned}
\eeq
In particular, the explicit evaluation of this expression gives
\beq
- \mathcal{Q}^{J_1 m_1;J_4 m_4}_{J_2 m_2; J_3 m_3; \Delta J-1, \Delta m, \Delta J, \rho=1} =
 \frac{1}{8 |\Delta J| (2 |\Delta J|+1)} \mathcal{C}^{\CJ_1 \CM_1}_{\CJ_2 \CM_2; |\Delta J|, \Delta m, \Delta J} \mathcal{C}^{\CJ_4 \CM_4}_{\CJ_3 \CM_3; |\Delta J|, \Delta m, \Delta J} \, ,
\label{eq:result_sum_J_FGterm}
\eeq
which is nothing but Eq.~\eqref{eq:intermediate_result_purely_scalar_su122}.
The special case $\Delta J=0$ can be treated separately by taking into account the singularity of the function $\mathcal{P}$ in $J=0.$ We obtain as a result 
Eq.~\eqref{eq:intermediate_result_case0_purely_scalar_su122}.

As for the purely scalar term, this case is a generalization of a computation performed in Appendix C of \cite{Baiguera:2020jgy}.
When the momenta are saturated to $m_i=-J_i,$ we recover Eq. (C.51) of that paper.

The analysis of these three examples concludes the study of the fundamental blocks for the terms mediated by the non-dynamical gauge field.
The mixed terms can be studied using the same technique, and their sum over $J$ still reduces to a boundary term.

\section{$\SU(1,1|1)$ algebra}
\label{app:su111}

The letters in the free $\mN=4$ SYM theory can be well-formulated in terms of oscillators $\mba_\alpha,\mbb_{\dot{\alpha}}$ and fermionic oscillators $\mbc^a$ with $\alpha,\dot{\alpha}=1,2$ and $a=1,2,3,4$. The oscillators satisfy 
\begin{equation}\label{eq:oscillator224}
	[\mba_{\alpha},\mba_\beta^\dagger] = \delta_{\alpha\beta}, \quad [\mbb_{\dot{\alpha}}, \mbb_{\dot{\beta}}^\dagger] = \delta_{\dalpha\dbeta}, \quad \{\mbc_a^\dagger,\mbc_b \} =\delta_{ab} \, .
\end{equation}
The corresponding number operators for each oscillator are defined as \begin{equation}
	a^\alpha \equiv \mba_\alpha^\dagger \mba_\alpha, \quad b^{\dalpha} \equiv \mbb_{\dalpha}^\dagger \mbb_{\dalpha}, \quad c^a \equiv \mbc_a^\dagger \mbc_a \, .
\end{equation}
We are following the notations in \cite{Harmark:2007px,Beisert:2004ry}. 

In the SU$(1,1|1)$ subsector, the relevant oscillators are two bosonic harmonic oscillators $\mba_1,\mbb_1$ and one fermionic harmonic oscillator $\mbc_3$. The other number operators due to the decoupling condition are
\begin{equation}
a^2=b^2 =c^1=c^2=0, \quad c^4=1 \, .
\end{equation}
The corresponding central charge constraint is
\begin{equation}
	a^1 - b^1 + c^3 = -1 \, .
\end{equation}
The global algebra contains the $\SU(1,1)$ generators $L_{0,\pm}$, the U$(1)$ R-symmetry generator $R$
\begin{equation}
	L_0 = \frac{1}{2} ( 1 + \mba_1^\dagger \mba_1 + \mbb_1^\dagger \mbb_1 ) \spa L_+ = \mba_1^\dagger \mbb_1^\dagger \spa L_- = \mba_1 \mbb_1, \quad 	R = \frac{1}{2} \mbc_3^\dagger \mbc_3 \, ,
\end{equation}
and the fermionic generators given by the supercharges
\begin{equation}
	Q = \mba_1 \mbc_3^\dagger \spa S = \mbb_1 \mbc_3 \, .
\end{equation}
The full SU$(1,1|1)$ algebra can then be worked out by the oscillators \eqref{eq:oscillator224}
\begin{equation}
	\label{su111_algebra}
	\begin{array}{c} \ds
		[L_0 , L_\pm ] = \pm L_\pm \spa [L_-,L_+]=2L_0
		\\[3mm] \ds
		\{ Q, Q^\dagger \} = L_0 + R \spa \{ S, S^\dagger \} = L_0 - R \spa \{ S, Q \} = L_-
		\\[3mm] \ds
		[L_0,Q]=-\frac{1}{2}Q \spa [L_0,S]=-\frac{1}{2}S \spa[Q, L_+ ] = S^\dagger \spa [S, L_+ ] = Q^\dagger  \, .
	\end{array}
\end{equation}

States of SU$(1,1|1)$ include one scalar mode and one fermion mode, including their 
$d_1$
descendants. 
\begin{equation}
	|\Phi_n\rangle = \frac{1}{n!} (\mba_1^\dagger \mbb_1^\dagger )^n \mbc_3^\dagger \mbc_4^\dagger |0\rangle
	\spa
	|\psi_n\rangle = \frac{1}{\sqrt{n!(n+1)!}} (\mba_1^\dagger \mbb_1^\dagger )^n \mba_1^\dagger \mbc_4^\dagger |0\rangle \, .
\end{equation}
These define the modes of the fields which can be used to represent the Lie algebra generators. These are shown as \cite{Baiguera:2020jgy}
\begin{align}\label{eq:SU111-fields}
	\begin{split}
	L_0 &= \sum_{n=0}^\infty \text{tr} \left[\left(n+\frac{1}{2} \right) \Phi^\dagger_n \Phi_n + (n+1) \psi^\dagger_n \psi_n \right]  \\
	L_+ &= L_-^\dagger =  \sum_{n=0}^\infty \text{tr}  \left[ (n+1)\Phi_{n+1}^\dagger \Phi_n +\sqrt{(n+1)(n+2)} \psi^\dagger_{n+1} \psi_n \right] \,.
	\end{split}
\end{align}
The Lie algebra relations of \eqref{eq:SU111-fields} are realized by Dirac brackets. Typically, an (anti-)commutator for generators $\CG_i$ of the algebra 
$
	[\CG_1 , \CG_2 \} = \CG_3
$
is represented as
\begin{equation}
	\{ \CG_1 , \CG_2 \}_D = i \CG_3
\end{equation}
%

\section{$\SU(1,2)$ and $\SU(1,2|2)$ algebras}\label{sec:SU12algebra}

\subsection{A review of SU$(1,2)$}\label{appen:SU12}

We are focusing on the SU$(1,2)$ subsector and its supersymmetric extension $\SU(1,2|2)$ in the $\mN=4$ SYM theory. 
As shown in \cite{Harmark:2007px}, the su$(1,2)$ sectors are constrained to satisfy
\begin{equation}
	c^1=c^2=c^3=c^4=1, \qquad b^2=0 \, .
\end{equation}
It is convenient to use the oscillators to construct generators of SU$(1,2)$ algebra.   
The subset of the generators of PSU$(2,2|4)$ algebra  relevant in this subsector are \cite{bars1990unitary}
\begin{align}\label{eq:Basissu21oscila}
	\begin{split}
		L^1_{1}&=J_0 = \frac{1}{2} (\mba^\dagger_1 \mba_1 - \mba^\dagger_2 \mba_2), \qquad L^1_2 =J_+ =  \mba_1^\dagger \mba_2, \quad L_1^2=J_- = \mba_2^\dagger \mba_1 \, , \\
		D &= \frac{1}{2} (\mba_1^\dagger \mba_1 +\mba_2^\dagger \mba_2 + 2\mbb_1^\dagger \mbb_1 ) +1 \, , \\
		P_{11} &= \mba^\dagger_1 \mbb_1^\dagger, \qquad P_{21}= \mba^\dagger_2 \mbb_1^\dagger, \qquad K_{11} = \mba_1 \mbb_1, \qquad K_{21} = \mba_2 \mbb_1 \, .
	\end{split}
\end{align}
There are eight generators in total, which can be mapped to the Gell-Mann matrices as the matrix representations of SU$(3)$. 
To be explicit, the matrices are  
\begin{eqnarray}\nonumber
	&& L_1^1 = \frac{1}{2}\left(
	\begin{array}{ccc}
		1 & 0 & 0 \\
		0 & -1 & 0 \\
		0 & 0 & 0
	\end{array}
	\right)
	\quad 
	L_2^1 = \left(
	\begin{array}{ccc}
		0 & 1 & 0 \\
		0 & 0 & 0 \\
		0 & 0 & 0
	\end{array}
	\right)
	\quad 
	L_1^2 = \left(
	\begin{array}{ccc}
		0 & 0 & 0 \\
		1 & 0 & 0 \\
		0 & 0 & 0
	\end{array}
	\right) \\ \nonumber
	&& P_{11}   = \left(
	\begin{array}{ccc}
		0 & 0 & i \\
		0 & 0 & 0 \\
		0 & 0 & 0
	\end{array}
	\right)
	\quad 
	P_{21}  = \left(
	\begin{array}{ccc}
		0 & 0 & 0 \\
		0 & 0 & i \\
		0 & 0 & 0
	\end{array}
	\right)
	\quad 
	D = \frac{1}{2} \left(
	\begin{array}{ccc}
		1 & 0 & 0 \\
		0 & 1 & 0 \\
		0 & 0 & -2
	\end{array}
	\right) \\
	&&K_{21}  = \left(
	\begin{array}{ccc}
		0 & 0 & 0 \\
		0 & 0 & 0 \\
		0 & i & 0
	\end{array}
	\right)
	\quad 
	K_{11} = \left(
	\begin{array}{ccc}
		0 & 0 & 0 \\
		0 & 0 & 0 \\
		i & 0 & 0
	\end{array}
	\right) \, .
\end{eqnarray}
The generator $D$ is related to the dimension operator $D_0$ of psu$(2,2|4)$ by $D= D_0 + \bar{L}_{1}^{1}$
where
\begin{equation}
	D_0 = 1+ \frac{1}{2} (\mba_1^\dagger \mba_1  +\mba_2^\dagger \mba_2 +\mbb_1^\dagger \mbb_1  +\mbb_2^\dagger \mbb_2), \quad \bar{L}_{1}^{1} = \frac{1}{2} (\mbb^\dagger_1 \mbb_1 - \mbb^\dagger_2 \mbb_2) \, .
\end{equation}
Thus the $\mbb_2$ oscillator completely gets decoupled from this sector. 
The two Cartan generators of SU$(1,2)$ algebra \eqref{eq:Basissu21oscila} are the dimension operator $D$ and $J_0$ respectively.  

There are two independent Casimir operators in SU$(1,2)$ group. 
It is useful to parametrize these Casimirs as \begin{eqnarray}\label{eq:twoCasimir}
	C_2 &=&  p+q +\frac{1}{3} (p^2+pq+q^2) \, , \\
	C_3 &=& \frac{1}{27} (p-q)(p+2q+3) (q+2p+3)  \, .
\end{eqnarray}
The unitary representations of SU$(1,2)$ were thoroughly explored in \cite{AIHPA_1965__3_1_13_0,bars1990unitary}.
It was  shown that the corresponding representations are parametrized generically by  five quantum numbers, which are $(p,q)$ to parametrize the two Casimir operators $C_2,C_3$, the spin $j$ of SU$(2)$ subalgebra spanned by $J_{\pm,0}$, the dimension $\Delta$ as the eigenvalue of $D$   and the eigenvalue $\mfm$ of U$(1)$ generator $J_0$ in the SU$(2)$ subalgebra \footnote{The dimension $ \Delta$ is equivalent to $\frac{3}{2}y$ in \cite{bars1990unitary}. }. 
The $D$ is analogous to  the hypercharge $Y$ in SU$(3)$ QCD theory.

The $J_{\pm,0}$ form the SU$(2)$ subalgebra of SU$(1,2)$, and its operations on states $\ket{pq;j\mfm \Delta}$ follow the conventional expressions
\begin{eqnarray} \label{eq:opSU2}
	J_0 \ket{pq;j\mfm \Delta} &=&  \mfm \ket{pq;j\mfm \Delta} \, , \\ \label{eq:opSU2-2}
	J_\pm \ket{pq;j\mfm \Delta} &=&  \sqrt{j(j+1)-\mfm (\mfm \pm1)} \ket{pq;j(\mfm \pm 1) \Delta} \, .
\end{eqnarray} 
The operations of Casimir operators and $D$ are stated above. 
The other operators we need to specify are $P_{a1}$ and $K_{a1}$. As shown in \cite{bars1990unitary}, it is convenient to introduce  a cubic function 
\begin{equation}\label{eq:cubiccoeff}
	G(x) = (x-x_1) (x-x_2) (x-x_3) = x^3 -(C_2+1) x-C_3 \, ,
\end{equation}
such that the action of generators on  states can be written as 
\begin{eqnarray} \nonumber
	K_{\pm  \frac{1}{2}}\ket{pq;j\mathfrak{m} \Delta}  &=& \sqrt{\frac{G(j+\Delta/3+1)}{2j+2}} \sqrt{\frac{j\pm \mfm +1}{2j+1}} \ket{pq;\left( j+\frac{1}{2}\right), \left(\mfm \pm \frac{1}{2}\right), \Delta+\frac{3}{2}} \\ \nonumber 
	&\pm& \sqrt{\frac{-G(-j+\Delta/3)}{2j+1}} \sqrt{\frac{j\pm \mfm }{2j}} \ket{pq;\left( j-\frac{1}{2}\right), \left(\mfm \pm \frac{1}{2}\right), \Delta+\frac{3}{2}} \, , \\ \nonumber 
	K_{\pm  \frac{1}{2}}^\dagger\ket{pq;j\mfm \Delta}  &=& \mp \sqrt{\frac{-G(-j+\Delta/3-1)}{2j+2}} \sqrt{\frac{j\pm \mfm +1}{2j+1}} \ket{pq;\left( j+\frac{1}{2}\right), \left(\mfm \pm \frac{1}{2}\right), \Delta-\frac{3}{2}} \\ \label{eq:updownop}
	&+& \sqrt{\frac{G(j+\Delta/3)}{2j+1}} \sqrt{\frac{j\pm \mfm }{2j}} \ket{pq;\left( j-\frac{1}{2}\right), \left(\mfm \pm \frac{1}{2}\right), \Delta-\frac{3}{2}} \, ,
\end{eqnarray}
where $K_{\pm \frac{1}{2}} = (P_{11},P_{21})$ and $K_{\pm \frac{1}{2}}^\dagger = (K_{21},K_{11})$.
Now the remaining tasks are to use  \eqref{eq:updownop} to figure out the allowed parameter region for $p,q,j,\mfm,\Delta$ such that the representation is unitary.

The landscape of SU$(1,2)$ representation theory was systematically explored in \cite{bars1990unitary}. 
There are representations such that $p,q$ can be complex numbers, which is analogous to the continuous representation of SU$(1,1)$.
There are various versions of discrete series representation which are distinguished by whether $p$ or $q$ or $p+q$ are integers. 
To our  interest, we only need to focus on the integer series, where $(p,q)$ are simultaneously integers. 
The states in the integer representation are parametrized by the lowest spin and dimension $(j_0,\Delta_0)$ \cite{bars1990unitary}. 

\subsubsection*{Alternative basis of SU$(1,2)$ algebra}
Our aim is to formulate the representation theory for letters of the form $d_1^nd_2^k \mO$ including gauge fields, fermions and scalars. 
The $d_{1,2}$ are considered to be descendants of two spatial directions of the effective $2+1$ dimensional quantum field theory.
Therefore, we would like to have a basis of SU$(1,2)$ algebra such that the two spatial directions are treated equally.   
The algebra \eqref{eq:Basissu21oscila} is then reorganized into the following oscillator representations 
\begin{eqnarray} \nonumber
	L_0 &=& \frac{1}{2} (D+J_0)=\frac{1}{2} (1+\mba_1^\dagger \mba_1 +\mbb_1^\dagger \mbb_1), \quad L_+=P_{11} = \mba_1^\dagger \mbb_1^\dagger, \quad L_{-} = K_{11}=\mba_1 \mbb_1 \, , \\ \nonumber
	\tilde{L}_0 &=&  \frac{1}{2} (D-J_0)= \frac{1}{2} (1+\mba_2^\dagger \mba_2 +\mbb_1^\dagger \mbb_1), \quad \tilde{L}_+=P_{21} = \mba_2^\dagger \mbb_1^\dagger, \quad \tilde{L}_{-} = K_{11}=\mba_2 \mbb_1 \, , \\ \label{eq:newbasisforSU21}
	J_+ &=& \mba_1^\dagger \mba_2, \qquad J_- =\mba_2^\dagger \mba_1 \, .
\end{eqnarray}

The new basis \eqref{eq:Basissu21oscila} is written to treat SU$(1,2)$ as two coupled SU$(1,1)$ algebra, which are generated by $L_{\pm,0}$ and $\tilde{L}_{\pm,0}$ respectively. 
We use $\tilde{L}_{\pm,0}$ to emphasize it is not anti-holomorphic part since it does not commute with the untilded generators.
The Cartan generators are $L_0$ and $\tilde{L}_0$, while the $d_{1,2}$ operators are identified with $L_+$ and $\tilde{L}_+$ respectively.
Each of SU$(1,1)$ itself is a $1+1$ dimensional field theory, which can be acquired by compactifying the other spatial direction. 
We would like to show how these generators act on the descendant states in SU$(1,2)$ subsector.

\subsection{SU$(1,2)$ with $\mN=2$ supersymmetry}\label{sec:SU122-represen}

The main task of our work is studying the SU$(1,2|2)$ subsector of $\mN=4$ SYM theory, which indicates that $\mN=2$ supersymmetry is added into the global SU$(1,2)$ symmetry.
As explained in \cite{Harmark:2007px}, the subsector contains the following letters: a scalar $\Phi=Z$, fermions $\zeta_{1,2}=\bar{\chi}_{5,7}$ and the gauge component $\bar{F}_+$. We label them by $\mW_I$ (see \eqref{eq:letter-label}).
We define the states $\ket{n,k,\mW_I}$ as
\begin{align}\label{eq:gaugecomponent-nkdesc}
	\begin{split}
		\ket{d_1^nd_2^k \Phi} &=\ket{n,k,\mW_1} =  \frac{1}{\sqrt{n!k!(k+n)!}}  (\mba_1^\dagger\mbb_1^\dagger)^n (\mba_2^\dagger \mbb_1^\dagger)^k \mbc_3^\dagger \mbc_4^\dagger \ket{0} \, ,  \\
		\ket{d_1^nd_2^k \zeta_1} &=\ket{n,k,\mW_2^1} = \frac{1}{\sqrt{n!k!(n+k+1)!}} (\mba_1^\dagger \mbb_1^\dagger)^n   (\mba_2^\dagger \mbb_1^\dagger)^k \mbb_1^\dagger  \mbc_2^\dagger  \mbc_3^\dagger  \mbc_4^\dagger  \ket{0} \, , \\
		\ket{d_1^nd_2^k \zeta_2} &=\ket{n,k,\mW_2^2} = \frac{1}{\sqrt{n!k!(n+k+1)!}} (\mba_1^\dagger \mbb_1^\dagger)^n   (\mba_2^\dagger \mbb_1^\dagger)^k \mbb_1^\dagger  \mbc_1^\dagger  \mbc_3^\dagger  \mbc_4^\dagger  \ket{0} \, , \\
		\ket{d_1^nd_2^k \bar{F}_+} &= \ket{n,k,\mW_3} = \frac{1}{\sqrt{n!k!(n+k+2)!}} (\mba_1^\dagger \mbb_1^\dagger)^n   (\mba_2^\dagger \mbb_1^\dagger)^k (\mbb_1^\dagger)^2 \mbc_1^\dagger  \mbc_2^\dagger  \mbc_3^\dagger  \mbc_4^\dagger  \ket{0} \, ,
	\end{split}
\end{align}
such that unit normalization is ensured for all of these states. 
Then we get the following actions of the generators listed in  Eq.~\eqref{eq:newbasisforSU21} on the states:
\begin{align}\label{eq:SU21onstates}
	\begin{split}
		L_+ \ket{n,k,\mW_I} &= \sqrt{(n+1)(n+k+I)} \ket{n+1,k,\mW_I} \, , \\
		L_{-} \ket{n,k,\mW_I} &= \sqrt{n(n+k+I-1)} \ket{n-1,k,\mW_I} \, , \\
		L_0 \ket{n,k,\mW_I} &= \left( n+ \frac{k+I}{2}\right) \ket{n,k,\mW_I} \, , \\
		\tilde{L}_+ \ket{n,k,\mW_I} &= \sqrt{(k+1)(n+k+I)} \ket{n,k+1,\mW_I} \, , \\ 
		\tilde{L}_{-} \ket{n,k,\mW_I} &= \sqrt{k(n+k+I-1)} \ket{n,k-1,\mW_I} \, , \\
		\tilde{L}_0 \ket{n,k,\mW_I} &= \left( k+ \frac{n+I}{2}\right) \ket{n,k,\mW_I} \, , \\
		J_+ \ket{n,k,\mW_I} &= \sqrt{k(n+1)} \ket{n+1,k-1,\mW_I} \, , \\
		J_- \ket{n,k,\mW_I} &=\sqrt{n(k+1)} \ket{n-1,k+1,\mW_I} \, .
	\end{split}
\end{align}
This basis is symmetric under the action of two descendant generators $L_+$ and $\tilde{L}_+$. 
We will show later that the coefficients obtained from the actions of generators on the states can be mapped to specific representations by identifying $(p,q)$ in Eq.~\eqref{eq:updownop}.  

In additional to the bosonic SU$(1,2)$ generators in \eqref{eq:newbasisforSU21}, there are following supersymmetry transformation generators in this sector
\begin{align}\label{eq:supercharge-oscil}
	Q_i= \mba_1 \mbc_i^\dagger, \qquad \tilde{Q}_i =\mba_2 \mbc_i^\dagger, \qquad S_i= \mbb_1 \mbc_i, \qquad (i=1,2) \, .
\end{align}
The R-symmetry acts as SU$(2)$ subalgebra, formed by   \begin{equation}
	R_{ij} = \mbc_i^\dagger \mbc_j - \frac{1}{2} \delta_{ij} \sum_{k=1,2} \mbc_k^\dagger \mbc_k \, .
\end{equation}
The corresponding R-charges are defined to be the eigenvalues of $R_0$
\begin{equation}
	R_0 = \frac{1}{2} (\mbc_2^\dagger \mbc_2- \mbc_1^\dagger \mbc_1) \, .
\end{equation}
At the end we also have an U$(1)_r$ symmetry $G$ written as 
\begin{equation}\label{eq:GU(1)}
	G= -1+ \mba_1^\dagger \mba_1 +\mba_2^\dagger \mba_2-\mbb_1^\dagger \mbb_1
	+\frac{3}{2}(\mbc_1^\dagger \mbc_1 + \mbc_2^\dagger \mbc_2) \, .
\end{equation}

The Lie algebra structures can be then derived by the help of oscillator (anti)-commutators \eqref{eq:oscillator224}. 
The non-trivial anti-commutators involving the supercharges and their hermitian conjugate are 
\begin{align}\label{eq:anticommu-QtildeQ}
\begin{split}
\{ Q_i,Q_j^\dagger\} &= R_{ij} + \frac{1}{3} \delta_{ij} (4L_0-2\tilde{L}_0+G) \, ,\\
\{ \tilde{Q}_i,\tilde{Q}_j^\dagger\} &= R_{ij} + \frac{1}{3} \delta_{ij} (-2L_0+4\tilde{L}_0+G) \, , \\
\{ S_i,S_j^\dagger\} &=- R_{ji} + \frac{1}{3} \delta_{ij} (2L_0+2\tilde{L}_0-G) \, , \\
\{ Q_i,\tilde{Q}_j^\dagger\} &=  \delta_{ij} J_-, \qquad \{ \tilde{Q}_i,Q_j^\dagger\} =  \delta_{ij} J_+ \, .
\end{split}
\end{align}
The tilded and untilded anti-commutators do not decouple in \eqref{eq:anticommu-QtildeQ}, indicating the two SU$(1,1)$ subalgebras are interacting. 
The anti-commutators in  \eqref{eq:anticommu-QtildeQ} generate the whole SU$(2)$ subalgebra $\{ J_\pm,L_0-\tilde{L}_0 \}$.
The other SU$(1,2)$ generators are generated by 
\begin{align}\label{eq:anticommu-QS}
\begin{split}
& \{Q_i, S_j \} = \delta_{ij} L_{-}, \quad \{Q_i^\dagger, S_j^\dagger \} =\delta_{ij} L_+ \, , \\
& \{\tilde{Q}_i, S_j \} = \delta_{ij} \tilde{L}_{-}, \quad \{\tilde{Q}_i^\dagger, S_j^\dagger \} =\delta_{ij} \tilde{L}_+ \, .
\end{split}
\end{align}

\begin{figure}
	\centering
	\begin{subfigure}{.5\textwidth}
		\centering
		\includegraphics[trim=3cm 15cm 8cm 4cm,width=0.96\linewidth]{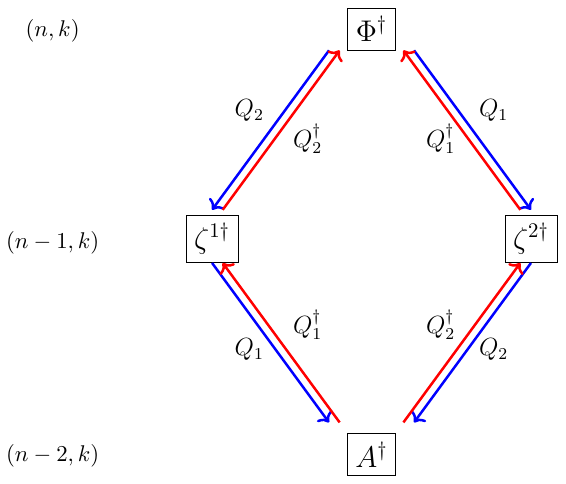}
		\caption{Actions by supercharge generator $Q_i$}
		\label{fig:R3-x-sub-letter}
	\end{subfigure}%
	\begin{subfigure}{.45\textwidth}
		\centering
		\includegraphics[trim=3cm 15cm 9cm 4cm,width=0.96\linewidth]{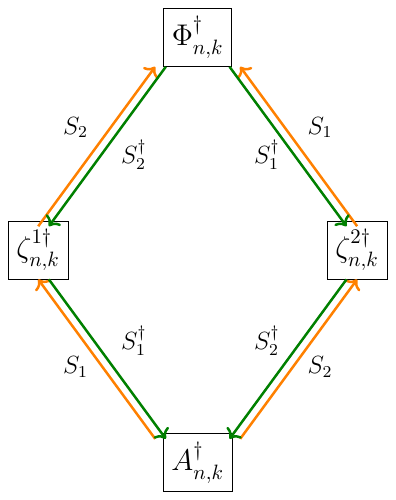}
		\caption{Actions by supercharge generator $S_i$}
		\label{fig:R2-x-sub-letter}
	\end{subfigure}
	\caption{\footnotesize Structures of letters as an $\mN=2$ supermultiplet.}
	\label{fig:diagram-letter}
\end{figure}

Based on the definition of states \eqref{eq:gaugecomponent-nkdesc} and supercharges \eqref{eq:supercharge-oscil}, the supercharge actions on states are 
\begin{align}\label{eq:SU122-super-onstates}
	\begin{split}
		Q_1^\dagger \ket{d_1^nd_2^k\zeta_2} &= \sqrt{n+1} \ket{d_1^{n+1}d_2^k\Phi}, \qquad
		Q_1^\dagger \ket{d_1^nd_2^k \bar{F}_+} = \sqrt{n+1} \ket{d_1^{n+1}d_2^k\zeta_1} \, , \\
		\tilde{Q}_1^\dagger \ket{d_1^nd_2^k\zeta_2} &= \sqrt{k+1} \ket{d_1^{n}d_2^{k+1}\Phi}, \qquad
		\tilde{Q}_1^\dagger \ket{d_1^nd_2^k\bar{F}_+} = \sqrt{k+1} \ket{d_1^{n}d_2^{k+1}\zeta_1} \, , \\ 
		S_1^\dagger \ket{d_1^{n}d_2^k\Phi} &= \sqrt{n+k+1} \ket{d_1^{n}d_2^k\zeta_2}, \qquad 
		S_1^\dagger \ket{d_1^{n}d_2^k\zeta_1} = \sqrt{n+k+2} \ket{d_1^{n}d_2^k\bar{F}_+} \, , \\
		Q_2^\dagger \ket{d_1^{n}d_2^k\zeta_1} &= \sqrt{n+1} \ket{d_1^{n+1}d_2^k\Phi},\qquad
		Q_2^\dagger \ket{d_1^{n}d_2^k \bar{F}_+} = - \sqrt{n+1} \ket{d_1^{n+1}d_2^k\zeta_2} \, , \\ 
		\tilde{Q}_2^\dagger \ket{d_1^{n}d_2^k\zeta_1} &= \sqrt{k+1} \ket{d_1^{n}d_2^{k+1}\Phi},\qquad
		\tilde{Q}_2^\dagger \ket{d_1^{n}d_2^k \bar{F}_+} = - \sqrt{k+1} \ket{d_1^{n}d_2^{k+1}\zeta_2} \, , \\
		S_2^\dagger \ket{d_1^{n}d_2^k\Phi} &= \sqrt{n+k+1} \ket{d_1^{n}d_2^k\zeta_1}, \qquad
		S_2^\dagger \ket{d_1^{n}d_2^k\zeta_2} = - \sqrt{n+k+2} \ket{d_1^{n}d_2^k \bar{F}_+} \, .
	\end{split}	
\end{align}
The $Q_i^\dagger$ are lifting the level of $d_1$ derivatives by $1$ while the $\tilde{Q}_i$ are lifting the level of $d_2$ derivatives. 
Besides, $S^\dagger$ generators do not affect the level of derivatives.
The corresponding actions on the letters are forming an $\mN=2$ vector multiplet shown in the Figure \ref{fig:diagram-letter}.

The set of U$(1)$ generators can be reorganized into the Dynkin labels $[s_1,r_1,q_1,r_2]$ as shown in \eqref{eq:Dynkin-U1Cartan}, which can be used to parametrize the representations of the supermultiplet.
We can thus specify the representation by the Dynkin labels of the highest weight state. These are 
\begin{align}\label{eq:Dynkin-SU122-Wi}
	\begin{split}
d_1^n d_2^k \Phi:\quad & L_0 = n+\frac{k+1}{2}, \quad \tilde{L}_0 = k+\frac{n+1}{2}, \quad R_0=0,\quad G=-1 \, , \\ 
d_1^n d_2^k \zeta^1:\quad & L_0 = n+\frac{k+2}{2}, \quad \tilde{L}_0 = k+\frac{n+2}{2}, \quad R_0=\frac{1}{2},\quad G=-\frac{1}{2} \, , \\
d_1^n d_2^k \zeta^2:\quad & L_0 = n+\frac{k+2}{2}, \quad \tilde{L}_0 = k+\frac{n+2}{2}, \quad R_0=-\frac{1}{2},\quad G=-\frac{1}{2} \, , \\
d_1^n d_2^k \bar{F}_+:\quad & L_0 = n+\frac{k+3}{2}, \quad \tilde{L}_0 = k+\frac{n+3}{2}, \quad R_0=0,\quad G=0 \, .
	\end{split}
\end{align}
Then the corresponding Dynkin labels are 
\begin{align}
\begin{split}
d_1^n d_2^k \Phi:\quad & [k-n,k,0,n+k+1] \, ,\\ 
d_1^n d_2^k \zeta^1:\quad &  [k-n,k,1,n+k+1] \, , \\
d_1^n d_2^k \zeta^2:\quad & [k-n,k+1,-1,n+k+2] \, , \\
d_1^n d_2^k \bar{F}_+:\quad & [k-n,k+1,0,n+k+2] \, .
\end{split}
\end{align}
The highest weight state annihilated by all the raising operator is the scalar $\Phi$. The corresponding Dynkin labels are $[0,0,0,1]$.

\subsection{Examples of integer representations of SU$(1,2)$}
\label{sec:SU12-rep-pq}
In the previous section, we have used the Dynkin labels to specify the representations of the letters of $\mN=4$ SYM in the SU$(1,2|2)$ subsector.
In this section, we would like to relate these letters to the representation theory of SU$(1,2)$ studied in the literature \cite{bars1990unitary}, which was reviewed in appendix \ref{appen:SU12}. 
Instead of Dynkin labels, we will study the representations where the letters of SU$(1,2|2)$ are transforming in terms of $(p,q)$ parameters. 
This is analogous to parametrize the representations of SU$(1,1)$ by the quantum number $j$. 
We will only consider the representations made by oscillators \cite{bars1990unitary}. 
Assume $(a,b)$ to be the number of $\mba,\mbb$ oscillators applied to construct the state, then the eigenvalues of $(J,D)$ will be easily read to be 
\begin{equation}
	(J,\Delta) = \left(\frac{a}{2}, \frac{a+2b+2}{2} \right) \, .
\end{equation}
The ground state operators in SU$(1,2|2)$ subsector do not have $\mba$ generators. Its corresponding dimension is thus denoted as $I$ satisfying $\Delta_0 = b+1 =I$. 
Comparing with the data in the integer representation \cite{bars1990unitary}, we can fix the $(p,q)$ to be 
\begin{equation}
	p=0, \quad q=I-3  \, .
\end{equation}
Furthermore, since all the descendants are generated by $L_+^n$ and $\tilde{L}_+^k$, one can read that there are in total $(n+k)$ more $\mba$ and $\mbb$ generators. The corresponding dimension and spin $J$ are thus 
\begin{equation}\label{eq:deltajnk}
	\Delta = \frac{3}{2}(n+k) + I, \quad j= \frac{a}{2} = \frac{n+k}{2} \, .
\end{equation} 
In the remaining part of this section, we will carefully compare the $(p,q)= (0,I-3)$ representations studied by \cite{bars1990unitary} and \eqref{eq:SU21onstates}. We will find they match exactly. 

\subsubsection*{$(p,q)=(0,I-3)$ representation}
We would like to work with the quantum numbers $(n,k)$ which take values on all the negative integers and represent the level of derivative generators $L_+,\tilde{L}_+$. The quantum numbers $(j,\mfm)$ of SU$(2)$ in \eqref{eq:opSU2}  are then rewritten as 
\begin{equation}\label{eq:jmfmnk}
	j= \frac{n+k}{2}, \quad \mfm= \frac{n-k}{2} \, .
\end{equation}
The corresponding action by $J_{\pm}$ are thus consistent with \eqref{eq:SU21onstates}.

The equations \eqref{eq:deltajnk} and \eqref{eq:jmfmnk} are adequate to match the $(p,q)=(0,I-3)$ representation of SU$(1,2)$ studied in \cite{bars1990unitary} with the symmetry actions in the new basis \eqref{eq:SU21onstates}. 
Then it is straightforward to check the actions by $L_0$ and $\tilde{L}_0$ are consistent with the definition in the new basis
\begin{equation}
	L_0 = \frac{1}{2} (	\Delta + \mfm) = n+\frac{k+I}{2}, \qquad \tilde{L}_0 = \frac{1}{2}(\Delta -\mfm) = k+\frac{n+I}{2} \, .
\end{equation} 

To check the actions by the others generators \eqref{eq:SU21onstates} are consistent with \eqref{eq:updownop}, we can first work out the polynomial $G(x)$ to be $$
G(x) = \left(x- \frac{I}{3} \right) \left(x+ \frac{2I}{3}-1 \right) \left(x+1 -\frac{I}{3}\right) \, .
$$
Then let's take the actions by $L_+$ as an example. We find 
\begin{align}
	\begin{split}
		\sqrt{\frac{G(j+\Delta/3+1)}{2j+2}} \sqrt{\frac{j+ \mfm +1}{2j+1}} &=  \sqrt{\frac{G(2j+\frac{I}{3}+1)}{2j+2}} \sqrt{\frac{j+ \mfm +1}{2j+1}} 
		= \sqrt{(n+k+I)(n+1)} \, .
	\end{split}
\end{align}
Actions by other generators can be extracted similarly from \eqref{eq:updownop}.
The other generators from \eqref{eq:SU21onstates} can be checked similarly. We can thus conclude the \eqref{eq:SU21onstates} is indicating the letter $\mW_I$ are transforming in the $(p,q)=(0,I-3)$ representations of SU$(1,2)$. 

\subsubsection*{Summary for field representations}
Just as the discrete representation of SU$(1,1)$ is parametrized by a half integer parameter $j$, the representations of SU$(1,2)$ are parametrized by $(p,q)$. As subsectors of $\mN=4$ SYM, the letters studied in this paper are all transforming in the integer representations \cite{bars1990unitary}. These are summarized below
\begin{itemize}
	\item \textbf{Gauge field}. 
	The gauge field $\bar{F}_+$ is parametrized by $(p,q)=(0,0)$ representation. 
	\item \textbf{Fermion}: The fermion $\zeta_a$ is parametrized by $(p,q)=(0,-1)$ representation. 
	\item \textbf{Scalar}: The scalar $\Phi$ is parametrized by $(p,q)=(0,-2)$ representation.
\end{itemize}
They together form an $\mN=2$ vector multiplet whose highest weight state is $\Phi$ with Dynkin labels $[0,0,0,1]$.

\subsection{Actions by SU$(1,2)$ symmetry}\label{sec:symmetry-action}

In this section, we would like to give an example showing how the action is invariant under the SU$(1,2)$ symmetry action. 
Without loss of generality, we will only consider the action by supercharges
in \eqref{eq:SU122-super-block}. 
Take $Q_1^\dagger$ as the example. 
Its action on the Hamiltonian is 
\begin{align}
\begin{split}
(Q_1^\dagger)_D H_{\text{int}} &= \frac{1}{2N} \sum_{n,k=0}^\infty \text{tr} \left[\Big((Q_1^\dagger)_D (B_2^\dagger)_{nk}  \Big)  (B_2)_{nk}  + 
\Big((Q_1^\dagger)_D (B_1^{1\dagger})_{nk}  \Big)  (B_1^1)_{nk} 
 \right]  \\
& + \frac{1}{2N} \sum_{n,k=0}^\infty \text{tr} \left[ (B_0^\dagger)_{nk}    \Big( (Q_1^\dagger)_D (B_0)_{nk}  \Big) + 
 (B_1^{2\dagger})_{nk}  \Big((Q_1^\dagger)_D  (B_1^2)_{nk}  \Big)
\right] \\
&=  \frac{1}{2N} \sum_{n,k=0}^\infty \text{tr} \left[-\sqrt{n+1} (B_1^{2\dagger})_{n+1,k} (B_2)_{nk} + \sqrt{n+1} (B_0^\dagger)_{n+1,k} (B_1^1)_{nk} \right]  \\
&+ \frac{1}{2N} \sum_{n,k=0}^\infty \text{tr} \left[-(B_0^\dagger)_{nk}  \sqrt{n}(B_1^{1\dagger})_{n-1,k}     + 
\sqrt{n}  (B_1^{2\dagger})_{nk}  (B_2)_{n-1,k}
\right] =0 \, .
\end{split}
\end{align}
The variation is exactly a telescopic sum and can be proven to vanish by shifting $n \to n-1$. 
The $Q_2^\dagger$ and $S^\dagger_i$ actions are similar. 
The supercharge $Q_1^\dagger$ is relating $(B_2^\dagger, B_1^{2\dagger})$ as vector multiplet and $(B_0^\dagger, B_1^{1\dagger})$ to be an $\mN=1$ chiral multiplet. 
The cancellation is happening between the corresponding $\mN=1$ supermultiplet. 

Since the bosonic SU$(1,2)$ generators \eqref{eq:SU12-generator-bysusy} are the anticommutators of supercharges, we can prove the invariance of Hamiltonian $H_{\text{int}}$ under SU$(1,2)$ generator actions simply by the group associativity.

\addcontentsline{toc}{section}{References}

\bibliography{newbib}
\bibliographystyle{newutphys}

\end{document}